\documentclass[11pt,a4paper]{article}

\usepackage{url,colortbl}
\usepackage{bm,amsmath,amssymb}
\usepackage[]{graphicx}
\usepackage[margin=2.5cm]{geometry}
\usepackage[ruled,linesnumbered]{algorithm2e}
\usepackage{todonotes}
\usepackage{authblk}
\graphicspath{{./Images/}}

\usepackage{xr}

\newcommand{\ind}{\mathbb{I}}

\begin{document}

\title{Bayesian uncertainty quantification for \\ data-driven equation learning}

\author{Simon Martina-Perez$^{1}$, Matthew J. Simpson$^{2}$ and Ruth E. Baker$^{1}$}

\affil[1]{Mathematical Institute, University of Oxford, Oxford, United Kingdom}
\affil[2]{School of Mathematical Sciences, Queensland University of Technology, Brisbane, Australia}
\date{}
\maketitle


\begin{abstract}
Equation learning aims to infer differential equation models from data. While a number of studies have shown that differential equation models can be successfully identified when the data are sufficiently detailed and corrupted with relatively small amounts of noise, the relationship between observation noise and uncertainty in the learned differential equation models remains unexplored. We demonstrate that for noisy datasets there exists great variation in both the structure of the learned differential equation models as well as their parameter values. We explore how to exploit multiple datasets to quantify uncertainty in the learned models, and at the same time draw mechanistic conclusions about the target differential equations. We showcase our results using simulation data from a relatively straight-forward agent-based model which has a well-characterised partial differential equation description that provides highly accurate predictions of averaged agent-based model behaviours in relevant regions of parameter space. Our approach combines equation learning methods with Bayesian inference approaches so that a quantification of uncertainty can be given by the posterior parameter distribution of the learned model.
\end{abstract}



\section{Introduction}

Many phenomena in nature arise as a result of complex interactions between individual agents at the microscale that give rise to emergent properties at the macroscale. Understanding the mechanistic basis for the observed macroscale behaviour, in order to gain fundamental insights into biological phenomena, is one of the key challenges in biology.

Mathematical models are well-placed to help provide such insights,  providing a rigorous framework where hypotheses can be generated, tested and refined. While interactions between individual agents can be naturally described by agent-based models (ABMs) that prescribe precise rules for the interactions between agents~\cite{Codling:2008:RWM,Othmer:1988:MDB,Othmer:1997:ABC,Plank:2012:MCC}, predicting the macroscale behaviour of ABMs can be a challenging task, since their governing equations are often intractable and stochastic simulations can be computationally expensive, often prohibitively so in the context of parameter sensitivity analysis or parameter inference~\cite{An:2017:OAC,Keeling:2009:EMS,Marino:2008:AMF,nardini2020learning,vanderVaart:2015:CAE}. This makes differential equation models an indispensable tool to describe the expected macroscale properties of the population. The benefits of differential equation models  include the fact that they are relatively fast to solve numerically, their different terms often carry a physical interpretation, and they can be explored using a range of analytical and numerical approaches. Understanding how such a model can be parametrised, then, can provide key insights into the system under consideration, and aid in making quantitative as well as qualitative predictions~\cite{Liepe:2014:AFF}.

Traditional approaches to mathematical modelling use experimentally-derived mechanistic hypotheses to derive differential equation models in which the various terms of a given model are designed to describe the hypothesised mechanisms for that scenario. Calibration of the model to data then involves finding the parameters that minimise the discrepancy between the model output and data. The ensuing, iterative process of testing and refining the model against further experimental data allows the original hypotheses to be refined, and so new insights gained. 

Equation learning methods take a different approach to model building, aiming to infer the dynamical systems model that best describes given time series data by leveraging statistical and machine learning tools to learn the appropriate terms of a differential equation model directly from the data. In particular, the PDE-FIND algorithm \cite{Rudy17,Silva:2020:DOP} takes as input quantitative data, together with a library of candidate terms for a partial differential equation (PDE) model, and aims to learn which terms to include in the PDE model, as well as their coefficients. Algorithm hyperparameters can be tuned to enable a balance between the requirement for a good model fit with the desire for a simple, interpretable model. 

Equation learning methods have rapidly gained popularity, mainly thanks to increases in computational power, and a number of other techniques to establish models from data now exist. For example, biologically informed neural networks~\cite{BINNs}, an extension of physically informed neural networks~\cite{Raissi:2019:PIN}, have been developed to learn the different terms of a PDE model without the need to specify a library of possible terms. Furthermore, a major advance has come from the use of techniques such artificial neural networks (ANNs)~\cite{ANNs} to accurately recover models from artificially generated noisy data from PDEs. 

\enlargethispage{0.2cm}

The fact that equation learning can discover previously undetected mechanisms, discriminate between competing models, or estimate biological quantities of interest that are difficult to measure experimentally, makes equation learning attractive to scientists working with real-world data. However, practitioners wishing to develop models that they can use in real-world settings require, in addition, a thorough quantification of uncertainty~\cite{Liepe:2014:AFF,Eriksson:2019:UQP,Arriola:2009:SAF,Kirk:2015:SBU,Komorowski:2011:SRA}. This need comes from the fact that the, often significant, noise in real-world data can impact the models predicted by equation learning methods, and hence the predictive capability of the models for unseen data or scenarios~\cite{Thijssen:2018:BDI,Toni:2010:SBM}. For example, Nardini \textit{et al.}~\cite{nardini2020learning} have recently shown, through the use of several case studies, that it is possible to infer differential equation models that describe noisy data generated by stochastic ABMs. However, the stochasticity in the ABM results in variability in the learned macroscale differential equation. This means that, for a particular realisation generated from a stochastic model, the learned differential equation is a point estimate of the underlying differential equation, and there is no quantification of uncertainty in the learned equation. 

Recently, some authors~\cite{Rudy17,rPCA} have begun to address this problem by analysing the robustness of PDE-FIND with noisy or sparse data. For example, Rudy \textit{et al.}~\cite{Rudy17} investigate how the learned PDE varies as the numerical solution of a ground truth PDE is corrupted by additive noise, while Li \textit{et al.}~\cite{rPCA} investigate how to increase the signal-to-noise ratio of a dataset prior to the use of equation learning techniques. While both works show that model parameters can be retrieved to within an impressive margin of error when the data are corrupted with relatively small amounts of noise, these approaches give no statistical quantification of the uncertainty in model predictions, nor do they address how to deal with significant noise levels. 

\begin{figure}[htbp]
    \centering
    \includegraphics[width=0.9\linewidth]{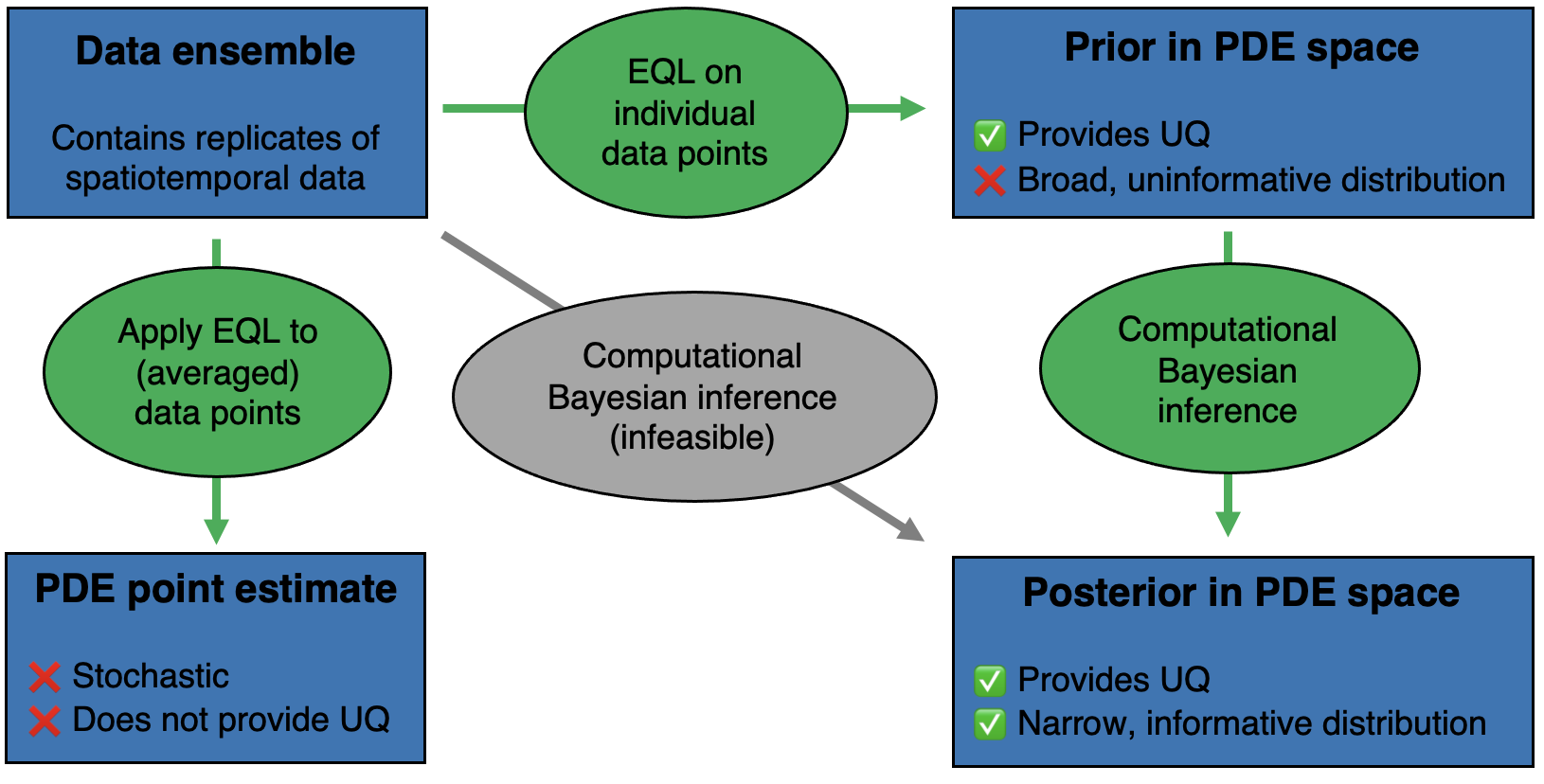}
    \caption{Proposed framework for uncertainty quantification in equation learning (EQL). An ensemble of noisy datasets are used to create an informative prior distribution, which can then be used to obtain a posterior parameter distribution for the learned PDE.}
    \label{PDE_diag}
\end{figure}

In this work we demonstrate that noise can significantly impact both the structure and the values of the parameters of learned differential equation models, rendering uncertainty quantification a crucial component of the equation learning process. As such, our overarching aim is to develop and showcase a method for uncertainty quantification in the context of equation learning, where we harness the immense computational efficiencies of PDE-FIND in learning point estimates of governing equations, together with the power of computational Bayesian inference in evaluating the level uncertainty in the learned equation. Figure~\ref{PDE_diag} shows our proposed framework for uncertainty quantification. We start from the basis that it is possible to collect an ensemble of spatiotemporal datasets from a given system, and develop an approach to understand how the data can be used to learn a governing equation while simultaneously estimating the uncertainty in that learned equation. 


\enlargethispage{0.2cm}

Our motivation is thus: on the one hand, PDE-FIND provides a computationally cheap method to obtain a point estimate for the governing equation from a single time series. However, when the data are noisy, the individual predictions are unreliable. This is seen on the left-hand side of Figure~\ref{PDE_diag}, where this approach is identified with obtaining a point estimate of the PDE. On the other hand, the field of computational Bayesian inference provides a number of methods to estimate, for a given model and data, posterior parameter distributions \textit{i.e.} it provides estimates of model parameters and quantifies the uncertainty in those estimates. In principle, computational Bayesian inference approaches could be used directly with the candidate library of the PDE-FIND method to estimate the posterior distribution of the library coefficients. However, due to the very large number of candidate terms in the PDE-FIND library, the computational cost associated with applying methods for computational Bayesian inference on the entire high-dimensional parameter space is generally prohibitive. Instead, we propose a framework that combines the strengths of each approach: first, we train the PDE-FIND algorithm on individual datasets from the ensemble to obtain an informed prior parameter distribution (top row of Figure~\ref{PDE_diag}). While this prior distribution will likely be relatively broad and uninformative of the uncertainty in the model, it can still be used to vastly reduce the dimensionality of the inference problem. Such a reduction in dimensionality makes it feasible to find an informative posterior distribution using computational Bayesian approaches (right-most column of Figure~\ref{PDE_diag}). 

In this work, we demonstrate the potential of this approach using synthetic data generated from a widely used ABM that describes the behaviour of a motile and proliferative cell population and can be coarse-grained to a mean-field PDE that accurately describes ABM dynamics in certain regions of parameter space. In Section~\ref{section:Methods}, we describe the ABM and discuss in detail its relation with a governing PDE, as well as the PDE-FIND algorithm. In Section~\ref{section:variability}, we demonstrate both that the PDE models learnt using PDE-FIND are intrinsically variable when the data are noisy, and that PDE-FIND can learn unphysical models. We provide an explanation for this in terms of the objective function of the PDE-FIND algorithm. In Section~\ref{section:BayesPDEFIND} we propose a method to combine methods for Bayesian inference with PDE-FIND in order to learn the structure of the governing PDE model, construct a prior parameter distribution for the PDE model, and infer the posterior parameter distribution of the learned PDE model. We conclude in Section~\ref{section:discussion} with a discussion of our results, and avenues for future research. Code for all algorithms can be found at \url{https://github.com/simonmape/UQ-for-pdefind}.


\section{Models and equation learning methodology}
\label{section:Methods}

We begin by describing the ABM and briefly outlining how to derive the corresponding coarse-grained PDE model, and then we provide details of the PDE-FIND algorithm.


\subsection{Agent-based model}

ABMs allow practitioners to investigate the collective behaviour of a population of individuals based on a description of the behaviour of individuals within that population. Here, in order to take into account the interactions between individuals of the population, we follow the volume-exclusion model presented in~\cite{Plank:2012:MCC,StochWound,Folks,Simpson:2007:SIC,Simpson:2009:MSS} for a population of agents that move and proliferate according to a discrete random walk model. This is a simple model that can be used to analyse a range of phenomena, including the collective migration of cells in a tissue, for example. 

We assume that agents occupy sites on a square lattice of spacing $\Delta$, so that their possible locations are $(i\Delta,j\Delta)$, where $(i,j)$ are integer coordinates, and volume exclusion entails that at most one agent can occupy a lattice site at any given time. We have $1\leq i\leq I$ and $1\leq j\leq J$, and throughout this work we take $I=200$ and $J=20$. A pseudo-one-dimensional initial condition is taken by initially populating all lattice sites with $90 \leq i \leq 110$, and leaving the rest of the lattice empty. We impose zero flux boundary conditions at all boundaries so that agents cannot leave the lattice, and use time step $\tau$ to advance the simulations through time, with $T=1000$ time steps in total for each simulation. Let $N(t)$ to denote the number of agents on the lattice at time $t$. The parameter $p_m\in[0,1]$ specifies the attempted movement probability of each agent in a time interval of duration $\tau$, and $\rho\in[-1,1]$ the left-right bias in movements. Similarly, the parameter $p_p\in[0,1]$ specifies the attempted proliferation probability of each agent in a time interval of duration $\tau$.

At each time step, $\tau$, a random sequential updating procedure is carried out: $N(t)$ agents are selected, one at a time, with replacement, and are allowed to attempt a movement or proliferation event. When an agent is selected, $S_1\sim U(0,1)$ is drawn. If $S_1 \leq p_p$ then the agent attempts to proliferate by placing a daughter agent into one of the randomly chosen nearest neighbour sites. If the target site is occupied then the proliferation event is aborted. If $p_p<S_1 \leq p_p+p_m$ then the agent attempts to move to one of its nearest neighbour lattice sites. A second random number $S_2\sim U(0,1)$ is drawn and the target site is chosen according to the rules in Table~\ref{move_table}. As for proliferation, if the target site is occupied then the movement event is aborted. If $S_1 > p_p+p_m$ then the agent does not attempt to move or proliferate. For convenience we take $\Delta=1$, $\tau=1$ and $p_m =1$ and consider the effects of varying $p_p$ and $\rho$. 


\bigskip
\begin{table}[h!]
	\centering
	\begin{tabular}{|c|c|c|c|} 
		\hline
		Move chosen & Target site & Probability & Where random number $S_2$ falls\\ [0.5ex] 
		\hline\hline
		vertically down & $(i,j-1)$ & $\frac{1}{4}$ & $0 \leq S_2 \leq \frac{1}{4}$\\
		\hline
		vertically down & $(i,j+1)$ & $\frac{1}{4}$ & $\frac{1}{4}\leq S_2 \leq \frac{1}{2}$\\\hline
		horizontally left & $(i-1,j)$ & $\frac{1-\rho}{4}$ & $\frac{1}{2} \leq S_2 \leq \frac{1}{2}+\frac{1-\rho}{4}$\\\hline
		horizontally right & $(i+1,j)$ & $\frac{1+\rho}{4}$ & $\frac{1}{2} + \frac{1-\rho}{4} \leq S_2 \leq 1$ \\
		\hline\hline
	\end{tabular}
	\caption{Algorithm by which an agent at site $(i,j)$ selects a target site to move into.}
	\label{move_table}
\end{table}


Let $C_{ij}^k(t)$ denote the occupancy of site $(i,j)$ at time $t$ in simulation $k$, so that $C_{ij}^k(t)=1$ if $(i,j)$ is occupied by an agent at time $t$ and $C_{ij}^k(t)=0$ if it is empty. We can average the site occupancy over the columns of the lattice, defining the mean occupancy of column $i$ at time $t$ in simulation $k$ as, for $1\leq{i}\leq{I}$, 
\begin{equation}
C^k_i(t) = \frac{1}{J}\sum_{j=1}^J C_{ij}^k(t),
\end{equation}
to give a one-dimensional averaged agent density profile for simulation $k$.


\subsubsection{Coarse-grained PDE model}

To make progress in deriving a coarse-grained PDE equivalent, we first note that the choice of a pseudo-one-dimensional initial condition means that we can consider deriving a one-dimensional PDE for $c(x,t)$, the density of agents at position $x$ at time $t$, without making explicit reference to the spatial coordinate $y$. We use $\langle{}C_i(t)\rangle{}$ to denote the average probability of occupancy of lattice site $i$ at time $t$, for $1\leq{i}\leq{I}$, where the average is taken over $K$ simulations:
\begin{equation}
\label{equation:Caverage}
\langle{}C_i(t)\rangle{}=\frac{1}{K}\sum_{k=1}^{K}C_i^k(t) = \frac{1}{J\cdot{K}}\sum_{k=1}^{K}\sum_{j=1}^J C_{ij}^k(t).
\end{equation}
We now consider the change in average occupancy of site $i$ over a time step of duration $\tau$ to write: 
\begin{eqnarray}
\langle{}C_i(t+\tau)\rangle{}-\langle{}C_i(t)\rangle{}&=&
\notag
\frac{(1+\rho)}{4}p_m\langle{}C_{i-1}(t)\rangle{}\left(1-\langle{}C_i(t\right)\rangle{})+ 
\frac{(1-\rho)}{4}p_m\langle{}C_{i+1}(t)\rangle{}\left(1-C_i(t)\right)\\
&&
\notag
-\frac{(1+\rho)}{4}p_m\langle{}C_i(t)\rangle{}\left(1-\langle{}C_{i+1}(t)\rangle{}\right)-
\frac{(1-\rho)}{4}p_m\langle{}C_i(t)\rangle{}\left(1-\langle{}C_{i-1}(t)\rangle{}\right)\\
&&
+\frac{1}{2}p_p\langle{}C_{i-1}(t)\rangle{}\left(1-\langle{}C_i(t)\rangle{}\right)+
\frac{1}{2}p_p\langle{}C_{i+1}(t)\rangle{}(t)\left(1-\langle{}C_i(t)\rangle{}\right),
\end{eqnarray}
where the first four terms on the right-hand side correspond changes in occupancy owing to agent movement, and the final two to agent proliferation. Note that in writing down this conservation statement, we have implicitly assumed that lattice site occupancies are independent, so that \textit{e.g.} the average probability that site $i$ is occupied and site $i\pm1$ is unoccupied can be written as $\langle{}C_i(t)\rangle{}\left(1-\langle{}C_{i\pm1}(t)\rangle{}\right)$. This is a standard assumption called the mean-field approximation~\cite{Baker:2010:CMF,Murrell:2004:OMC,Raghib:2011:MME}. 

We then identify $\langle{}C_i(t)\rangle{}$ with the continuous density $c(x,t)$, Taylor expand the resulting equation and take limits as $\Delta,\tau\to0$, to arrive at the following PDE:
\begin{equation}
c_t = Dc_{xx} - V[c(1-c)]_x + P [c(1-c)],
\label{mean-field}
\end{equation}
where 
\begin{equation}
D=\lim_{\Delta,\tau\to0} \frac{p_m\Delta^2}{4\tau},
\qquad
V=\lim_{\Delta,\tau\to0}\frac{p_m\Delta\rho}{2\tau}
\qquad\text{and}\qquad
P=\lim_{\tau\to0}\frac{p_p}{\tau}, 
\end{equation}
and the subscripts $x$ and $t$ denote partial derivatives. For the full derivation and details, we refer the reader to~\cite{Folks,Simpson:2007:SIC,Simpson:2009:MSS}. 

Identification of the ABM with a coarse-grained macroscale PDE model motivates us to investigate the performance of equation learning methods trained on data generated by the ABM, since the PDE accurately describes the time-evolution of the expected value of the density profile, and so we can evaluate the performance of equation learning methods against Equation~\eqref{mean-field}. Note that in order for Equation~\eqref{mean-field} to provide an accurate description of the averaged dynamics of the ABM, we require that the assumption of lattice-site occupancy independence (\textit{i.e.} the mean-field assumption) approximately holds. Typically, this requires $p_{p}$ and $|\rho|$ to be small relative to $p_{m}$~\cite{StochWound,Baker:2010:CMF,Simpson:2011:CMF}. 


\subsubsection{Comparison of the ABM and PDE model predictions}\label{section:cases}

As test cases for learning the governing equations from data, we explore three different parameter regimes in the model, which each correspond to a biologically relevant setting: in Case I, we consider agents moving without bias and without proliferation ($\rho=p_p=0$); in Case II, we consider agents moving with bias but without proliferation ($\rho=0.075$ and $p_p = 0$); and in Case III, we consider agents moving without bias but with proliferation ($\rho=0$ and $p_p = 0.001$). Table~\ref{table:cases} outlines these different cases, along with a statement of the corresponding coarse-grained PDE model.


\begin{table}[h!]
\centering
\begin{tabular}{|c||c|c|c|c|} 
\hline
Case & $p_{m}$ & $\rho$ & $p_p$ & Coarse-grained PDE\\ [0.5ex] 
\hline\hline
Case I: no bias, no proliferation & \cellcolor[gray]{0.7}\textbf{1.0} & 0.0 & 0.0 & $c_t = 0.25 c_{xx}$\\
\hline
Case II: bias, no proliferation & \cellcolor[gray]{0.7}\textbf{1.0} & \cellcolor[gray]{0.7}\textbf{0.075} & 0.0 &$c_t = 0.25 c_{xx} -0.0375 [c(1-c)]_x$  \\
\hline
Case III: proliferation, no bias & \cellcolor[gray]{0.7}\textbf{1.0} & 0.0 & \cellcolor[gray]{0.7}\textbf{0.01} & $c_t = 0.25 c_{xx} + 0.01[c(1-c)]$\\
\hline\hline
\end{tabular}
\caption{The mean-field PDE models describing evolution of the mean population density over time for the three example cases used in this work.}
\label{table:cases}
\end{table}

\begin{figure}[htbp]
\centering
\includegraphics[width=0.85\textwidth]{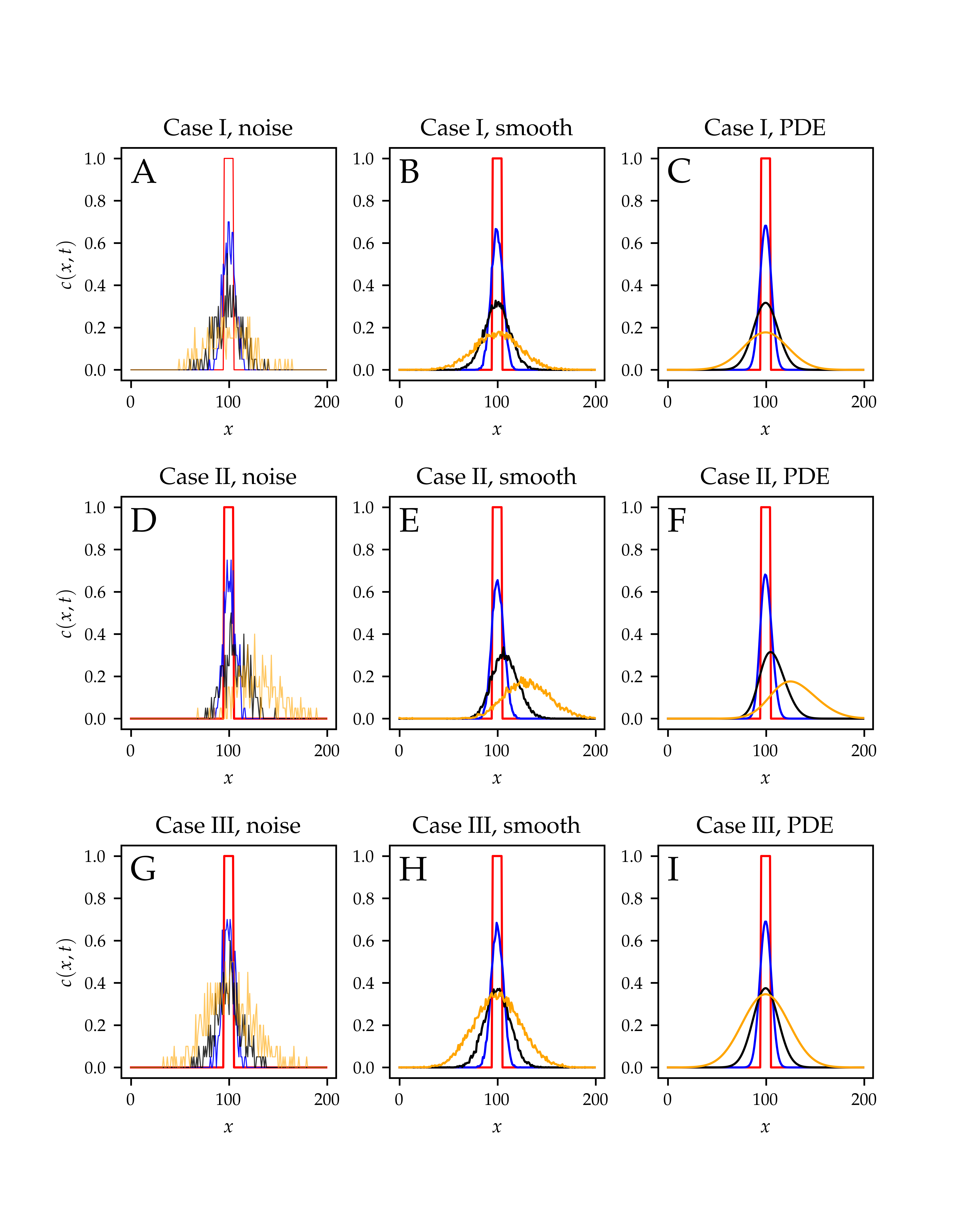}
\caption{Typical one-dimensional density profiles. Plots in the left-hand column are from a single realisation of the model ($K=1$), plots in the centre column are averaged over fifty simulations ($K=50$), and plots in the right-hand column are solution of the corresponding coarse-grained PDE. In each plot we show the density at times $t=0,50,150,500$ (red, blue, black, orange).}
\label{fig:data_traces}
\end{figure}

Figure~\ref{fig:data_traces} shows results from simulation of the ABM, averaged over different numbers of realisations, alongside solution of the corresponding PDE model. The PDE model is solved numerically using the PyPDE package~\cite{Zwicker2020}, which solves the PDE using the method of lines by discretising space using the grid on which spatial data for the ABM has been collected. The resulting ODEs are solved using a fourth-order Runge-Kutta method on the domain $x\in [0,200]$ with space discretisation $\Delta x = 10^{-3}$, and constant time discretisation $\Delta t = 10^{-4}$. 

We make two observations. First, we note that the solutions of the PDE models accurately predict the dynamics of the ABM in the chosen parameter regimes (see also Figure S1 of the Supplementary Information), so that we have a ``ground truth'' PDE against which to benchmark the equation learning methodology. Second, at early times the density profiles are very similar for the three different cases, but at later times differences due to the effects of bias and proliferation are clearly discernible. This observation implies that the equation learning methodology will require data on sufficiently long timescales in order to be able to accurately learn the correct PDE model.


\subsection{Equation learning: PDE-FIND}

In the following, assume that we have time series data for an unknown function $u(x,t)$ on a grid of $n$ points in time and $m$ points in space. This data is stored in a matrix $U\in \mathbb{R}^{n\times m}$. We assume that the data are a noisy discretisation of a function $u(x,t)$, the solution of an unknown PDE, and the aim is to learn the PDE that best describes the governing equation of the observed data. Henceforth, and to avoid confusion, we will write $U(x,t)$ for the observed data, $u(x,t)$ for the learned PDE, and $c(x,t)$ for the solution to the coarse-grained PDE defined in Equation~\eqref{mean-field}. We follow Rudy \textit{et al.}~\cite{Rudy17,Silva:2020:DOP} in assuming that the PDE governing $u(x,t)$ is given in the following form
\begin{equation}
u_t = \mathcal{N}(u,u_x,u_{xx},\dots),
\end{equation}
where $\mathcal{N}$ is a nonlinear function of $u(x,t)$ and its partial derivatives. Furthermore, it is assumed that $\mathcal{N}$ is a linear combination of a finite number of distinct library terms so that we can write
\begin{equation}
\label{PDEFINDEQ}
u_t = \sum_{i=1}^{N_\ell} \mathcal{N}_i(u,u_x,u_{xx},\dots)\xi_i,
\end{equation}
for coefficients $\xi_i$, where $i=1,\ldots,N_\ell$. By design, we will specify that $\mathcal{N}$ has polynomial nonlinearities, as is common in many equations in the natural sciences, and we note that Equation~\eqref{mean-field} falls within this class of PDEs. The aim of PDE-FIND is then to select, from the large library of terms $\mathcal{N}_i$, for $i=1,\ldots,N_\ell$, a small subset of relevant terms. 

The first step of the PDE-FIND pipeline is to numerically approximate both sides of Equation~\eqref{PDEFINDEQ}. This is done by estimating derivatives of the data with respect to space and time. The standard PDE-FIND implementation in \cite{Rudy17} takes finite difference approximations when the data contains little noise and polynomial differentiation when data are very noisy. The data and its derivatives are combined in a matrix $\Theta(U)$, where each column of $\Theta$ contains all of the values of a particular candidate function across the entire $n\times m$ grid. For example, if the candidate library consists of all polynomials up to degree two and non-mixed derivatives up to second order, $N_\ell=9$ and $\Theta(U)$ will look like
\begin{equation}
\label{exampletheta}
\Theta(U) = \left[1,U,U^2,U_x,UU_x, U^2U_x,U_{xx},UU_{xx},U^2U_{xx}\right].
\end{equation}
As a result, if there are $N_\ell$ terms in the candidate library, $\Theta(U)$ is a $n \times m \times N_\ell$ matrix. 

The left-hand side of Equation~\eqref{PDEFINDEQ} is similarly approximated, and we obtain a linear matrix equation representing the PDE evaluated at the data points:
\begin{equation}
U_t = \Theta(U)\boldsymbol{\xi},
\end{equation}
where $\boldsymbol{\xi}=[\xi_1,\ldots,\xi_{N_\ell}]^T$. Taking the same example for $\Theta(U)$ as in Equation~\eqref{exampletheta}, this matrix equation is of the form

\begin{equation}
    \begin{pmatrix}
    U_t(x_0,t_0)\\
    U_t(x_1,t_0)\\
    U_t(x_2,t_0)\\
    \vdots\\
    U_t(x_{n-1},t_m)\\
    U_t(x_n,t_m)
    \end{pmatrix}
    =
    \begin{pmatrix}
    1 & U(x_0,t_0) & U^2(x_0,t_0) &  \dots & U^2U_{xx}(x_0,t_0)\\
    1 & U(x_1,t_0) & U^2(x_1,t_0) & \dots & U^2U_{xx}(x_1,t_0)\\
    1 & U(x_2,t_0) & U^2(x_2,t_0) & \dots & U^2U_{xx}(x_2,t_0)\\
    \vdots & \vdots & \vdots & \ddots & \vdots\\
    1 & U(x_{n-1},t_m) & U^2(x_{n-1},t_m) &  \dots & U^2U_{xx}(x_{n-1},t_m)\\
    1 & U(x_{n},t_m) & U^2(x_{n},t_m)  & \dots & U^2U_{xx}(x_{n},t_m)\\
    \end{pmatrix}
    \begin{pmatrix}
    \xi_1\\
    \xi_2\\
    \xi_3\\
    \vdots\\
    \xi_8\\
    \xi_9
    \end{pmatrix}.
\end{equation}

\noindent Note how this representation shows that each row in the matrix equation represents the governing dynamics behind the data at one point in time and space. The values of the coefficients $\xi_i$ determine the form of the PDE, and so the aim is to learn the coefficients $\xi_i$ in some sense ``optimally''. 

Following Rudy \emph{et al.}~\cite{Rudy17}, we will assume $\Theta$ to be overspecified, meaning that the dynamics can be represented as linear combinations of the columns of $\Theta$. However, many PDEs in the natural sciences contain only a few terms. Therefore, we wish to learn a \emph{sparse} vector $\boldsymbol{\xi}=[\xi_1,\ldots,\xi_{N_\ell}]^T$ as a solution of Equation~\eqref{PDEFINDEQ}. This is done in PDE-FIND by considering the optimisation criterion
\begin{equation}
\label{equation:ridgeregression}
\overline{\boldsymbol{\xi}} = \text{argmin}_{\boldsymbol{\xi}}\big(\Vert \Theta(U,Q)\boldsymbol{\xi} - U_t\Vert_2^2 + \lambda \Vert \boldsymbol{\xi}\Vert_2^2\big),
\end{equation}
for the coefficient vector $\boldsymbol{\xi}$, where $\lambda \in \mathbb{R}_{>0}$ is a free parameter that penalises large coefficients. This is the method of \emph{ridge regression}. We note here that the term $\Vert \boldsymbol{\xi}\Vert_2^2$ can be replaced with $\Vert \boldsymbol{\xi}\Vert_1^2$, which corresponds to performing LASSO~\cite{nardini2020learning,ANNs}. The optimal choice of implementation is largely problem-dependent, and various choices for the regularisation method have been compared in the literature~\cite{Rudy17,Silva:2020:DOP,Kaiser:2018:SIO,Zheng:2019:AUF}, although no method has been proven to be definitively preferred over another. 

The default implementation of PDE-FIND as proposed by Rudy \emph{et al.}~\cite{Rudy17} supplements the ridge regression problem with a \emph{sequential thresholding} procedure in which a solution to Equation~\eqref{equation:ridgeregression} is found, and a hard threshold is performed on the regression coefficients by eliminating all library terms that have coefficients smaller than some pre-specified parameter $d_{\text{tol}}$. This process is then repeated on the remaining library terms until all coefficients are larger than $d_{\text{tol}}$, or until a maximum number of iterations has been reached. The sequential thresholding process is undertaken to enforce sparsity as the solution to the ridge regression problem in Equation~\eqref{equation:ridgeregression} may contain several small, but non-zero values. The combined algorithm is called Sequential Thresholding Ridge regression (STRidge). For more details and motivation of the method we refer to~\cite{Rudy17}, and for completeness we summarise the PDE-FIND method in Algorithm~\ref{STRidge}.


\begin{algorithm}[h]
\label{STRidge}
\SetAlgoLined
\SetAlgoNoLine
\medskip
\KwIn{Library matrix, $\Theta(U)$, time derivative of data, $U_t$, and STRidge parameters $\lambda$ and $d_{\text{tol}}$, maximum number of iterations, $\textit{iters}$}
\KwOut{Sparse vector $\boldsymbol{\xi}$}
  Set $B=\{j: j=1,\ldots,N_\ell\}$\;
  \While{$\text{iters}\geq 0$}{
  Set $\Theta[:,B]$ to be the matrix consisting of all columns of $\Theta(U)$ for which the coefficient $c_j$ has index $j\in{B}$\;
  Solve the sparse regression problem including only coefficients in $B$, that is, compute $\hat{\boldsymbol{\xi}}[B] = \text{argmin}_{\bar{\boldsymbol{\xi}}}\Vert \Theta[:,B]\bar{\boldsymbol{\xi}} - U_t\Vert_2^2 + \lambda \Vert \bar{\boldsymbol{\xi}}\Vert_2^2$\;
  Update $B$: set \\ $B = \{j: \text{entry of } \hat{\boldsymbol{\xi}}[B] \text{ corresponding to coefficient } c_j \text{ has magnitude at least } d_{\text{tol}}\}$\; 
  $\textit{iters} \leftarrow \textit{iters}-1$\;}
  Update $\xi$: \vspace{-0.4cm}
  \[
  \begin{array}{l}
  \text{for $j\notin{B}$ set the $j$-th entry of $\boldsymbol{\xi}$ to be zero;}\\ 
  \text{for $j\in{B}$ set the $j$-th entry of $\boldsymbol{\xi}$ to be the entry of $\hat{\boldsymbol{\xi}}[B]$ corresponding to coefficient $c_j$.}
  \end{array}
  \]
  \caption{STRidge}
\end{algorithm}


\subsubsection{Application of PDE-FIND to the ABM data}\label{section:AMB_data}

We first generate ABM data for Cases I, II and III. For each, we generate two datasets with different noise levels, denoting them $\mathcal{D}_r^i$ where $r=\text{I},\text{II},\text{III}$ denotes the case and $i=1,2$ denotes the dataset / noise level. For Dataset 1 ($i=1$) the density profiles are generated using single realisations of the ABM and averaging (so that $K=1$ and the data are relatively noisy), whereas for Dataset 2 ($i=2$) the density profiles are generated using 50 realisations of the ABM and averaging (so that $K=50$ and the data contain relatively little noise). For each realisation of the ABM, we simulate for $T=1000$ time steps and subsample the data at every other time point so that each dataset contains information for $n=500$ time points and $m=200$ space points. Each dataset contains $N_s$ samples, each an average over $K$ realisations of the ABM. We use the standard implementation of PDE-FIND~\cite{Rudy17} and use polynomial differentiation at fourth order to evaluate both the time and space derivatives.

We select a library of candidate terms that includes all polynomial terms up to order two and up to the second derivative. Table~\ref{table:true_coefficients} shows the values of the coarse-grained PDE coefficients, according to Equation~\eqref{mean-field}. These are the values of the coefficients that we would expect the PDE-FIND algorithm to return for perfect spatio-temporal data. In Table~\ref{table:true_coefficients}, and the rest of this work, we use the notation $c_{i}$ for the coefficient of term $i$ in the learned PDE.


\bigskip
\begin{table}[h!]
\centering
\begin{tabular}{|c||c|c|c|c|c|c|c|c|c|} 
\hline
 & $c_{1}$ & $c_{u}$ & $c_{u^2}$ & $c_{u_x}$ & $c_{u\cdot u_x}$ & $c_{u^2\cdot u_{xx}}$ & $c_{u_{xx}}$ & $c_{u\cdot u_{xx}}$ & $c_{u^2\cdot u_{xx}}$\\ [0.5ex] 
\hline\hline
Case I & 0.0 & 0.0 & 0.0 & 0.0 & 0.0 & 0.0 & \cellcolor[gray]{0.7}\textbf{0.25} & 0.0 & 0.0\\
\hline
Case II & 0.0 & 0.0 & 0.0 & \cellcolor[gray]{0.7}\textbf{-0.0375} & \cellcolor[gray]{0.7}\textbf{0.075} & 0.0 & \cellcolor[gray]{0.7}\textbf{0.25} & 0.0 & 0.0 \\
\hline
Case III & 0.0 & \cellcolor[gray]{0.7}\textbf{0.01} & \cellcolor[gray]{0.7}\textbf{-0.01} & 0.0 & 0.0 & 0.0 & \cellcolor[gray]{0.7}\textbf{0.25} & 0.0 & 0.0\\
\hline\hline
\end{tabular}
\caption{Coefficients of the coarse-grained PDEs describing evolution of the mean population density over time for the three example cases used in this work. The coefficients correspond to the coarse-grained PDEs described in Table~\ref{table:cases}.}
\label{table:true_coefficients}
\end{table}


\section{Sources of variability and model misspecification}\label{section:variability}

In this section, we showcase three different, but related, directions in the uncertainty quantification of the learned differential equations. In Section~\ref{subsection:variability}, we demonstrate the intrinsic variability of the learned coefficients in the presence of observation noise. We evaluate how uncertainty changes as the noise level is varied and suggest that even when using state-of-the-art denoising approaches a need for uncertainty quantification remains. Although increasing the signal-to-noise ratio helps, regardless of the method there is still variability that needs to be quantified. In particular, this is important in biological applications where observations are often very noisy and practitioners rarely have access to very large amounts of data. In Section~\ref{section:hyperparameters}, we investigate the impact of varying the regularisation hyperparameter in PDE-FIND with a view to asking whether this can be optimised to reduce uncertainty. We find that, while this is possible, parameter estimates are still uncertain and this uncertainty needs to be quantified. Finally, in Section~\ref{section:comparison}, we demonstrate that a key issue with PDE-FIND is that it aims to fit the time derivative of the solution and does not take into account the fit of the observed density to the data, leading to unphysical predictions. We demonstrate how to mitigate these issues in Section~\ref{section:BayesPDEFIND} through the use of Bayesian methods where we can evaluate uncertainty in a framework that optimises the fit of the model density profile to the data.

In order to quantify the variability in results from the application of PDE-FIND, we introduce a statistic which we term the \emph{identification ratio}. Assume that for each sample, $s$, in the observed dataset (which contains $N_s$ averaged density profile samples), we have used PDE-FIND to produce an estimate of the library coefficients, $\boldsymbol{\xi}$, using STRidge, and denote this estimate $\hat{\boldsymbol{\xi}}^s$. For each term $i$ in the library, we define the identification ratio, $a_i$, as
\begin{equation}
a_i=\frac{1}{N_{s}} \sum_{s=1}^{N_{s}}\ind\left(\hat{\boldsymbol{\xi}}^s_i \neq 0\right),
\end{equation}
where $\ind{}$ represents the indicator function and $\hat{\boldsymbol{\xi}}^s_i$ is the $i$-th entry of $\hat{\boldsymbol{\xi}}^s$. Therefore $a_i$ quantifies how often the term $\mathcal{N}_i$ from Equation~\eqref{PDEFINDEQ} is included in the PDE-FIND predictions. When $a_i$ is close to unity, the term is identified across many samples as being relevant for the dynamics and, conversely, when $a_i$ is close to zero, the term is identified in only a small minority of samples as being relevant.


\subsection{Variability of relevant PDE-FIND coefficients with noisy observations} \label{subsection:variability}

We first demonstrate that a naive application of PDE-FIND on noisy synthetic data yields variable and unreliable parameter estimates. For this application, we do not carry out hyperparameter tuning, but simply use widely adopted parameter settings to learn the coefficients. For each of the two datasets associated to each of Case I, Case II and Case III, where Dataset $1$ averages over $K=1$ realisations and Dataset $2$ averages over $K=50$ simulations, we train the PDE-FIND algorithm using Algorithm~\ref{STRidge} (STRidge) with fixed hyperparameter settings\footnote{These settings were used in the context of estimating the diffusion parameter in a random walk model in the Supplementary Information of~\cite{Rudy17}.} $\lambda = 10^{-2}$ and $d_{\text{tol}} = 10^{-3}$. For each of the resulting datasets, we also compute the corresponding identification ratios (Table~\ref{a_table}) to quantify the extent of identification of the different terms in the model, and compare the performance of PDE-FIND on the different case studies. In this case, $N_s = 1000$ samples.


\bigskip
\begin{table}[h!]
	\centering
	\begin{tabular}{|c||c|c|c|c|c|c|c|c|c|} 
		\hline
		Experiment & $c_{1}$ & $c_{u}$ & $c_{u^2}$ & $c_{u_x}$ & $c_{u\cdot u_x}$ & $c_{u^2\cdot u_{xx}}$ & $c_{u_{xx}}$ & $c_{u\cdot u_{xx}}$ & $c_{u^2\cdot u_{xx}}$\\ [0.5ex] 
		\hline\hline
		 $\mathcal{D}_\text{I}^1$ & 0.001 & 0.0 & 0.002 & 0.0 & 0.008 & 0.008 & \cellcolor[gray]{0.7}\textbf{0.826} & 0.199 & 0.05\\
		\hline
		 $\mathcal{D}_\text{I}^2$ & 0.0 & 0.0 & 0.0 & 0.0 & 0.0 & 0.0 & \cellcolor[gray]{0.7}\textbf{1.0} & 0.0 & 0.0\\
		\hline
		 $\mathcal{D}_\text{II}^1$ & 0.0 & 0.0 & 0.0 & \cellcolor[gray]{0.7}\textbf{0.999} & \cellcolor[gray]{0.7}\textbf{0.0} & 0.0 & \cellcolor[gray]{0.7}\textbf{0.012} & 0.002 & 0.0\\
		\hline
		 $\mathcal{D}_\text{II}^2$ & 0.0 & 0.0 & 0.0 & \cellcolor[gray]{0.7}\textbf{1.0} & \cellcolor[gray]{0.7}\textbf{0.0} & 0.0 & \cellcolor[gray]{0.7}\textbf{0.0} & 0.0 & 0.0\\
		\hline
		 $\mathcal{D}_\text{III}^1$ & 0.013 & \cellcolor[gray]{0.7}\textbf{0.659} & \cellcolor[gray]{0.7}\textbf{0.482} & 0.002 & 0.002 & 0.007 & \cellcolor[gray]{0.7}\textbf{0.571} & 0.01 & 0.014\\
		\hline
         $\mathcal{D}_\text{III}^2$ & 0.0 & \cellcolor[gray]{0.7}\textbf{1.0} & \cellcolor[gray]{0.7}\textbf{0.0} & 0.0 & 0.0 & 0.0 & \cellcolor[gray]{0.7}\textbf{1.0} & 0.0 & 0.0\\
		\hline
		Subsampling $\mathcal{D}_\text{I}^1$ & 0.0 & 0.0 & 0.003 & 0.001 & 0.002 & 0.007 & \cellcolor[gray]{0.7}\textbf{0.998} & 0.365 & 0.063\\
		\hline\hline
	\end{tabular}
	\caption{Identification ratios for different datasets. Bold fonts indicate terms that we anticipate in the learned PDEs based on the results from ABM coarse-graining. Recall that Case I includes non-biased motility and  no proliferation, Case II includes motility bias but no proliferation, whilst Case III inclues non-biased motility and proliferation. Dataset $1$ contains averages over $K=1$ realisations whilst Dataset $2$ contains averages over $K=50$ simulations.}
	\label{a_table}
\end{table}


\subsubsection{Case I}

\begin{figure}[htbp]
    \centering
    \includegraphics{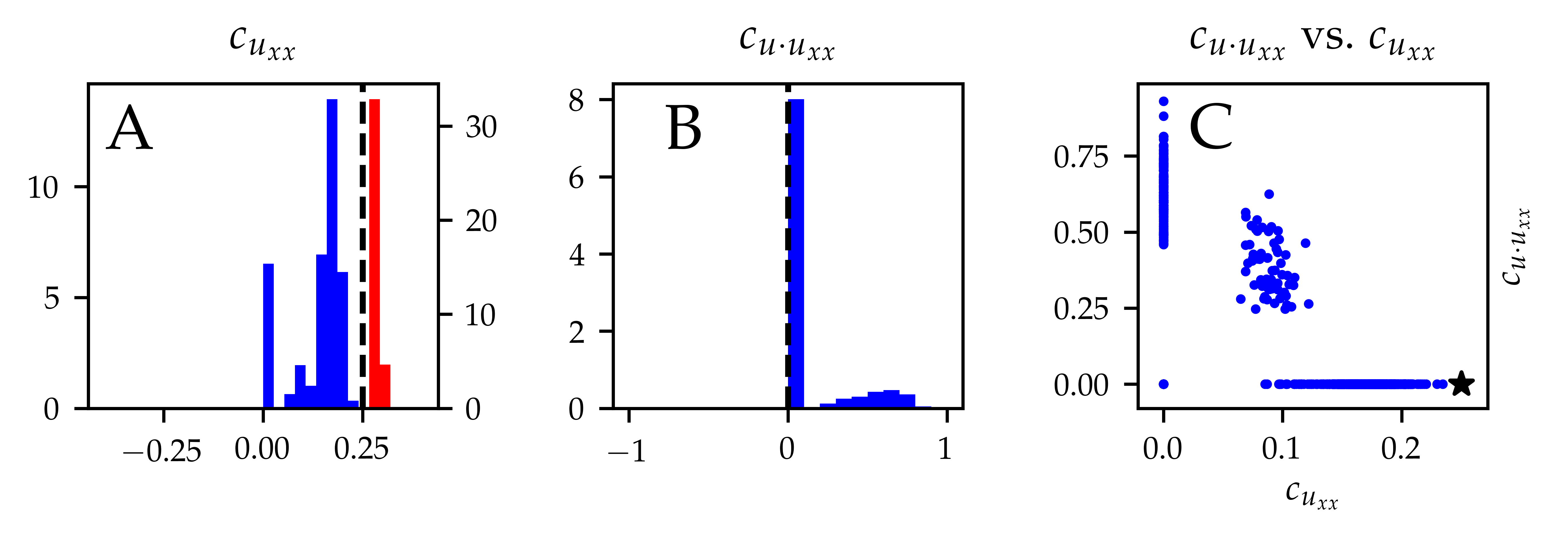}
    \caption{Histograms showing the empirical distribution of relevant PDE-FIND coefficients for Case I. A: histograms for $c_{u_{xx}}$ generated using $\mathcal{D}_\text{I}^1$ (blue) and $\mathcal{D}_\text{I}^2$ (red) compared to the true parameter value (black line). B: histogram for $c_{u\cdot{u}_{xx}}$ generated using $\mathcal{D}_\text{I}^1$ compared to the true value (black line). C: joint distribution of $c_{u_{xx}}$ and $c_{u\cdot{u}_{xx}}$ generated using $\mathcal{D}_\text{I}^1$ compared to the true parameter (black star).}
    \label{fig:CaseI_hists}
\end{figure}

For Case I, recall that the true PDE model contains only the term $u_{xx}$ with coefficient $0.25$, hence in noise-free scenarios we anticipate that $c_{u_{xx}}$ should be non-zero and all other coefficients should be zero. Table~\ref{a_table} shows that the two terms identified regularly by PDE-FIND on $\mathcal{D}_\text{I}^1$, the high-noise dataset, are $u_{xx}$ and $uu_{xx}$, with identification ratios of $0.826$ and $0.199$, respectively. On $\mathcal{D}_\text{I}^2$, the low-noise dataset, $u_{xx}$ is consistently identified and no other terms are identified. However, there is significant variability in the learned coefficients between different samples from the same dataset (Figure~\ref{fig:CaseI_hists}). In addition, for the high-noise dataset, $\mathcal{D}_\text{I}^1$, the coefficients of $u_{xx}$ and $uu_{xx}$ are correlated (Figure~\ref{fig:CaseI_hists}C). In some cases, PDE-FIND identifies just one of the two terms, and in others it identifies a combination of the two. This result highlights that potentially the wrong PDE can be learnt from noisy data, partly due to the fact that different PDEs can give rise to similar predictions. Since all of the ABM data samples have the same corresponding coarse-grained PDE, this highlights the inability of PDE-FIND to confidently learn the governing PDE from noisy data.


\subsubsection{Case II}

For Case II, where motility is biased, Table~\ref{a_table} shows that the two terms identified regularly by PDE-FIND on the high-noise dataset $\mathcal{D}_\text{II}^1$ are $u_x$ and $u_{xx}$, with identification ratios of $0.999$ and $0.012$, respectively. Note that the true model should also contain the term $uu_x$, but PDE-FIND fails to identify this term across all the samples of the dataset. This term arises in the coarse-grained PDE as a result of volume exclusion (incorporated into the ABM through the requirement that at most one agent can occupy a lattice site at any instant in time). Therefore we infer in this case that the data are insufficient to identify the impact of volume exclusion. This is most likely a result of the initial conditions and / or the timescale over which data are collected since the density is relatively low across the domain and so crowding likely unimportant.

\begin{figure}[htbp]
    \centering
    \includegraphics{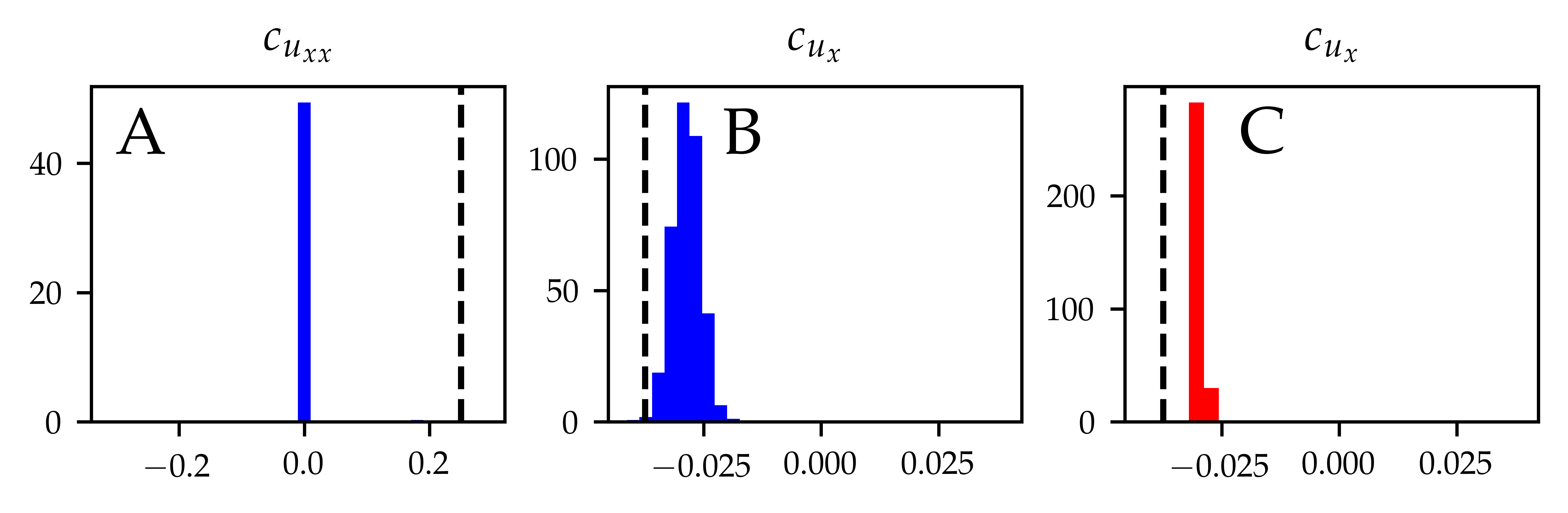}
    \caption{Histograms showing the empirical distribution of relevant PDE-FIND coefficients for Case II, in which motility is biased but there is no proliferation. A: histogram for $c_{u_{xx}}$ generated using $\mathcal{D}_\text{II}^1$ (blue) compared to the true parameter value (black line). B: histogram of $c_{u_x}$ generated using $\mathcal{D}_\text{II}^1$ (blue) compared to the true parameter value (black line). C: histogram of $c_{u_x}$ generated using $\mathcal{D}_\text{II}^2$ (red) compared to the true parameter value (black line).}
    \label{fig:CaseII_hists}
\end{figure}

For the low-noise dataset $\mathcal{D}_\text{II}^2$ only $u_x$ is identified, with an identification ratio of $1.0$. The histograms in Figure~\ref{fig:CaseII_hists} reveal a significant amount of variability in the learned parameters. For instance, for both $\mathcal{D}_\text{II}^1$ and $\mathcal{D}_\text{II}^2$, the parameters are distributed far away from the true parameter value. The variability appears to decrease with the noise level, and the distribution of estimated parameters moves towards the true parameter value, however the $u_{xx}$ coefficient is ``lost'' in the process.


\subsubsection{Case III}

\begin{figure}[htbp]
    \centering
    \includegraphics{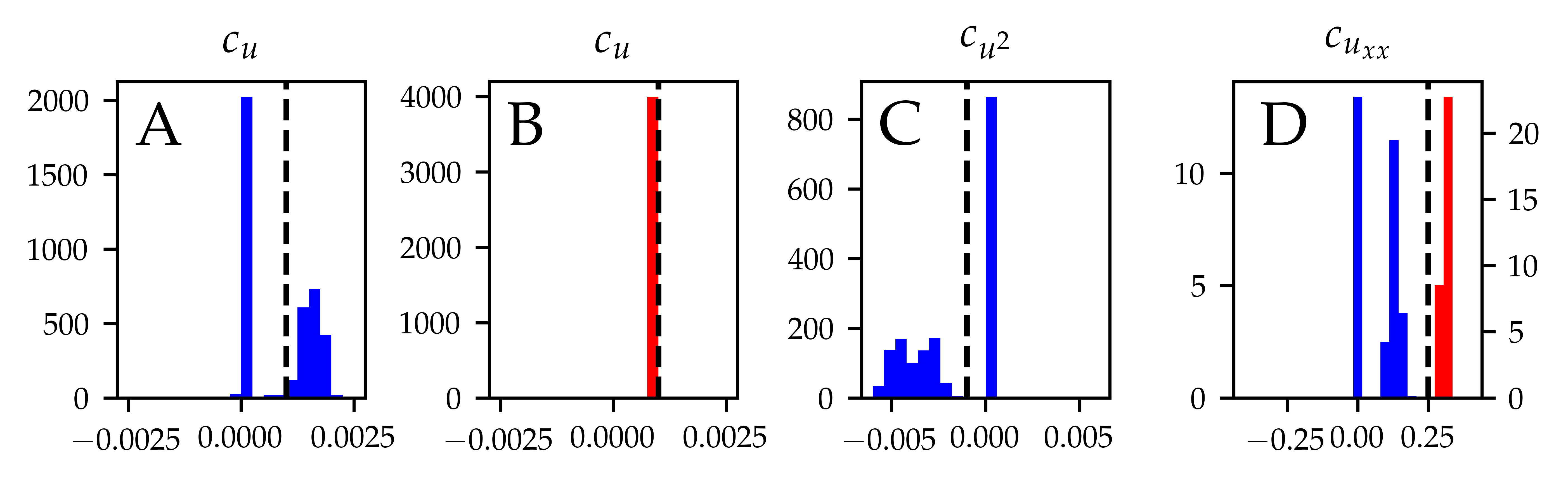}
    \caption{Histograms showing the empirical distribution of relevant PDE-FIND coefficients for Case III, which includes random motility and proliferation. A: histogram for $c_u$ trained on $\mathcal{D}_\text{III}^1$ (blue) compared to the true parameter value (black line). B: histogram for $c_u$ trained on $\mathcal{D}_\text{III}^2$ (red) compared to the true parameter value (black line). C: histogram for $c_{u^2}$ trained on $\mathcal{D}_\text{III}^1$ (blue) compared to the true parameter value (black line). D: histogram for $c_{u_{xx}}$ trained on $\mathcal{D}_\text{III}^1$ (blue) and $\mathcal{D}_\text{III}^2$ (red) compared to the true parameter value (black line).}
    \label{fig:CaseIII_hists}
\end{figure}

For Case III, which includes proliferation, Table~\ref{a_table} shows that the terms identified regularly by PDE-FIND on the high-noise dataset $\mathcal{D}_\text{III}^1$ are $u$, $u^2$ and $u_{xx}$ with identification ratios equal to 0.659, 0.482 and 0.571, respectively. Note that this means that all terms we would expect to appear in the PDE are identified. However, as shown in detail in Figure S6 of the Supporting Information, there is a correlation between the learned coefficients of $u$ and $u^2$, which points towards non-identifiability~\cite{Maiwald:2016:DTM,Raue:2009:SAP}. For the low-noise dataset $\mathcal{D}_\text{III}^2$, the parameters identified are $u$ and $u_{xx}$, both with identification ratio equal to 1.0. The histograms in Figure~\ref{fig:CaseIII_hists} reveal that the variability in the learned coefficients decreases as the noise level in the data is decreased. However, this does not mean that the model is increasingly well identified as the noise is decreased -- the term $u^2$ is not identified by PDE-FIND for the low-noise dataset $\mathcal{D}_\text{III}^2$, which contradicts the mean-field analysis. 


\subsubsection{Methods to decrease the noise levels}

To investigate the impact of noise on PDE-FIND, we carried out two further studies in which the noise in the data is reduced. First, we investigated whether choosing a more coarse spatial grid improves the PDE-FIND predictions. Choosing a more coarse spatial discretisation results in a smoother density profile, however greater errors are incurred in the approximation of the spatial derivatives and fewer data points are available. For this experiment, we subsampled the data along the $x$-dimension by averaging the occupancy over multiple columns at a time. Mathematically, from the empirical densities, $C_i$, at each time point, we subsample over intervals containing $2B$ lattice sites, estimating the average occupancies $\tilde{C}_i$ for $1\leq i\leq I/(2B)$, as
\begin{equation}
    \tilde{C}_i = \frac{1}{2B}\sum_{\ell = 2B(i-1)+1}^{2Bi}C_\ell.
\end{equation}
Table~\ref{a_table} summarises the identification ratios found for the high-noise dataset $\mathcal{D}_\text{I}^1$, where motility is unbiased and there is no proliferation, and we take $B=2$. We see that with spatial subsampling the identification ratio of $c_{u_{xx}}$ increases significantly, although there is no marked improvement in the identification ratios of other terms in the model. We conclude that even if this method of noise reduction allows the correct coefficients to be identified more frequently, there remains a need to mitigate the fact that many other terms are spuriously identified by PDE-FIND.  

Second, we investigated other means to reduce observation noise. In applications of PDE-FIND to real-life data, practitioners will typically not be able to control for the amount of observation noise in the way that is done in the numerical experiments of this work, hence methods to smooth data may be useful in allowing identification of the PDE model. In Supplementary Information Section S3 we explore two practically appealing methods, convolution with a Gaussian kernel and an implementation of principal component analysis for equation learning by Li \emph{et al.}~\cite{rPCA}. Our results show that even with these well-established techniques for reducing the influence of noise, predictions remain variable with coefficients highly correlated, and that uncertainty quantification remains necessary for a reliable application of PDE-FIND to realistic biological data. 


\subsection{Role of the regularisation hyperparameter}
\label{section:hyperparameters}


Recall that in Algorithm~\ref{STRidge}, a free parameter $\lambda$ controls the level of penalty incurred by choosing large coefficients in the solution of Equation~\eqref{PDEFINDEQ}. It is well known that the choice of regularisation parameter is nontrivial because it modulates the amount of sparsity that is enforced on the estimated coefficients. The issue of how to choose the optimal value of this hyperparameter in the context of ABMs was addressed recently by Nardini \textit{et al.}~\cite{nardini2020learning}, where cross-validation is discussed, amongst other options. As a test case to investigate the effects of algorithm hyperparameters on the uncertainty of learned coefficients, we perform cross-validation on the dataset $\mathcal{D}_\text{I}^1$ and then apply PDE-FIND using the optimal value of $\lambda$ found. To do this, we apply the grid search implementation of cross-validation suggested in \cite{nardini2020learning}, as detailed in Supplementary Information Section S4, to arrive at an optimal value of $\lambda=0.5$. We note here that this value is problem-dependent, and whenever a new dataset is being investigated a different value of $\lambda$ will generally be appropriate.


\begin{figure}[htbp]
\centering
	\includegraphics{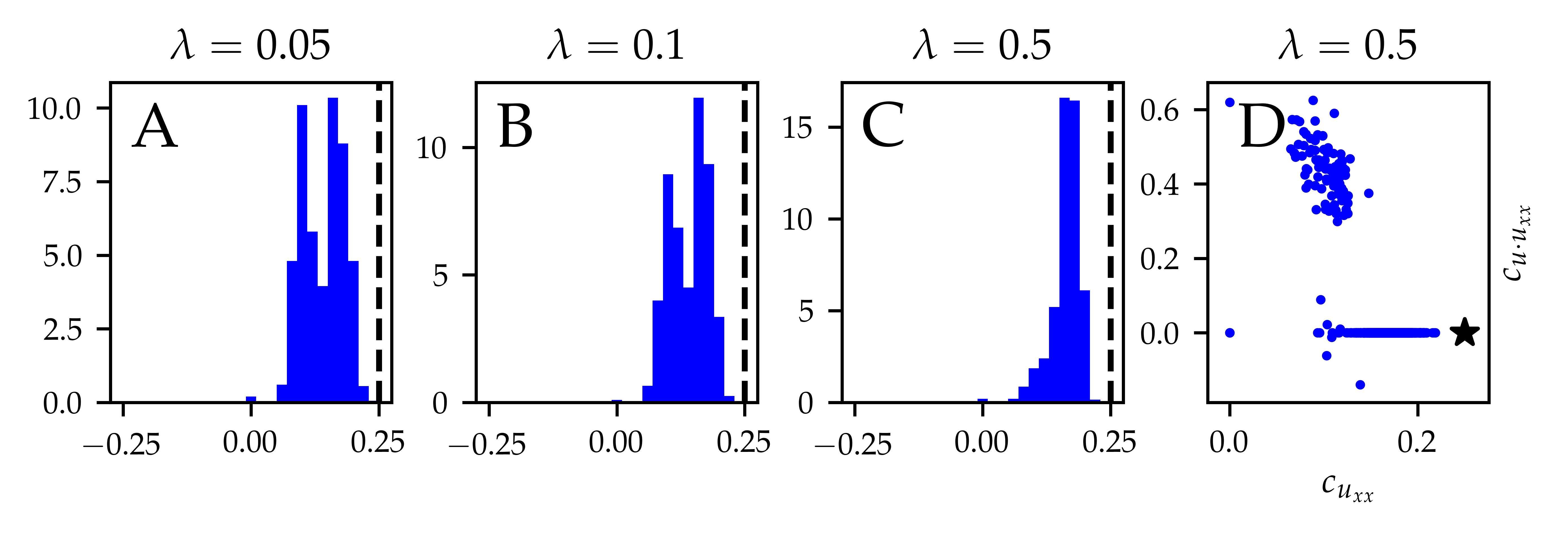}
	\caption{Empirical distributions of relevant PDE-FIND coefficients learnt generated using $\mathcal{D}_\text{I}^1$, which consists of unbiased motility only. A: histogram of $c_{u_{xx}}$ coefficients generated using $\lambda =0.01$, compared to the true value 0.25 (black line). B: histogram of $c_{u_{xx}}$ coefficients generated using $\lambda =0.05$, compared to the true value 0.25 (black line). C:  histogram of $c_{u_{xx}}$ coefficients generated using the optimal value of $\lambda =0.5$, compared to the true value 0.25 (black line). D: empirical joint distribution of $c_{u_{xx}}$ and $c_{u\cdot{u}_{xx}}$ coefficients generated using the optimal value of $\lambda = 0.5$, compared to the true parameters $(c_{u_{xx}},c_{u\cdot{u}_{xx}})=(0.25,0.0)$ (black star).}
	\label{lambdas}
\end{figure}


While the results presented in Supplementary Information Section S4 show that cross-validation improves the performance of PDE-FIND dramatically, as the number of misspecified coefficients decreases sharply when the regularisation parameter is optimised, cross-validation does not provide a sufficient solution to manage the uncertainty associated with variability in the predicted coefficients. Figure~\ref{lambdas} shows that, even with the optimal value of the regularisation coefficient, there is still much uncertainty in the coefficients, as the support of the histogram is large. While the atom at zero has nearly vanished, uncertainty quantification is still necessary because the empirical distribution still indicates a large degree of variability. Moreover, Figure~\ref{lambdas} shows that at the optimal value of the tuning parameter, $\lambda$, the coefficients $c_{u_{xx}}$ and $c_{u\cdot{u_{xx}}}$ still have a nontrivial joint distribution, implying that even with an optimal choice of the regularisation parameter, Bayesian methods are needed to analyse the joint behaviour of these two coefficients.


\subsection{Comparison of model predictions}
\label{section:comparison}

\begin{figure}[htbp]
    \centering
    \includegraphics{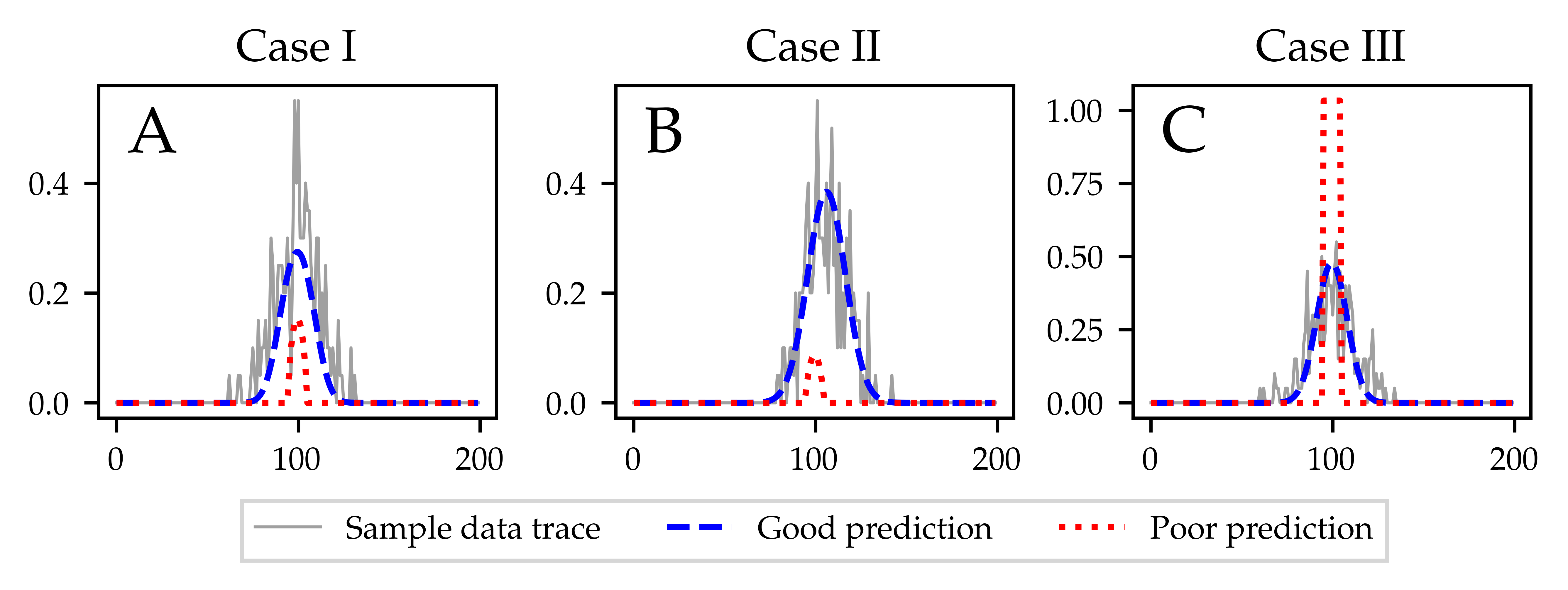}
    \caption{Comparison of predictions made by misspecified PDEs that are learned through the application of PDE-FIND to noisy data. A: Case I -- using $c_{u_{xx}} = 0.104$, $c_{u\cdot{u}_{xx}} = 0.26$ (blue dash) and $c_{u^2\cdot{u}_{xx}} = 1.517$ (red dots). B: Case II -- using $c_{u_x} = -0.0264$, $c_{u_{xx}} = 0.20$ (blue dash) and $c_{u\cdot{u}_{xx}} = 0.641$ (red dots). C: Case III -- using $c_{u_{xx}} = 0.12$ (blue dash) and $c_{u} = 0.0130$ (red dots). All other coefficients set are set to zero.}
    \label{fig:comparing_predictions}
\end{figure}

We now provide an explanation for the poor performance of PDE-FIND on the ABM data. The PDE-FIND algorithm solves a sparse regression problem to fit linear combinations of spatial derivatives to the time derivative. When data contains little-to-no noise, the temporal and spatial derivatives can be accurately estimated, and so the relationship between spatial and temporal derivatives can be inferred from observed data. In this context, comparing model predictions by their performance with respect to the $L^2$-loss in the learned temporal derivative retrieves the ground truth\footnote{For functions $f$ and $g$, the $L^2$-loss is given by $\left\Vert f-g\right\Vert_2^2 = \int \left[f(x)-g(x)\right]^2\text{d}x$.}. However, when the data are noisy, a number of different linear combinations of the spatial derivatives can result in an $L^2$-loss comparable to (or better than) those of the ground truth PDE (Figure~\ref{loss_landscape}). In Supplementary Information Section S5 (Figures S11--S13), we demonstrate this by selecting, for each dataset, two instances where the learned equations contain different terms to the coarse-grained PDE yet in both cases the temporal derivatives reproduce the observed temporal derivative qualitatively. However, there is no guarantee that such a match in the temporal derivative is sufficient to yield solutions that resemble the observed data when the PDE is numerically evaluated. We illustrate this in Figure~\ref{fig:comparing_predictions} where, for each of Cases I, II, and III, we select two sets of coefficients learned by PDE-FIND: one where the solution of the corresponding PDE resembles a typical data trace, and one where the solution of corresponding PDE bears little resemblance to typical observed data traces. 


In summary, what this striking difference in predictive capabilities of the learned PDEs reveals is that coefficients that optimise Equation~\eqref{equation:ridgeregression} do not necessarily perform well in terms of their ability to predict the evolution of the spatio-temporal density profiles. We illustrate this in further detail in Figure~\ref{loss_landscape}. We take the dataset $\mathcal{D}_\text{I}^1$ which consists of unbiased motility and no proliferation, and each sample in the dataset consists of an average over $K=1$ simulations from the ABM. First, we average over all $N_s=1000$ samples in the dataset to obtain the density profile $\langle{}C_i(t)\rangle{}$ as in Equation~\eqref{equation:Caverage}. The two coefficients consistently identified for dataset $\mathcal{D}_I^1$ are $c_{u_{xx}}$ and $c_{u\cdot{u}_{xx}}$, which gives the PDE
\begin{equation}
u_t=c_{u\cdot{u}_{xx}}uu_{xx}+c_{u_{xx}}u_{xx}.
\label{equation:lossPDE}
\end{equation}
We integrate this PDE numerically over a grid of values of $c_{u_{xx}}$ and $c_{u\cdot{u}_{xx}}$, and then evaluate the $L^2$-loss between the time derivative of the PDE model and that of the averaged ABM data, $\langle{}C_i(t)\rangle{}$ (Figure~\ref{loss_landscape}A), as well as the difference between the density predicted by the PDE model and that of the averaged ABM data (Figure~\ref{loss_landscape}B). We estimate the sum of the $L^2$-loss between the PDE and ABM data at five time points as
\begin{equation}
d(X^{\text{obs}},X^{\text{sim}}) = \sum_{j=1}^5 \left\Vert X_{50j}^{\text{obs}} - X_{50j}^{\text{sim}}\right\Vert_2,
\label{eq:spatialL2}
\end{equation}
where, for example when comparing density profiles, 
\begin{eqnarray}
\label{equation:ABM_L2}
X^{\text{obs}}_{50j}&=&\left[\langle{}C_1(50j)\rangle{},\langle{}C_2(50j)\rangle{},\ldots,\langle{}C_{200}(50j)\rangle{}\right]^T, \\
\label{equation:PDE_L2}
X_{50j}^{\text{sim}}&=&\left[u(\Delta,50j),u(2\Delta,50j),\ldots,u(200\Delta,50j)\right]^T,
\end{eqnarray} 
and when comparing time derivatives,
\begin{eqnarray}
X^{\text{obs}}_{50j}&=&\left[\langle{}{C_1}_t(50j)\rangle{},\langle{}{C_2}_t(50j)\rangle{},\ldots,\langle{}{C_{200}}_t(50j)\rangle{}\right]^T, \\
X_{50j}^{\text{sim}}&=&\left[u_t(\Delta,50j),u_t(2\Delta,50j),\ldots,u_t(200\Delta,50j)\right]^T,
\end{eqnarray}
with $\langle{}{C_i}_t(50j)\rangle{}=\left(\langle{}C_i(50j)\rangle{}-\langle{}C_i(50j+\tau\right)\rangle{})/\tau$ for $j=1,\ldots,5$ where $\tau$ is the time step of the ABM simulation algorithm. The blue shading in Figure~\ref{loss_landscape} shows the $L^2$-loss in each case, and we also plot the PDE-FIND-estimated coefficients on the same axes for each of the $N_s=1000$ samples of the dataset.


\begin{figure}[htbp]
    \centering
    \includegraphics[width=\linewidth]{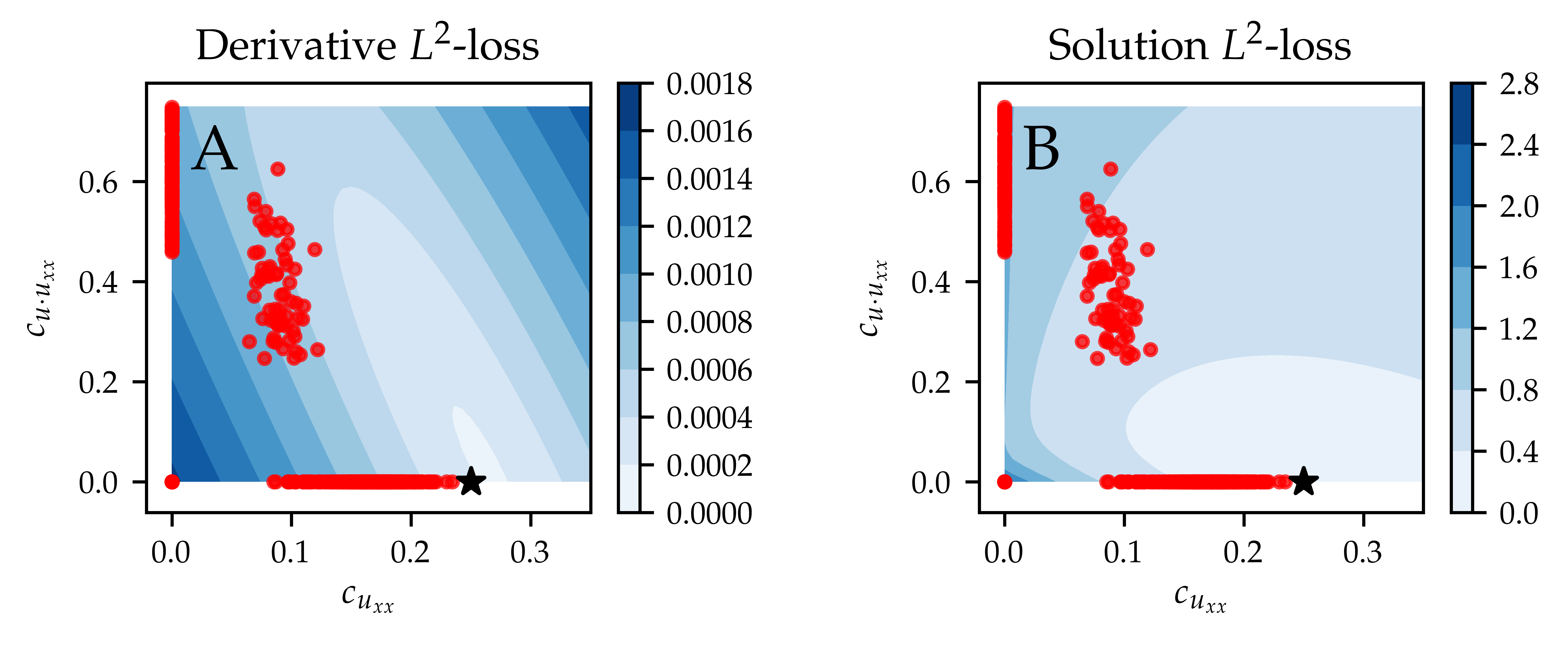}
	\caption{Heatmaps showing differences between averaged ABM data and the PDE solution in Equation~\eqref{equation:lossPDE}. A: $L^2$-loss landscape of the time derivative, estimated using Equation~\eqref{eq:spatialL2}. B: $L^2$-loss landscape for the density profile, estimated using Equation~\eqref{eq:spatialL2}. In each plot the pairs of coefficients estimated by applying PDE-FIND to each of the $N_s=1000$ samples of the dataset are plotted using red dots, and the parameter values used to generated the ABM data are indicated using a black star.}
	\label{loss_landscape}
\end{figure}


Figure~\ref{loss_landscape} demonstrates that there are significant differences between the loss landscapes of the different error metrics. While both loss landscapes show a minimum around the parameter set used in the ABM simulations ($c_{u_{xx}}=0.25$ and $c_{u\cdot{u_{xx}}}=0.0$, black stars in Figure~\ref{loss_landscape}), the derivative loss landscape in Figure \ref{loss_landscape} is unable to distinguish different regions of $(c_{u_{xx}},c_{u\cdot{u_{xx}}})$ parameter space. For example, for there are multiple PDE-FIND parameter sets (red dots) that lie close to the $L^2$-loss contours of 0.0010 and 0.0012 -- some of these sit on the horizontal axis (where $c_{u_{xx}}$ is non-zero and $c_{u\cdot{u_{xx}}}$ is zero), whereas others sit on the vertical axis of the plot (where $c_{u_{xx}}$ is zero and $c_{u\cdot{u_{xx}}}$ is non-zero). On the other hand, the $L^2$-loss for the density profiles provides much more useful information -- as one moves further from the input parameter values ($c_{u_{xx}}=0.25$ and $c_{u\cdot{u_{xx}}}=0.0$, black star in Figure~\ref{loss_landscape}) we see increasing errors between the density profile predicted by solution of the PDEs and the averaged ABM data. Figure S14 of the Supplementary Information demonstrates that this issue is further compounded as the noise in the data increases.

In summary, in this section we have shown that the density profile loss landscape is much more informative about the underlying PDE model than the derivative loss landscape. We exploit this observation in the following section, where we propose a method which we term ``Bayes-PDE-FIND'' to tackle the issues relating to both the uncertainty in the predictions of PDE-FIND for noisy data and also quantify the uncertainty in PDE-FIND predictions.


\section{Bayes-PDE-FIND}\label{section:BayesPDEFIND}

In this section, we propose an approach that harnesses the likelihood-free method of approximate Bayesian computation to quantify the uncertainty in estimates provided by PDE-FIND. In brief, our method involves the application of PDE-FIND to multiple datasets in order to define a prior distribution for the coefficients of the PDE library terms, $\mathcal{N}_i$ for $i=1,\ldots,N_\ell$, followed by application of Bayesian approaches for estimation of the posterior parameter distribution.
 

\subsection{Approximate Bayesian computation}

The goal of Bayesian parameter estimation is to update prior beliefs about model parameters $\boldsymbol\theta$ encoded in a prior distribution $\pi(\boldsymbol\theta)$. In this context, $\boldsymbol\theta$ constitutes the coefficients, $\xi_i$, of the library terms $\mathcal{N}_i$ for $i=1,\ldots,N_\ell$. The updating process is dependent on observations $\mathcal{D}_\mathrm{obs}$, which in this work are the noisy, averaged data from the ABM, as detailed in Section~\ref{section:AMB_data}. The mathematical model is the PDE defined in Equation~\eqref{PDEFINDEQ}, which then defines a likelihood $P(\mathcal{D}_\text{obs}\,|\,\boldsymbol{\theta})$. In Bayesian statistics, the likelihood is combined with the prior distribution to give the posterior distribution: 
\begin{equation}
P(\boldsymbol{\theta}\,|\,\mathcal{D}_\text{obs})\propto
P(\mathcal{D}_\text{obs}\,|\,\boldsymbol{\theta})\pi(\boldsymbol\theta).
\end{equation}
Such posterior distributions provide information as to the uncertainty in parameter estimates that are learned from observed data, and also allow practitioners to understand the range of realistic parameter values that can produce observed data. The likelihood $P(\mathcal{D}_\text{obs}\,|\,\boldsymbol{\theta})$ defines the probability density of the observations $\mathcal{D}_{\text{obs}}$ given the model parameters $\boldsymbol{\theta}$. In this context, the observed data $\{U(x,t)\}$ at space points $x = x_1, x_2, \dots, x_N$ and time points $t = t_1, t_2, \dots, t_N$ are obtained from a stochastic ABM. The solution $u(x,t;\boldsymbol{\theta})$ of the PDE model is an approximation of the mean of the ABM data, \textit{i.e.} $u(x,t;\boldsymbol{\theta}) \approx \mathbb{E}_{\boldsymbol{\theta}}[(U(x,t)]$.To define a classical likelihood, one would need to first assume that the mean of the ABM data is exactly given by the PDE solution and prescribing the distribution of ABM outputs around the PDE model mean. However, for a general ABM, the distribution for the deviation from the mean is unknown. In some cases, one might choose to make a simplifying assumption, such as a Gaussian approximation.  However, in the small data limit considered in EQL applications, such an assumption is unreasonable. It will depend on the details of the ABM as to the extent to which individual realisations vary from their mean, which is for the purposes of inference, unknown. As we prefer to avoid placing unnecessary assumptions on the process, we opt instead for a likelihood-free approach.  We provide further mathematical insight and justification for avoiding likelihood-based methods in Supplementary Information section S7.

Approximate Bayesian computation is a popular likelihood-free tool to estimate the posterior parameter distribution~\cite{Sunnaker13,Toni:2009:ABC}. It approximates the likelihood, $P(\mathcal{D}_\text{obs}\,|\,\boldsymbol{\theta})$, using repeated simulation of the model, and acceptance of the parameter $\boldsymbol{\theta}$ requires that the output of the model, $\mathcal{D}_\text{sim}(\boldsymbol\theta)$, is in some sense close enough to the data, $\mathcal{D}_\mathrm{obs}$. The ABC posterior can be written
\begin{equation}
P_\text{ABC}(\boldsymbol{\theta}\,|\,\mathcal{D}_\text{obs})\propto
P\left(d(\mathcal{D}_\text{obs},\mathcal{D}_\text{sim})<\varepsilon\,|\,\boldsymbol{\theta}\right)\pi(\boldsymbol\theta),
\end{equation}
where $d$ is a distance function that quantifies the difference between the data, $\mathcal{D}_\text{obs}$, and model output, $\mathcal{D}_\text{sim}$. In this work, our aim is to use ABC to estimate the posterior distribution of the PDE model coefficients, $\xi_i$ for $i=1,\ldots,N_\ell$, for a given dataset. 

Despite the apparent simplicity of the ABC method, unless an informative prior is used to constrain the space of possible parameters and informative summary statistics can be found, the application of ABC methods to models with high-dimensional parameter space and output space is generally computationally prohibitive. In particular, this high computational cost means that the direct application of ABC methods to estimate the coefficients $\xi_i$, for $i=1,\ldots,N_\ell$, in Equation~\eqref{PDEFINDEQ} is essentially infeasible unless it is possible to construct an informed prior distribution. Here, we propose a method that uses the predictions of PDE-FIND  to construct an informed prior so that ABC can then be used to estimate the library coefficients $\xi_i$. 


\subsection{Using PDE-FIND to define a prior distribution for ABC}\label{section:prior}

Assume that for each sample, $s$, in the observed dataset (which contains $N_s$ averaged density profile samples), we have used PDE-FIND (as defined in using Algorithm~\ref{STRidge}) to produce an estimate, $\hat{\boldsymbol{\xi}}^s$, of the parameters $\boldsymbol{\xi}$. Recall that for each term $i$ in the library, we have defined the identification ratio, $a_i$, as
\begin{equation}
\nonumber
a_i=\frac{1}{N_{s}} \sum_{s=1}^{N_{s}}\ind\left(\hat{\boldsymbol{\xi}}_i^s \neq 0\right),
\end{equation}
to quantify how often the term $\mathcal{N}_i$ from Equation~\eqref{PDEFINDEQ} is included in the PDE-FIND predictions. 

To make progress in specifying a prior distribution for the library coefficients, $\xi_i$, we first threshold, using parameter $0<\delta<1$, so that we can define $A=\{i: a_i > \delta\}$ as the set of coefficients that are identified by PDE-FIND in more than a fraction $\delta$ of the $N_s$ samples of the dataset. We then eliminate from the library all terms for which $a_i<\delta$, \textit{i.e.} we set the marginal prior for coefficient $\xi_i$ to be $\pi_i\equiv0$. This achieves a first step of coefficient selection by eliminating variables for which the initial PDE-FIND screen indicates low confidence. On the other hand, for $i\in{A}$, there is still a need to investigate which coefficients to include in the final model, and a prior must be carefully chosen to explore which parameters to include, and which parameters to eliminate.

Spike-and-slab models are powerful tools to perform variable selection in regression problems \cite{Ishwaran05,Xu15,Mitchell88,Lempers70}. The main idea of a spike-and-slab type prior is that it defines a two-point mixture distribution in which coefficients are mutually independent. Each mixture is made up of a flat distribution with large support (the slab) and a degenerate distribution at zero (the spike). In early formulations, the slab was modeled as a uniform distribution over some region of parameter space \cite{Mitchell88,Lempers70}, whereas in more recent work, inference is performed on hyperparameters of the marginal distributions \cite{Ishwaran05,Xu15}. Samples of the hyperparameters yielding a high variance will lead to sampling parameters far away from zero, whereas samples of the hyperparameters yielding a low variace will sample close to zero. In this way, the aim is to explore parameter space by iteratively sampling over the hyperparameters and the values for the coefficients using Gibbs sampling. In this work, we wish to exploit the simplicity of the earliest slab-and-spike models, which use a Dirac measure at zero to enforce sparsity, whilst utilising as much information as possible from the PDE-FIND screen in defining the prior distribution without violating the likelihood principle.

We follow the hierarchical Bayesian group LASSO model with an independent spike and slab type prior for each coefficient \cite{Xu15}. We set the group size in the model of Xu~\textit{et al.} \cite{Xu15} equal to one, so that the prior $\pi_i(\xi_i)$ for each coefficient $\xi_i$ is given by
\begin{align*}
    \xi_i \vert\mu,\sigma_i^2 &\sim (1-a_i)\delta_0 + a_i \mathcal{N}(\mu_i,\sigma_i^2),\\
    \sigma_i^2 &\sim \mathcal{IG}(\alpha_i,\beta_i),
\end{align*}
where $\mathcal{IG}$ is the inverse-gamma distribution with parameters $\alpha_i,\beta_i$ that define the shape of the prior on $\sigma_i^2$. This is the standard choice for modeling the distribution of the hypervariances. In the approach of Xu~\textit{et al.} \cite{Xu15}, $\mu=0$. In this work, we can use knowledge of the coefficients gained through our initial PDE-FIND screen to inform the $\mu_i$. To do so, we randomly divide the ABM data in half and use one subset in the PDE-FIND screen to inform the prior (\textit{exploration subset}), and the other subset to perform inference (\textit{inference subset}). We first set $a_i$ equal to the $i$-th identification ratio and $\mu_i$ equal to the $i$-th sample mean of the PDE-FIND coefficients trained on the exploration subset. Since the hyperprior for the variances $\sigma_i^2$ allows for large values of $\sigma_i^2$, this prior is not overly restrictive, since values far away from the sample mean can be sampled. Second, we tune $\alpha_i,\beta_i$ using the exploration subset so that the variance is on average the same order of magnitude as the PDE-FIND coefficients. This is crucial: a large (small) variance in parameters that are typically small (large) will fail to sample from the relevant regions of parameter space. The exact values of $\mu_i, \alpha_i, \beta_i$ are given in Supplementary Information Section S9.

These considerations now imply that the prior for $\boldsymbol{\xi}$ is given by
\begin{equation}
\boldsymbol{\pi} = \bigotimes_{i =1}^{N_\ell}\Big\{\ind(i \in A)\pi_i\xi_i + \ind(i \not \in A)\cdot\delta_0\Big\}.
\label{equation:prior}
\end{equation}

The key advantage in specifying this prior distribution is that we are now only required to perform Bayesian inference for a reduced model that has a much lower dimensional parameter space (equal to $|A|\ll{N_\ell}$) compared to the original model (with parameter space of dimension $N_\ell$). This is because PDE-FIND promotes sparsity and so the majority of coefficients will have identification ratio, $a_i$, close to zero. Such terms can be confidently eliminated from the model and so not considered in the ensuing parameter estimation process. On the other hand if $a_i\approx1$ this means PDE-FIND has consistently identified the $i$-th term as relevant for the dynamics. As such, we can be confident that $\mathcal{N}_i$ should be included in the model, however uncertainty in estimates of $\xi_i$ must still be quantified. Finally, if $a_i$ is neither close to zero or one, which means that PDE-FIND has included the $i$-th term in the library for a non-trivial number of samples in the dataset, Bayesian approaches can be used to investigate the joint posterior distribution of the $i$-th coefficient, $\xi_i$, with the rest of the model terms by considering the performance of models that both include and exclude the $i$-th term.


\subsection{The Bayes-PDE-FIND algorithm}

We now outline the Bayes-PDE-FIND approach. In essence, we apply the PDE-FIND algorithm to each sample $s=1,\ldots,N_s$ of the dataset under consideration, and use the results to formulate a prior distribution for ABC as described in Section~\ref{section:prior} and, in particular Equations 4.2-4.3. We then apply ABC to estimate the posterior parameter distribution, noting that the computational cost of ABC is much reduced through the use of PDE-FIND to generate an informed prior distribution -- in effect, we use PDE-FIND to reduce the target PDE in Equation~\eqref{PDEFINDEQ} to
\begin{equation}
\label{PDEFINDEQ_reduced}
u_t = \sum_{i\in{A}} \mathcal{N}_i(u,u_x,u_{xx},\dots)\xi_i,
\end{equation}
with a prior distribution over the $\xi_i$, for $i\in{A}$, given by Equation~\eqref{equation:prior}. 

Importantly, when we apply ABC to estimate the posterior parameter distribution, we use a low-noise dataset, $\langle{}C_i(t)\rangle{}_\text{LN}$ for $1\leq{i}\leq{I}$, created by averaging over the original $N_s$ samples of dataset as the observed data, $\mathcal{D}_\text{obs}$, where sample $s$ is calculated as 
\begin{equation}
\langle{}C_i(t)\rangle{}_s=\sum_{k=1}^{K}C_i^{k,s}(t) = \frac{1}{J\cdot{K}}\sum_{k=1}^{K}\sum_{j=1}^{J} C_{ij}^{k,s}(t),
\end{equation}
that is, each sample $s$ consists of column-averaged data from $K$ simulations of the ABM, and so, for $1\leq{i}\leq{I}$,
\begin{equation}
\label{equation:LNdata}
\langle{}C_i(t)\rangle{}_\text{LN}=\sum_{s=1}^{S}\langle{}C_i(t)\rangle{}_s.
\end{equation}
For the distance function, $d$, we again use an averaged estimate of the $L^2$-difference between the ABM data and the PDE solution, here defined as
\begin{equation}
\label{equation:ABCdistance}
d(\mathcal{D}_\text{obs},\mathcal{D}_\text{sim}(\boldsymbol\theta)) = \sum_{j=1}^5 \left\Vert (\mathcal{D}_\text{obs})_{50j} - \mathcal{D}_\text{sim}(\boldsymbol\theta)_{50j}\right\Vert_2,
\end{equation}
where, for $j=1,\ldots,5$,
\begin{eqnarray}
\label{equation:ABM_L2}
(\mathcal{D}_\text{obs})_{50j}&=&\left[\langle{}C_1(50j)\rangle{}_\text{LN},\langle{}C_2(50j)\rangle{}_\text{LN},\ldots,\langle{}C_{200}(50j)\rangle{}_\text{LN}\right]^T, \\
\label{equation:PDE_L2}
\mathcal{D}_\text{sim}(\boldsymbol\theta)_{50j}&=&\left[u(\Delta,50j;\boldsymbol{\theta}),u(2\Delta,50j;\boldsymbol{\theta}),\ldots,u(200\Delta,50j;\boldsymbol{\theta})\right]^T,
\end{eqnarray} 
and $u(x,t;\boldsymbol{\theta})$ is the solution to Equation~\eqref{PDEFINDEQ_reduced} with parameter set $\boldsymbol{\theta}$. Note that we choose in Equation~\eqref{equation:ABCdistance} to compare the solutions at a wide range of time points to capture the behaviours of the data over different time scales. We summarise the Bayes-PDE-FIND algorithm in Algorithm~\ref{algorithm:Bayes-PDE-FIND}.


\bigskip

\begin{algorithm}[h]
\label{algorithm:Bayes-PDE-FIND}
\SetAlgoLined
\SetAlgoNoLine
\medskip
\KwIn{time series dataset consisting of $N_s$ samples; PDE-FIND hyperparameters $\lambda$ and $d_\text{tol}$; PDE library $\mathcal{N}_i$ for $i=1,\ldots,N_\ell$; minimum identification ratio $\delta>0$.}
\KwOut{Posterior distribution over coefficients $\xi_i$, for $i=1,\ldots,N_\ell$, of library PDE.}
    \For{$s=1,\ldots,N_s$}{
        Compute $\hat{\boldsymbol{\xi}}^s$ using Algorithm~\ref{STRidge} with sample $s$ from the dataset\;
    }
    \For{$i \in 1,\dots,N_\ell$}{
        Compute the identification ratio \[a_i = \dfrac{1}{N_{s}} \sum_{s=1}^{N_{s}} \ind\left(\hat{\boldsymbol{\xi}}^s_i \neq 0\right).\]}
        
  Compute $A = \{i: a_i > \delta\}$, and define the prior distribution $\boldsymbol{\pi}$ as in Equation \ref{equation:prior}:
  
\begin{equation}
    \boldsymbol{\pi} = \bigotimes_{i =1}^{N_\ell}\Big\{\ind(i \in A)\pi_i\xi_i + \ind(i \not \in A)\cdot\delta_0\Big\}.
\end{equation}

Perform approximate Bayesian computation using observed data $\mathcal{D}_\text{obs}=\left[\langle{}C_1(t)\rangle{}_\text{LN},\ldots,\langle{}C_{200}(t)\rangle{}_\text{LN}\right]$ to obtain the posterior distribution. \medskip

 \caption{Bayes-PDE-FIND}
\end{algorithm}


\subsection{Results}

The aim of this section is to showcase how the Bayes-PDE-FIND algorithm can be used to significantly improve the quality of the learned PDE model. Recall that the aim is to reduce the uncertainty surrounding which coefficients to include in the model, to reduce uncertainty in the estimated model parameters, and to improve the posterior predictive capability of the model by finding a posterior parameter distribution that takes into account properties of the observed density profiles. For each of the noisy-data test cases, that is, datasets $\mathcal{D}_\text{I}^1$, $\mathcal{D}_\text{II}^1$ and $\mathcal{D}_\text{III}^1$, we apply Algorithm~\ref{algorithm:Bayes-PDE-FIND}, using the Pakman package~\cite{Pak2020} for the ABC step. 

Recall that for Case I, where the ABM contains only unbiased motility, the only two coefficients regularly identified using PDE-FIND are $c_{u_{xx}}$ and $c_{u\cdot{u_{xx}}}$ (see Table \ref{a_table}). This means that all library terms except for $u_{xx}$ and $uu_{xx}$ can be confidently excluded, and for the ABC process we consider the PDE model
\begin{equation}
u_t=c_{u_{xx}}u_{xx}+c_{u\cdot{u}_{xx}}uu_{xx},
\end{equation}
and aim to infer the $(c_{u_{xx}},c_{u\cdot{u}_{xx}})$ posterior parameter distribution. For Case II, which contains biased motility and no proliferation, the only coefficients regularly identified using PDE-FIND are $c_{u_x}$ and $c_{u_{xx}}$ (see Table \ref{a_table}) and so for the ABC process we consider the PDE model
\begin{equation}
u_t=c_{u_{xx}}u_{xx}+c_{u_{x}}u_x,
\end{equation}
and aim to infer the $(c_{u_{xx}},c_{u_{x}})$ posterior parameter distribution. Note that PDE-FIND failed to identify one relevant model term in Case II, which is $uu_x$, most likely either due to the significant noise in the samples of the dataset and / or the timescale over which the data are collected. For Case III, which contains both unbiased motility and proliferation, the only coefficients that are regularly identified are $c_u$, $c_{u^2}$ and $c_{u_{xx}}$ (see Table \ref{a_table}) and so for the ABC process we consider the PDE model
\begin{equation}
u_t=c_uu+c_{u^2}u^2+c_{u_{xx}}u_{xx},
\end{equation}
and aim to infer the $(c_u,c_{u^2},c_{u_{xx}})$ posterior parameter distribution. 


We explore the results of using Bayes-PDE-FIND with a form of ABC which is known as ABC-rejection sampling. At each each step, $\boldsymbol\theta^*$ is sampled from the prior distribution $\pi(\boldsymbol\theta)$, and the PDE given in Equation~\eqref{PDEFINDEQ} is integrated in time using parameters $\boldsymbol\theta^*$ to yield simulated data $\mathcal{D}_\text{sim}(\boldsymbol\theta^*)$. The simulated data, $\mathcal{D}_\text{sim}(\boldsymbol\theta^*)$, is compared to the observed data, $\mathcal{D}_\text{obs}$, according to a distance function $d(\mathcal{D}_\text{obs},\mathcal{D}_\text{sim}(\boldsymbol\theta^*))$ provided by the practitioner. Given a \emph{tolerance} $\varepsilon>0$, the sampled $\boldsymbol{\theta}$ is accepted into the posterior distribution whenever $d(\mathcal{D}_\text{obs},\mathcal{D}_\text{sim}(\boldsymbol\theta^*))<\varepsilon$. We provide full details of the ABC-rejection algorithm used in this work in Supplementary Information Section S8. 

\begin{figure}[htbp]
	\centering
	\includegraphics[width=\linewidth]{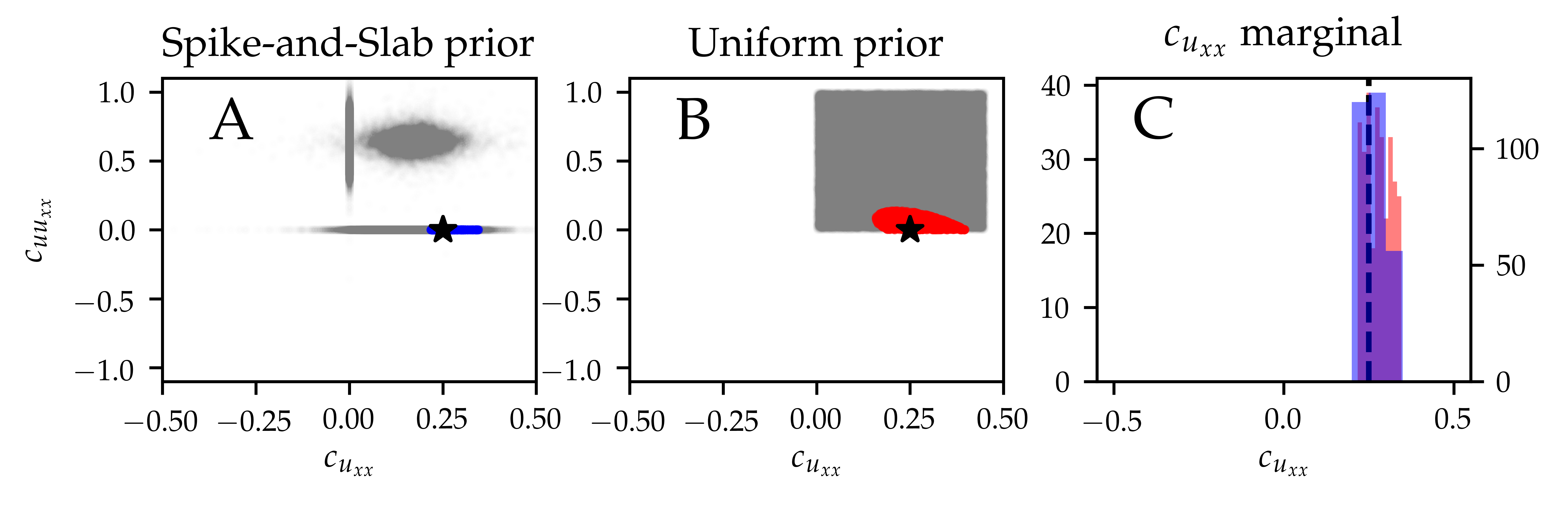}
	\caption{Posterior distributions obtained using dataset $\mathcal{D}_\text{I}^1$. A: spike-and-slab joint posterior distribution (blue) together with the true parameter values (black star) and spike-and-slab prior (grey). B: joint posterior distribution, using a uniform prior (red) together with the true parameter values (black star) and uniform prior (grey). C: marginal distribution of $c_{u_{xx}}$ generated using the spike-and-slab prior (blue) and a uniform prior (red) together with the true parameter value (black dashed line).}
	\label{Figure:CaseIposteriors}
\end{figure}

We set $\varepsilon = 0.15$ for Case I, and $\varepsilon=0.25$ for Case II and Case III and run ABC rejection until a total of $300$ parameters have been sampled in each of the cases. This is equivalent to an acceptance rate of approximately $15\%$ in each of the cases. The inferred posterior distributions are shown in Figures~\ref{Figure:CaseIposteriors},~\ref{Figure:CaseIIposteriors} and~\ref{Figure:CaseIIIposteriors} and we refer to Supplementary Information Section S8 for a full overview of the inferred pairwise marginal posterior distributions of the parameters in Case III. In each of Figures~\ref{Figure:CaseIposteriors},~\ref{Figure:CaseIIposteriors} and~\ref{Figure:CaseIIIposteriors} we also show, for comparison, the posterior obtained by applying ABC rejection sampling using a broad, uniform prior on the pre-selected coefficients. For Case I we take a uniform prior where $c_{u_{xx}} \sim \mathcal{U}(0,0.5)$ and $c_{u\cdot u_{xx}} \sim \mathcal{U}(0,0.8)$; for Case II, $c_{u_{x}} \sim \mathcal{U}(-0.05,0)$ and $c_{u_{xx}} \sim \mathcal{U}(0,0.5)$; for Case III $c_{u} \sim \mathcal{U}(0,0.005)$, $c_{u^2} \sim \mathcal{U}(-0.005,0))$ and $c_{u_{xx}} \sim \mathcal{U}(0,0.5)$. We note that the choice of uniform prior in such cases is non-trivial as it requires some prior knowledge on the part of the practitioner about the relevant parameter ranges. 


\subsubsection{Case I}

In the case of unbiased motility only (Case I, dataset $\mathcal{D}_\text{I}^1$, Figure~\ref{Figure:CaseIposteriors}) we see that use of the spike-and-slab prior distribution (Figure~\ref{Figure:CaseIposteriors}A) results in a posterior distribution with the true parameter value contained in the support of the posterior. Moreover, the sparsity enforced by the prior ensures that the correct PDE structure, with only $u_{xx}$ included, is selected. In contrast, with a uniform prior on $(c_{u_{xx}},c_{u\cdot{u_{xx}}})$ (Figure~\ref{Figure:CaseIposteriors}B) although the true parameter value is contained in the support of the posterior distribution, the correct PDE structure is generally not established, with both $u_{xx}$ and $uu_{xx}$ terms contained in the PDE model.


\subsubsection{Case II}

\begin{figure}[htbp]
	\centering
	\includegraphics[width=\linewidth]{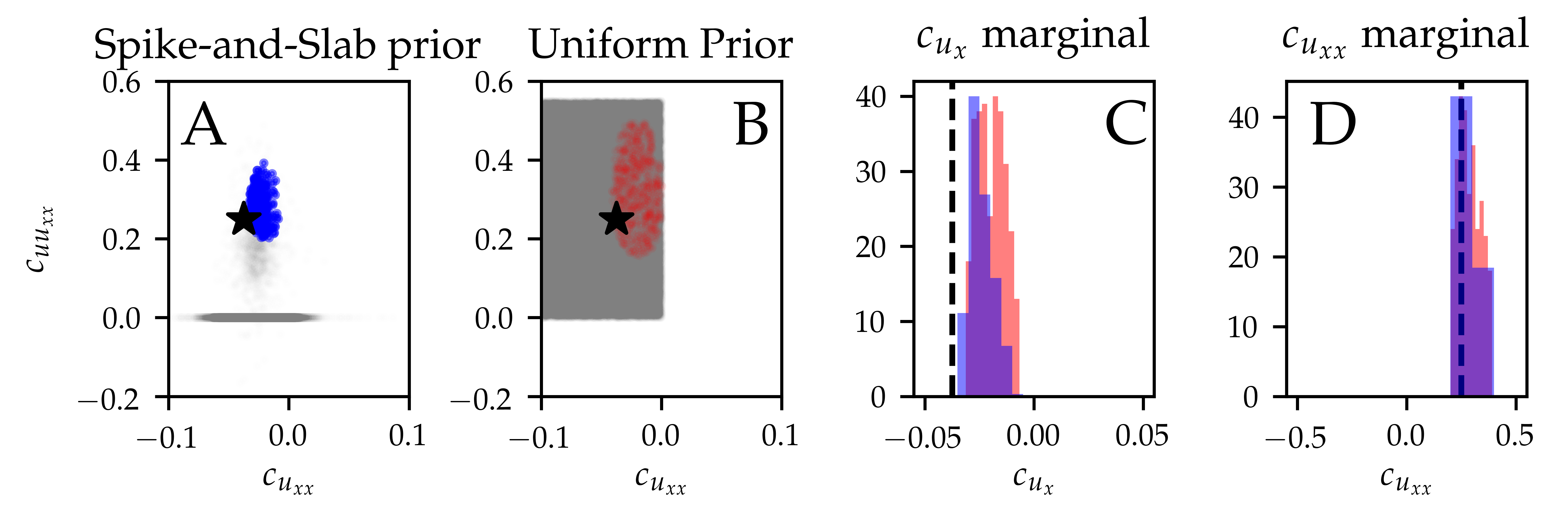}
	\caption{Posterior distributions obtained using dataset $\mathcal{D}_\text{II}^1$. A: joint spike-and-slab posterior distribution (blue) together with the true parameter values (black star) and spike-and-slab prior (grey). B: joint posterior distribution, using a uniform prior distribution (red) together with the true parameter values (black star) and uniform prior (grey). C: marginal distributions of $c_{u_{x}}$ generated using the spike-and-slab prior (blue) and a uniform prior (red) together with the true parameter value (black dashed line). D: marginal distributions of $c_{u_{xx}}$ generated using spike-and-slab prior (blue) and a uniform prior (red) together with the true parameter value (black dashed line).}
	\label{Figure:CaseIIposteriors}
\end{figure}

In the case of biased motility (Case II, dataset $\mathcal{D}_\text{II}^1$, Figure~\ref{Figure:CaseIIposteriors}), both $c_{u_x}$ and $c_{u_{xx}}$ are non-zero for all parameter values accepted into the spike-and-slab posterior distribution. This is a striking result given an identification ratio of just $0.012$ for $c_{u_{xx}}$ after the application of PDE-FIND. We remark that the posteriors in Figure~\ref{Figure:CaseIIposteriors} show that in the presence of model misspecification -- recall that the term $uu_x$ was not identified by PDE-FIND -- the posterior distribution may be biased. In this case it entails that the true parameter values are not contained in the support of the posterior distribution. 


\subsubsection{Case III}

\begin{figure}[htbp]
	\centering
	\includegraphics[width=\linewidth]{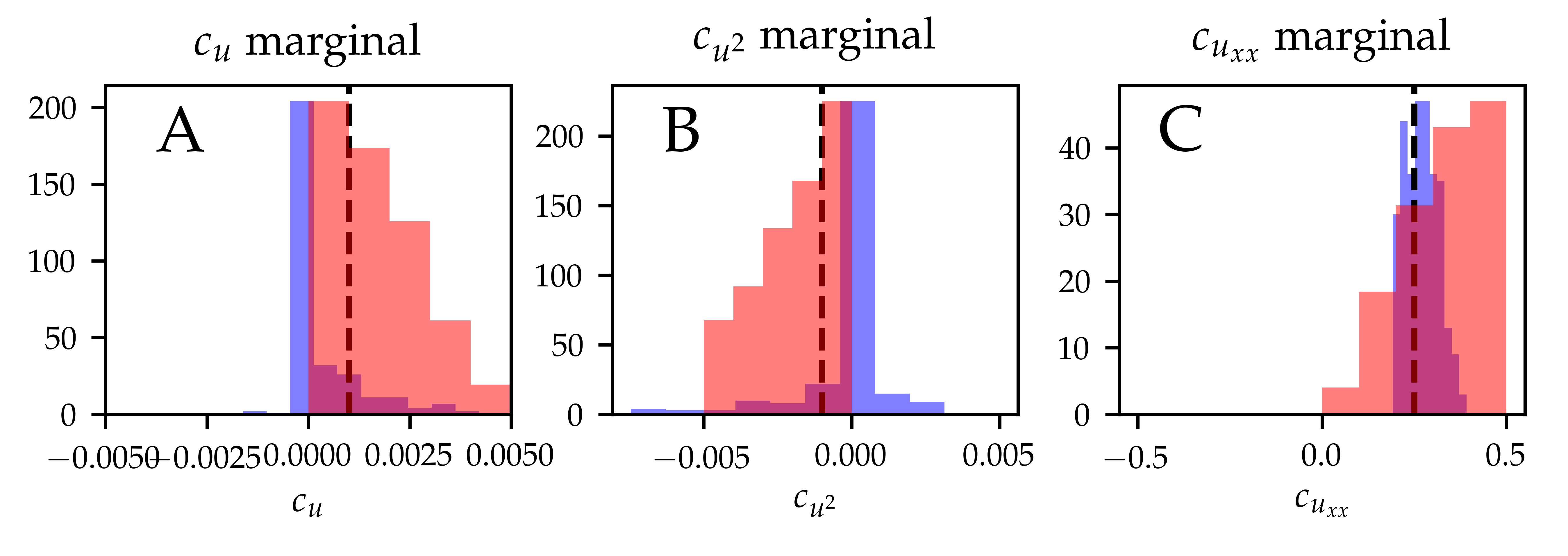}
	\caption{Marginal posterior distributions obtained dataset $\mathcal{D}_\text{III}^1$. A: marginal posterior distributions of $c_{u}$ generated using the spike-and-slab prior (blue) and a uniform prior (red) together with the true parameter value (black dashed line). B: marginal posterior distributions of $c_{u^2}$ generated using the spike-and-slab prior (blue) and a uniform prior (red) together with the true parameter value (black dashed line). C: marginal distributions of $c_{u_{xx}}$ generated using the spike-and-slab prior (blue) and a uniform prior (red) together with the true parameter value (black dashed line).}
	\label{Figure:CaseIIIposteriors}
\end{figure}

In the case of unbiased motility and proliferation (Case III, dataset $\mathcal{D}_\text{III}^1$, Figure~\ref{Figure:CaseIIIposteriors}), the posterior obtained using the PDE-FIND prior still contains some accepted parameter samples with either $c_{u}$ or $c_{u^2}$ equal to zero, demonstrating that there is potential non-identifiability of these terms given the data. However, all parameter samples in the posterior have non-zero $c_{u_{xx}}$, a significant result given an identification ratio of $0.57$ for $c_{u_{xx}}$. We note that the support of the spike-and-slab posterior contains the true parameter value for $c_{u_{xx}}$, even though it is not in the support of the empirical distribution of the PDE-FIND coefficients from the exploration subset.


\bigskip


\subsubsection{Computational performance}
To highlight the performance of our method, we compare the computational cost and the accuracy of our method against alternative options. In addition to the spike-and-slab model and uniform priors used to generate the posterior distributions (which we call \textit{informed spike-and-slab} and \textit{sparse uniform}, we also consider a spike-and-slab prior with mean $0$ in the slab for each coefficient (\textit{i.e.} the PDE-FIND screen only informs the prior through the identification ratios), which we call \textit{naive spike-and-slab}, as well as a uniform prior on all library coefficients (\textit{i.e.} the classic Bayesian scenario, where no variable selection is performed), which we refer to as \textit{classic Bayesian}. In each of the experiments, we perform ABC rejection to sample 300 parameters from the ABC posterior with the same thresholds as used in Case I-III and record the time taken to complete. Inference is done on a Lenovo desktop computer using 6 Intel(R) i5-8500T cores with clock speed 2.10GHz. Table \ref{table:performance} shows the time taken for each of the experiments alongside with their acceptance rates.
\begin{table}[h!]
    \label{table:performance}
	\centering
		\begin{tabular}{|c|c|c|c|} 
		\hline
		Method & Case I & Case II & Case III \\ [0.5ex] 
		\hline\hline
		Informed spike-and-slab & 1657 $s$, 24.13\% &  3337 $s$, 16.93\% & 2896 $s$, 19.66\%\\ 
		\hline
		Naive spike-and-slab & 8523 $s$, 6.68\% & 40867 $s$, 1.19\% & 40128 $s$, 1.08\%\\
		\hline
		Sparse uniform & 40658 $s$, 1.78\% & 81975 $s$, 0.5\% & 86213 $s$, 0.5\%\\
		\hline
		Classic Bayesian & Did not converge & Did not converge & Did not converge \\
		\hline\hline
	\end{tabular}
	\caption{Comparison of computational time and acceptance probabilities for alternatives to Bayes-PDEFIND prior. Using an informed prior significantly outperforms all alternatives.}
\end{table}
We note that the computational time of the spike-and-slab models is significantly lower than any of the uniform prior models and that using an informed spike-and-slab prior offers a substantial speed-up in computational time than using a naive spike-and-slab approach. The approach using a pre-screened uniform implementation failed to yield a sparse set of coefficients, thus showing unacceptable accuracy in learning the correct equations. The classic Bayesian analysis did not finish sampling within $1.5\cdot 10^6 s$ (approximately two weeks). By calculating the dimensionality of the space, we estimate that the acceptance probability should be expected to be of order $10^{-2} \%$, which confirms that performing a classical Bayesian analysis in such a case is inappropriate. We highlight that many applications of EQL methods will have even larger libraries, making the computational time of naive uniform priors exponentially longer. The posteriors from both naive and informed spike-and-slab models are qualitatively similar across all cases and both identify the correct regions of parameter space.

\subsubsection{Posterior predictive check}

\begin{figure}[htbp]
    \centering
    \includegraphics[width=\linewidth]{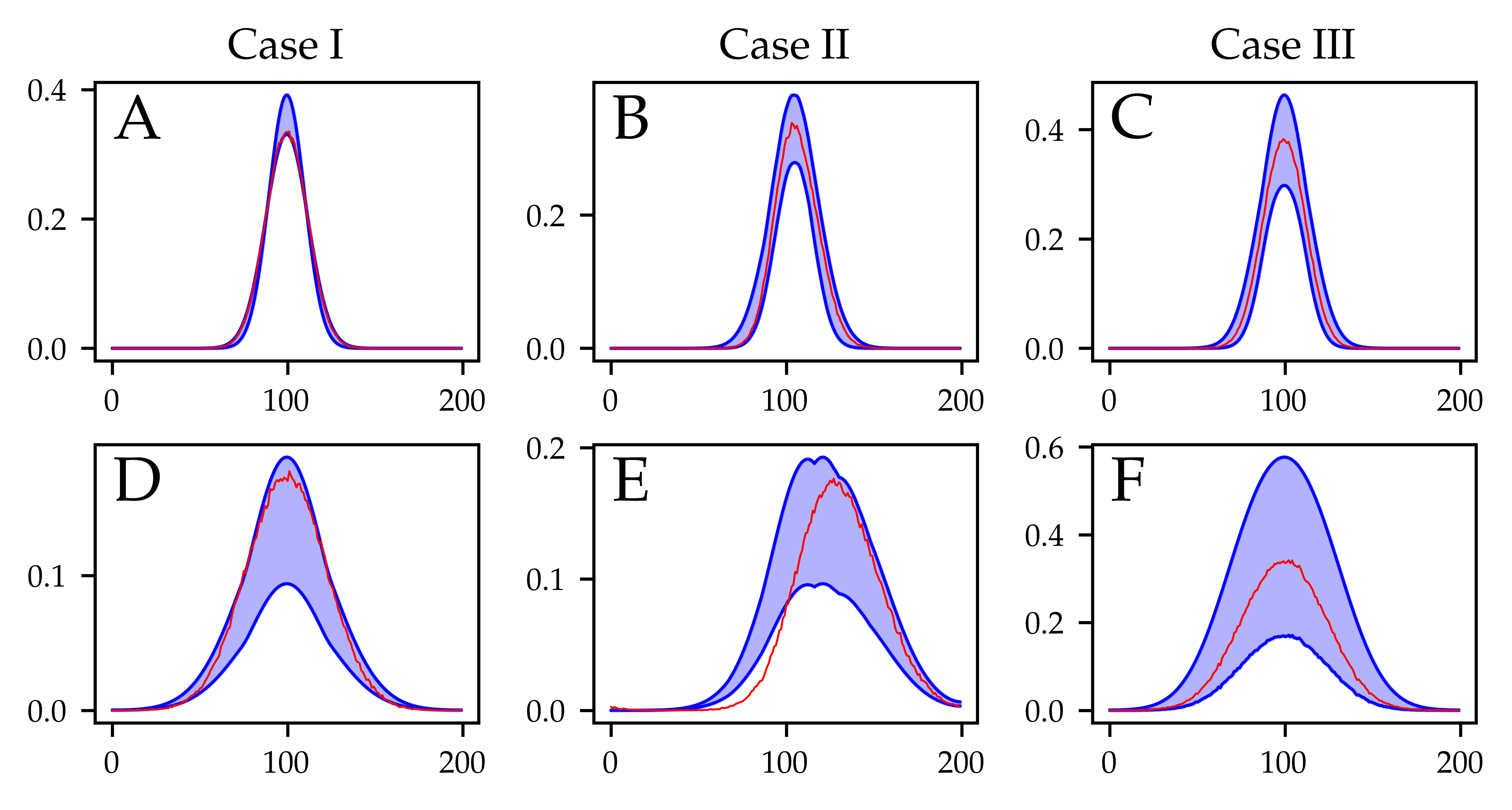}
    \caption{Posterior predictive check. We plot the $5\%$ and $95\%$ percentiles of the density distribution at each spatial location. Top row: evaluation of the interpolative capability of the PDE model using an average over all samples in the dataset (\text{i.e.} using $\langle{}C_i(t)\rangle{}_\text{LN}$, for $i=1,\ldots,200$, red) at $t=250$. Bottom row: evaluation of the extrapolative capability of the PDE model using an average over all samples in the dataset at $t=1000$ (red).}
    \label{fig:posterior_check}
\end{figure}

In summary, Figures~\ref{Figure:CaseIposteriors}--\ref{Figure:CaseIIIposteriors} highlight that the use of PDE-FIND in combination with ABC rejection can significantly reduce the uncertainty associated with the PDE coefficients. In Cases I and II, uncertainty regarding which parameter to include in the model is completely removed, as the posterior is has support only on the diffusion parameter axis in Case I and on a region where both $c_{u_x}$ and $c_{u_{xx}}$ are nonzero in Case II. In contrast, a uniform prior over all coefficients that have sufficiently large identification ratio (those for which $A_i>\delta$) does not enforce sparsity, which means that the resulting PDE models can be misspecified and / or contain greater complexity than is necessary to accurately predict the data. To further assess the quality of the resulting posterior parameter distributions, we carried out a posterior predictive check (Figure \ref{fig:posterior_check}). For each parameter sample accepted into the posterior distribution, we used numerical integration, as detailed in the Methods section, to obtain a prediction for the density at $t=250$ to assess how well the model interpolates the data, and a prediction for the density at $t=1000$, which is $t=750$ beyond the time horizon used to train PDE-FIND on the inference subset. We then plot the $5\%$ and $95\%$ quantiles of the output distributions and overlay them with a representative sample from each of the datasets for Cases I, II and III. The results shown in Figure \ref{fig:posterior_check} demonstrate that the PDE model predictions can both interpolate and extrapolate the data well. We conclude that even in the presence of model misspecification, such as in Case II, it is possible to obtain a posterior with reasonable predictive power, although as the time horizon is extended beyond the time horizon of the training data, the misspecification becomes apparent in the systematic prediction error. To highlight the increased accuracy, we compare to results shown in Figure~\ref{fig:comparing_predictions}, where the integrated model solutions fail to resemble the empirical data when PDE-FIND is used in isolation.


\section{Discussion and outlook}\label{section:discussion}

The aim of this work was to develop and showcase a framework to perform uncertainty quantification for  equation learning methods in the context of noisy spatio-temporal data. In essence, our approach harnesses equation learning methodologies to generate an informed prior distribution, that is then used within a Bayesian framework to estimate both the model structure and posterior parameter distribution. The framework was developed in the context of the PDE-FIND equation learning methodology and ABC rejection sampling, but it is sufficiently general that it could be extended and applied in the context of both other equation learning and Bayesian inference methodologies. 

The motivation for developing such a framework stems from the fact that equation learning methodologies such as PDE-FIND typically make variable predictions, both in terms of model structure and parameters, in the presence of noise, which is common for datasets in the life and biomedical sciences. Incorporating uncertainty quantification through the use of Bayesian statistics approaches provides a means to quantify the uncertainty in both model structure and parameter values, and understand how such uncertainty propagates into model predictions. 

We showcased our methodology in the context of noisy spatio-temporal data generated using a canonical ABM that has seen widespread use in the modelling of motile and proliferative cell populations, and which has a corresponding PDE model that can make accurate predictions of averaged ABM output. We used datasets generated in three different parameter regimes (that incorporate different cellular behaviours) to show how to combine the advantages of the PDE-FIND algorithm (efficiency and the ability to learn simple, interpretable models) with those of ABC (ability to quantify uncertainty in parameter estimates and model predictions). 

There are a number of ways in which our approach can be further improved going forward. For example, we saw that our approach may return models that fail to capture important features of the ABM, either as a result of learning the structure of the model incorrectly or inaccurate estimation of model parameter values. In the context of Case II, which models biased cell motility, these inaccuracies arise partly as a result of the data used and stem from the choice of initial condition, and the timescale over which the data are collected. Ultimately, the extensibility and robustness of any data-driven method are limited by the information contained in the data. Where the initial PDE-FIND screen fails to identify what are believed to be relevant terms in the PDE, it is likely that the supplied data does not provide enough information to discriminate between different forms of the PDE. In such a scenario, different experimental designs, such as a different initial condition or longer simulation time, might distinguish some for the terms of the system under consideration. We highlight that the flexibility of ABMs allows one to explore different behaviours in the model under varying experimental conditions. By analysing the effect of these variations on the resulting predictions, one can obtain important insights into how much information is contained in the data about the governing laws. This may be helpful in informing experiment design \textit{in vitro}, so that the experimental design can provide as much information as possible. Nonetheless, the standard choice of PDE-FIND library means that it is possible for the PDE-FIND algorithm to return PDE models in which the density does not satisfy a conservation equation of the form
\begin{equation}
u_t = -\nabla\cdot\boldsymbol{F}+S(u),
\end{equation}
where $\boldsymbol{F}$ is the flux and $S$ is the net proliferation rate, and hence for the models to make unphysical predictions (as occurs in Case II). A possible solution to this specific problem could be to encode the terms in the candidate library in flux form. More generally, however, it is not obvious how to balance the wish to include constraints in specifying terms in the candidate library whilst at the same time avoiding over-constraining the space of possible output PDE models. Bayesian approaches may prove useful this respect. In the case of severe model mis-specification due to incompleteness of the supplied library, Bayes-PDE-FIND offers several possibilities. In some scenarios, the learned PDE will interpolate the data well and extrapolate to new settings, even though it contains library terms that are different from the ground truth. Such a PDE can still be used for the purposes of simulation and inference, since the benefit of such a PDE model is that it is fast to solve, which is often crucial when performing inference. When the learned PDE terms performs poorly in interpolating or extrapolating, Bayes-PDEFIND returns a quantification of the error between model solutions and observed data. This may offer real-world insights: the library terms are usually provided by practitioners to reflect hypothesised mechanisms in the system under consideration. That those terms fail to explain the data provides a motivation to reconsider which mechanisms should form part of the model.

Secondly, our approach could be improved through the use of more efficient ABC samplers, such as ABC sequential Monte Carlo samplers that target the posterior distribution by evolving the prior distribution through a series of intermediate distributions. This requires the development of proposal distributions that can maintain the sparsity of PDE coefficients, as encoded by the PDE-FIND informed prior, so that the learned PDEs retain a simple and interpretable structure.



\textbf{Data access:} All code to generate synthetic data, as well as code used to analyse the data is available on Github at \url{https://github.com/simonmape/UQ-for-pdefind}.

\textbf{Author contributions:} S.M.P. created the code, produced all figures and carried out the analysis; R.E.B. and M.J.S. helped design, supervised and coordinated the study. S.M.P. wrote the paper, on which all other authors commented and revised. All authors gave final approval for publication.

\textbf{Competing interests:} There are no competing interests.

\textbf{Funding:} S.M.P. is supported by an EPSRC/UKRI Doctoral Training Award. M.J.S. is supported by the Australian Research Council (DP200100177). R.E.B. acknowledges funding from the BBSRC via BB/R00816/1 and would like to thank the Royal Society for a Wolfson Research Merit Award.

\textbf{Acknowledgements:} The authors would like to thank the referees for their comments.



\bibliographystyle{RS}
\bibliography{BIBL}


\end{document}


\title{Supplementary Information: Bayesian uncertainty \\ quantification for data-driven equation learning}

\author{Simon Martina-Perez, Matthew J. Simpson and Ruth E. Baker}

\date{}

\maketitle


\section{Comparison of averaged behaviours of the ABM with the solution of the coarse-grained PDEs}

To further demonstrate the excellent agreement between the numerical solution of the coarse-grained PDE, Equation (2.4), and the observed densities, we compare solutions of the coarse-grained PDE with averaged ABM data. The PDE model is solved numerically using the PyPDE package~[31], which solves the PDE using the method of lines, where space is discretised using the grid on which spatial data for the ABM is collected. The resulting ODEs are solved using a Runge-Kutta solver. For each of Case I, II, and III, we average over $K=1000$ realisations of the ABM. We numerically integrate the corresponding coarse-grained PDEs to find solutions at times $t=0,50,150,500$ and plot the observed averaged densities on the same axes in Figure S1. 


\begin{figure}[htbp]
    \centering
    \includegraphics{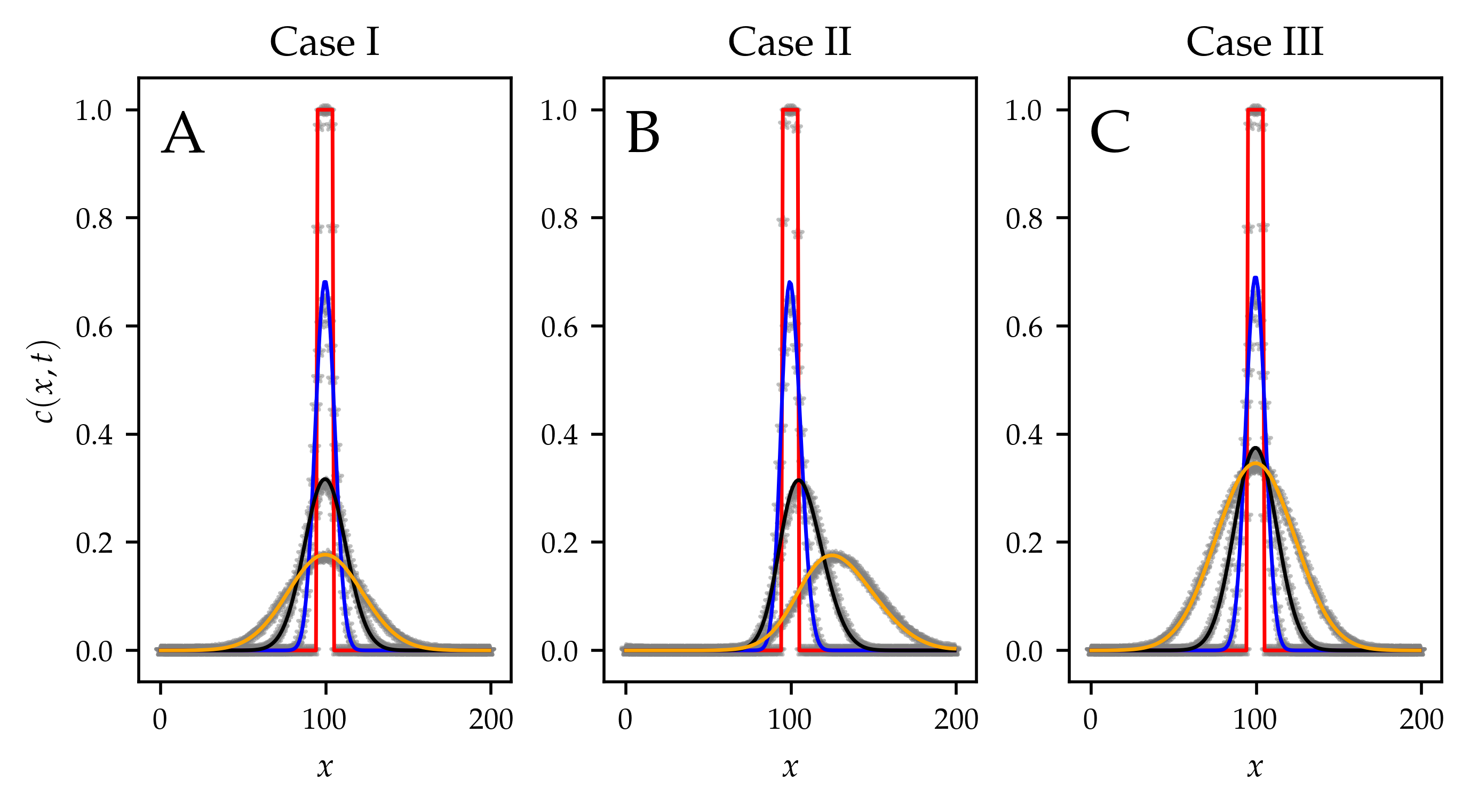}
    \caption{Plots comparing averaged ABM data ($K=1000$ realisations) with numerical solution of the coarse-grained PDE at times $t= 0,50,150,500$. Parameters are as stated in Section ii of the main text.}
    \label{fig:PDE_and_data_comparison}
\end{figure}


\clearpage
\pagebreak

\section{Variability of PDE-FIND coefficients}


\subsection{Identification ratios}

We report the identification ratios for the problems considered in the main text together with the additional experiments performed for this Supplementary Information in Table S1.


\begin{table}[h!]
	\centering
	\begin{tabular}{|l||c|c|c|c|c|c|c|c|c|} 
		\hline
		Experiment & $c_{1}$ & $c_{u}$ & $c_{u^2}$ & $c_{u_x}$ & $c_{u\cdot u_x}$ & $c_{u^2\cdot u_{x}}$ & $c_{u_{xx}}$ & $c_{u\cdot u_{xx}}$ & $c_{u^2\cdot u_{xx}}$\\ [0.5ex] 
		\hline\hline
		STRidge $\mathcal{D}_\text{I}^1$ & 0.001 & 0.0 & 0.002 & 0.0 & 0.008 & 0.008 & \cellcolor[gray]{0.7}\textbf{0.826} & 0.199 & 0.05\\
		\hline
		STRidge $\mathcal{D}_\text{I}^2$ & 0.0 & 0.0 & 0.0 & 0.0 & 0.0 & 0.0 & \cellcolor[gray]{0.7}\textbf{1.0} & 0.0 & 0.0\\
		\hline
		STRidge $\mathcal{D}_\text{II}^1$ & 0.0 & 0.0 & 0.0 & \cellcolor[gray]{0.7}\textbf{0.999} & \cellcolor[gray]{0.7}\textbf{0.0} & 0.0 & \cellcolor[gray]{0.7}\textbf{0.012} & 0.002 & 0.0\\
		\hline
		STRidge $\mathcal{D}_\text{II}^2$ & 0.0 & 0.0 & 0.0 & \cellcolor[gray]{0.7}\textbf{1.0} & \cellcolor[gray]{0.7}\textbf{0.0} & 0.0 & \cellcolor[gray]{0.7}\textbf{0.0} & 0.0 & 0.0\\
		\hline
		STRidge $\mathcal{D}_\text{III}^1$, $d_{\text{tol}} = 0.005$ & 0.234 & \cellcolor[gray]{0.7}\textbf{0.2} & \cellcolor[gray]{0.7}\textbf{0.009} & 0.005 & 0.001 & 0.007 & \cellcolor[gray]{0.7}\textbf{0.993} & 0.012 & 0.059\\
		\hline
		STRidge $\mathcal{D}_\text{III}^1$, $d_{\text{tol}} = 0.01$ & 0.013 & \cellcolor[gray]{0.7}\textbf{0.659} & \cellcolor[gray]{0.7}\textbf{0.482} & 0.002 & 0.002 & 0.007 & \cellcolor[gray]{0.7}\textbf{0.571} & 0.01 & 0.014\\
		\hline
		STRidge $\mathcal{D}_\text{III}^2$, $d_{\text{tol}} = 0.005$ & 0.0 & \cellcolor[gray]{0.7}\textbf{1.0} & \cellcolor[gray]{0.7}\textbf{0.0} & 0.0 & 0.0 & 0.0 & \cellcolor[gray]{0.7}\textbf{1.0} & 0.12 & 0.0\\
		\hline
		STRidge $\mathcal{D}_\text{III}^2$, $d_{\text{tol}} = 0.01$ & 0.0 & \cellcolor[gray]{0.7}\textbf{1.0} & \cellcolor[gray]{0.7}\textbf{0.0} & 0.0 & 0.0 & 0.0 & \cellcolor[gray]{0.7}\textbf{1.0} & 0.0 & 0.0\\
		\hline
		rPCA $\mathcal{D}_\text{I}^1$ & 0.0 & 0.408 & 0.69 & 0.0 & 0.01 & 0.001 & \cellcolor[gray]{0.7}\textbf{0.384} & 0.143 & 0.089\\
		\hline
		STRidge Convolution $\mathcal{D}_\text{I}^1$ & 0.0 & 0.003 & 0.222 & 0.0 & 0.0 & 0.001 & \cellcolor[gray]{0.7}\textbf{0.765} & 0.008 & 0.017\\
		\hline
		STRidge Convolution $\mathcal{D}_\text{II}^1$ & 0.0 & 0.002 & 0.109 & \cellcolor[gray]{0.7}\textbf{0.882} & \cellcolor[gray]{0.7}\textbf{0.028} & 0.005 & \cellcolor[gray]{0.7}\textbf{0.181} & 0.002 & 0.007\\
		\hline
		STRidge Convolution $\mathcal{D}_\text{III}^1$ & 0.0 & \cellcolor[gray]{0.7}\textbf{0.922} & \cellcolor[gray]{0.7}\textbf{0.086} & 0.0 & 0.0 & 0.001 & \cellcolor[gray]{0.7}\textbf{0.046} & 0.002 & 0.026\\
		\hline
		STRidge  Subsampling $\mathcal{D}_\text{I}^1$ & 0.0 & 0.0 & 0.003 & 0.001 & 0.002 & 0.007 & \cellcolor[gray]{0.7}\textbf{0.998} & 0.365 & 0.063\\
		\hline\hline
	\end{tabular}
	\caption{Values of identification ratios for all experiments used in this work. Terms that arise in the corresponding coarse-grained PDE are highlighted in grey and using \textbf{bold} font.}
	\label{SI:a_table}
\end{table}


\subsection{Empirical distribution of all learned PDE-FIND parameters}

In this section we report the pairwise distributions of all learned PDE-FIND parameters in Case I, Case II and Case III, both in the low-noise and high-noise regimes. For convenience, the terms in the library are ordered as: $1$, $u$, $u^2$, $u_x$, $u{u_x}$, $u^2{u_x}$, $u_{xx}$, $u{u_{xx}}$ and $u^2{u_{xx}}$.

\medskip


\begin{figure}
    \centering
    \includegraphics[width=\linewidth]{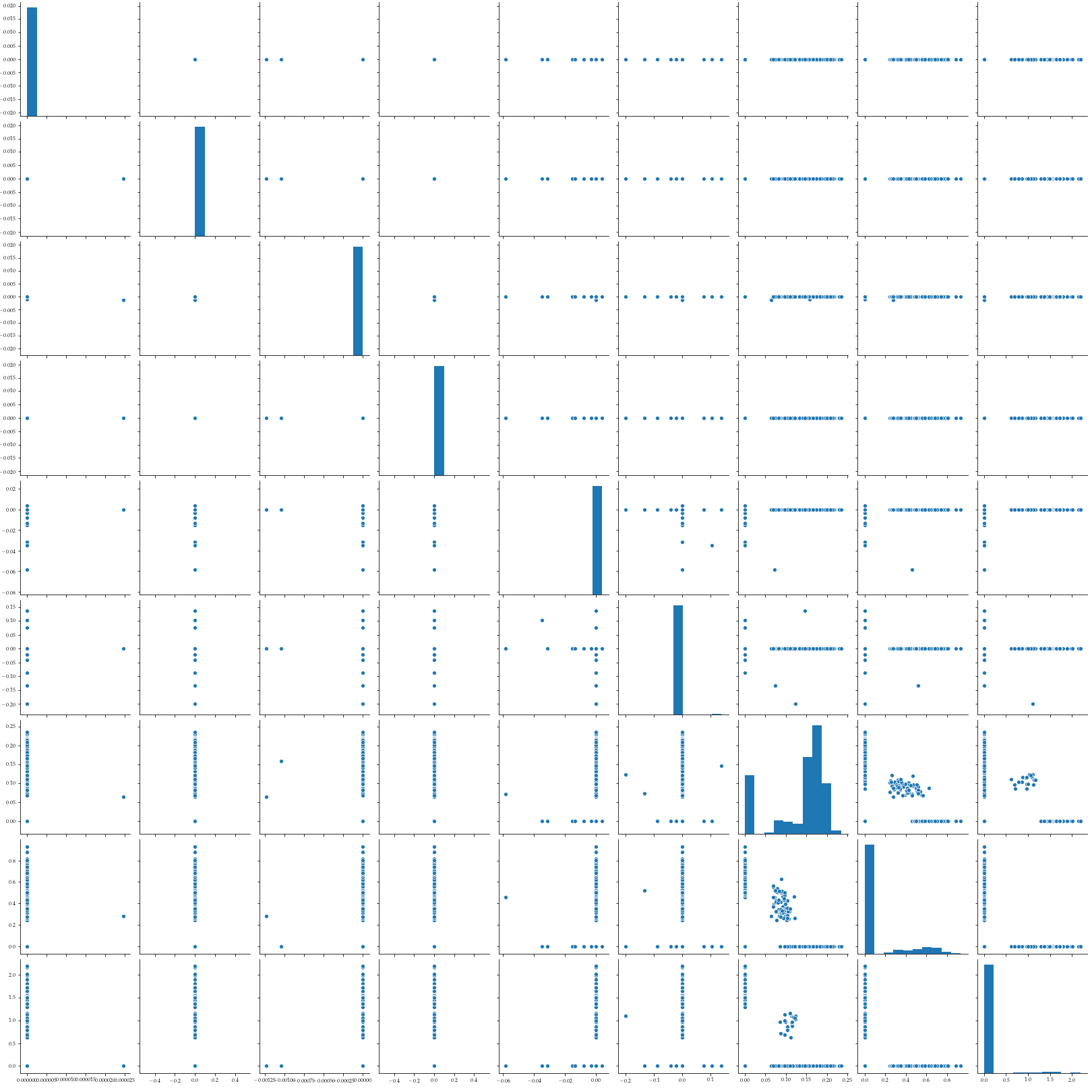}
    \caption{Pairwise correlation plots for coefficients estimated using $\mathcal{D}_\text{I}^1$.}
\end{figure}


\begin{figure}
    \centering
    \includegraphics[width=\linewidth]{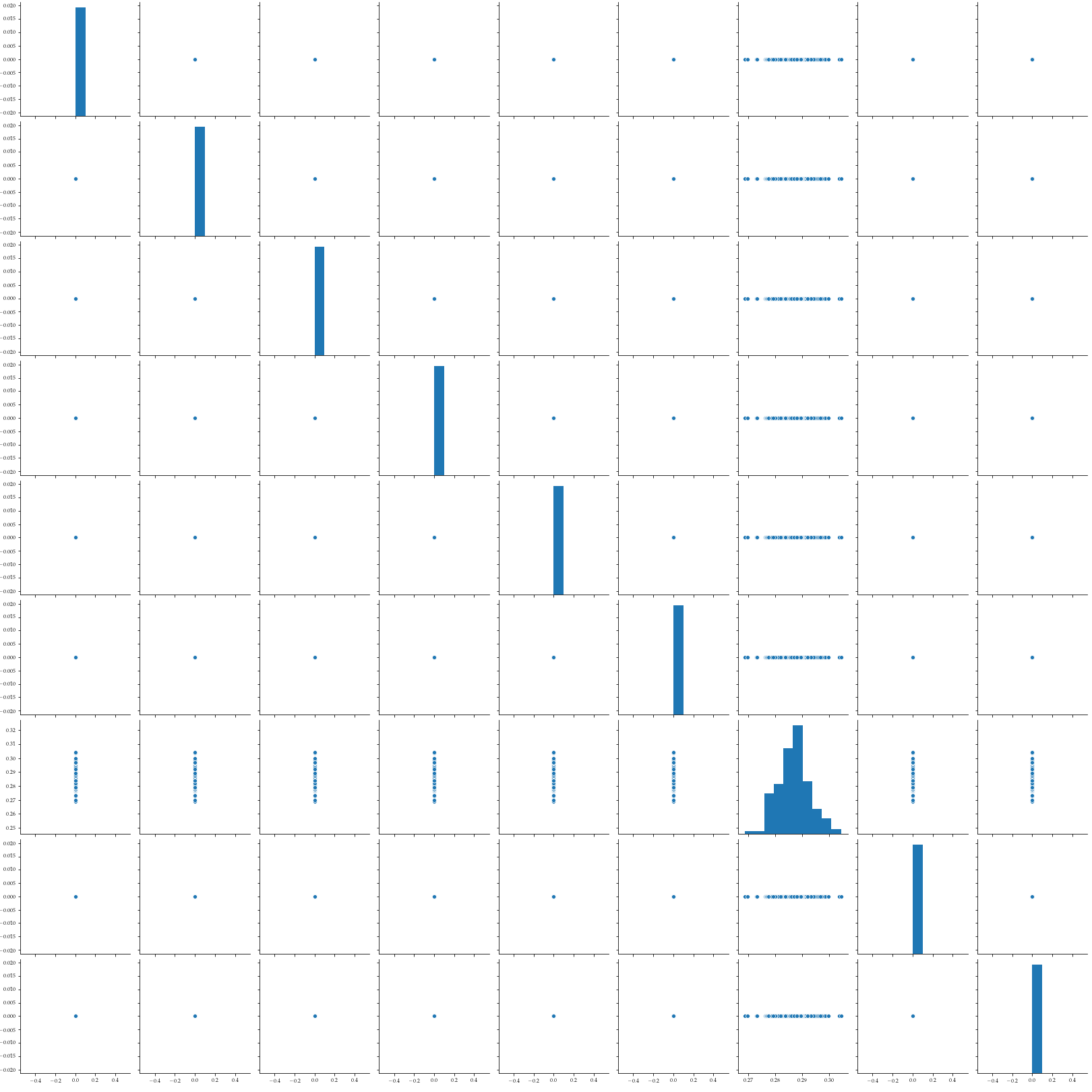}
    \caption{Pairwise correlation plots for coefficients estimated using $\mathcal{D}_\text{I}^2$.}
\end{figure}


\begin{figure}
    \centering
    \includegraphics[width=\linewidth]{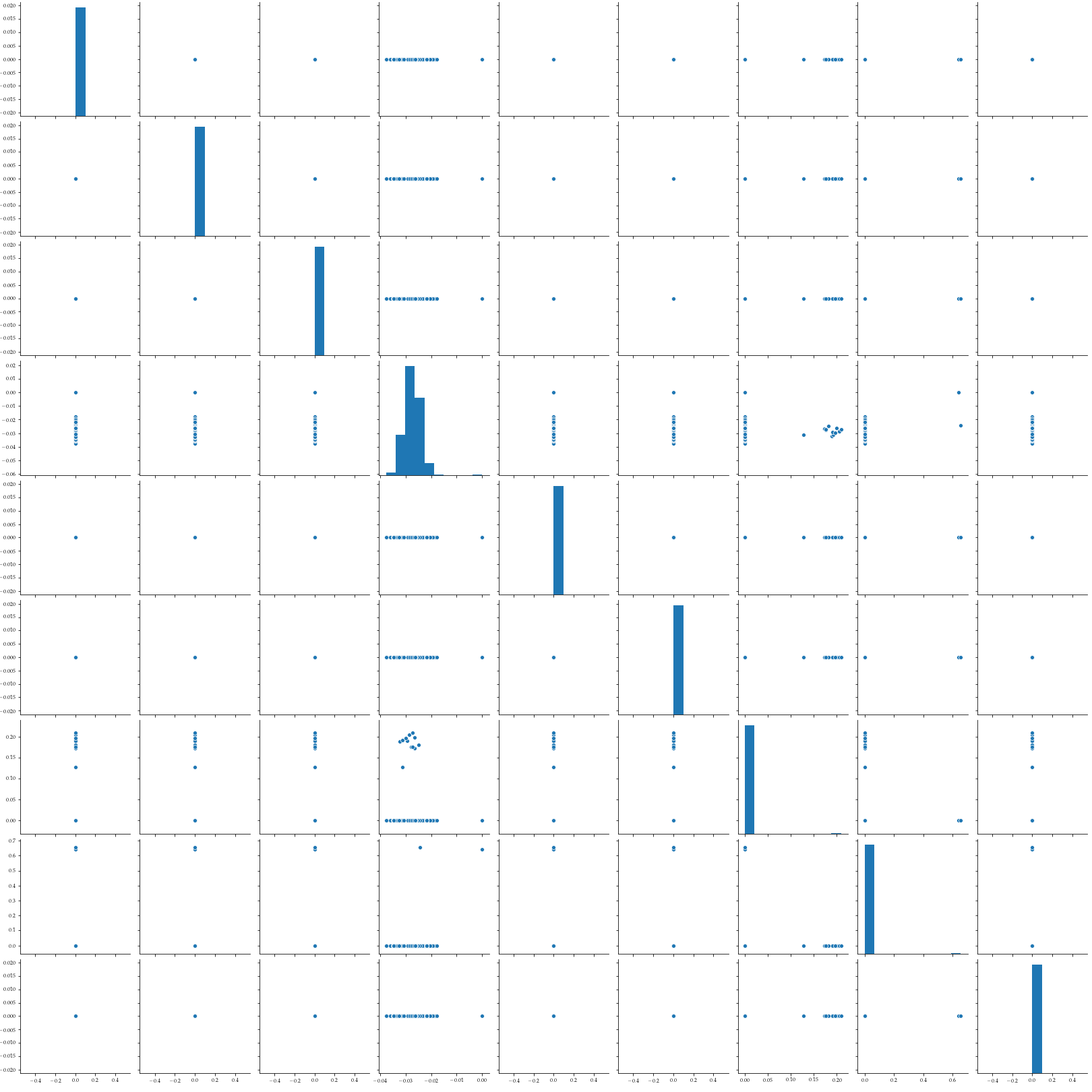}
    \caption{Pairwise correlation plots for coefficients estimated using $\mathcal{D}_\text{II}^1$.}
\end{figure}


\begin{figure}
    \centering
    \includegraphics[width=\linewidth]{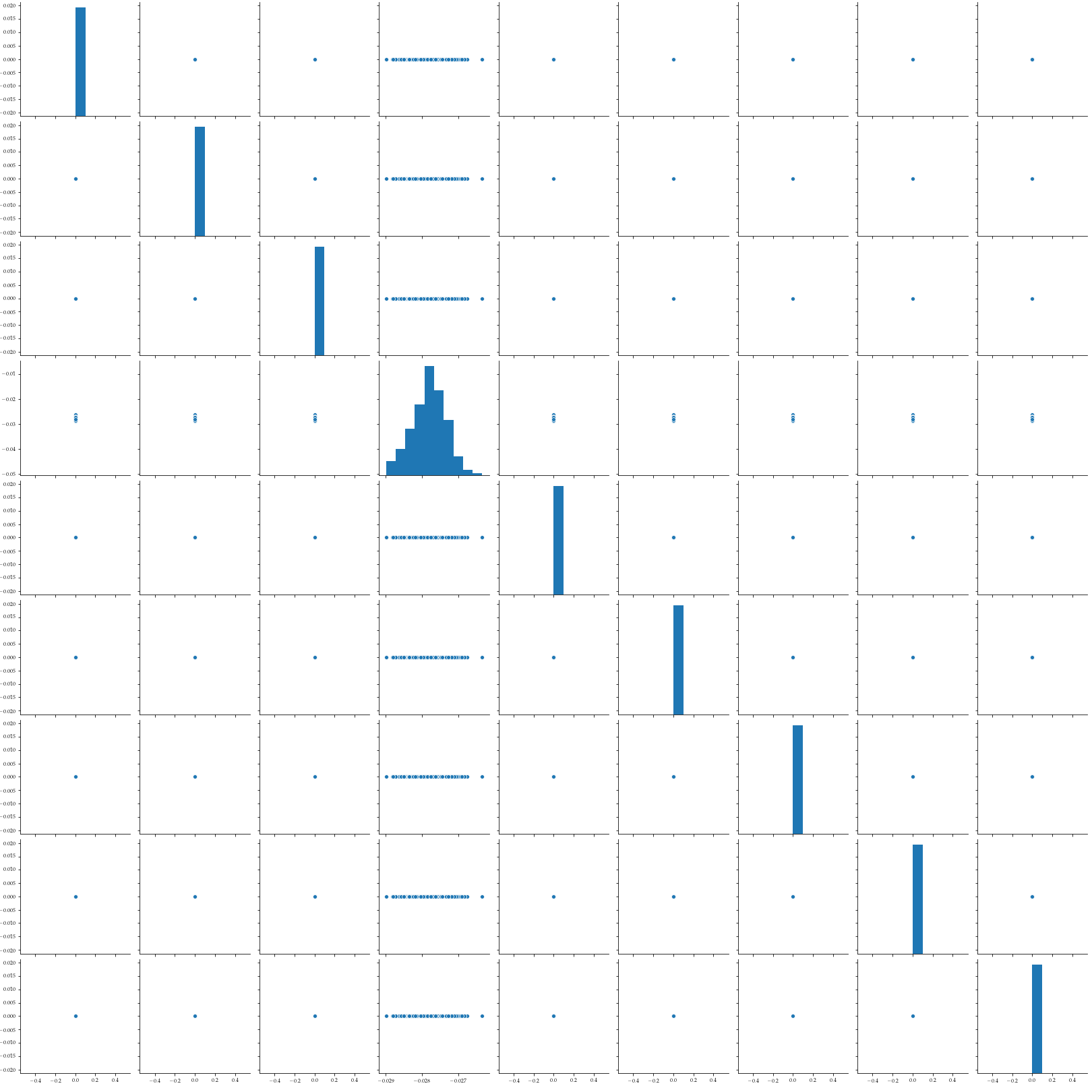}
    \caption{Pairwise correlation plots for coefficients estimated using $\mathcal{D}_\text{II}^2$.}
\end{figure}


\begin{figure}
    \centering
    \includegraphics[width=\linewidth]{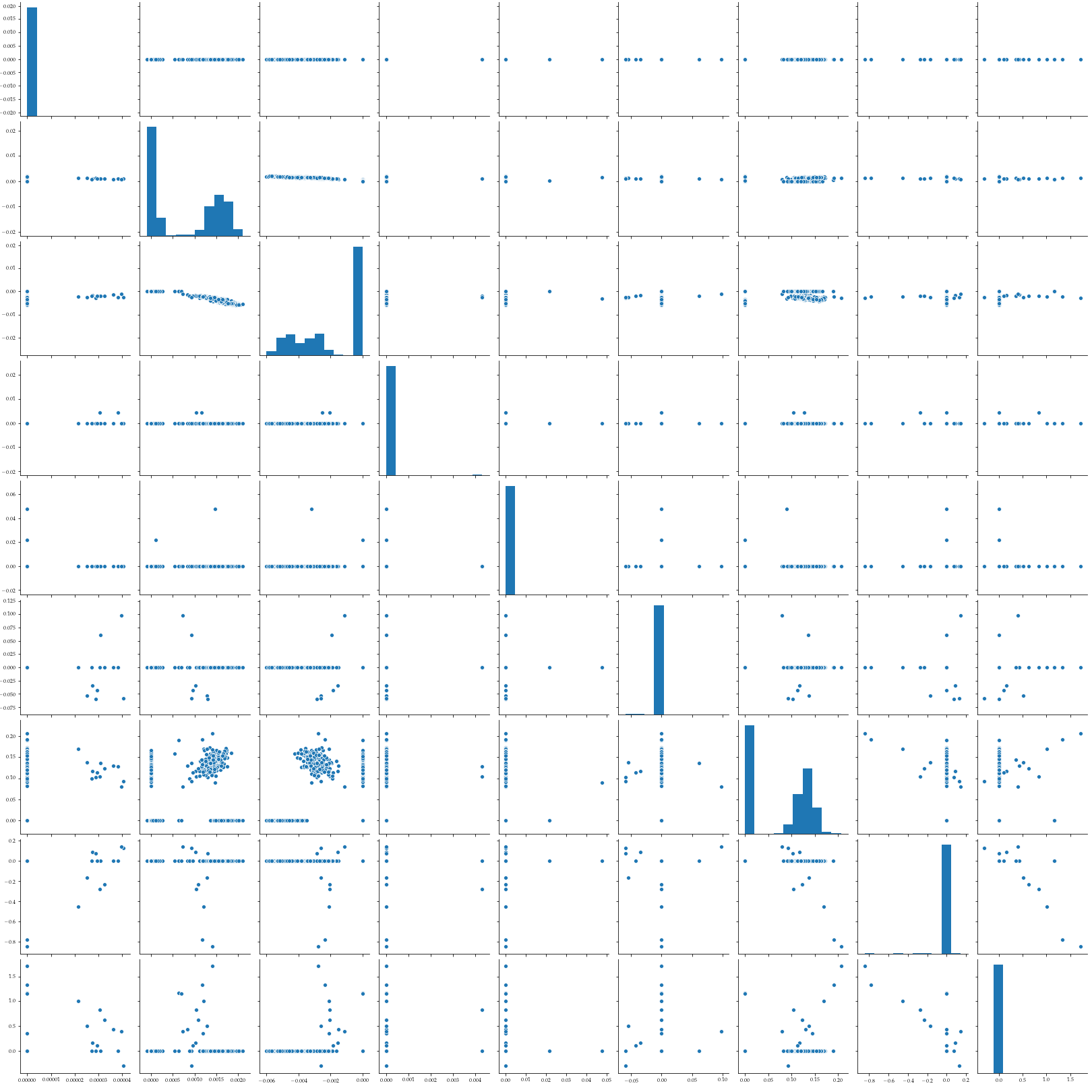}
    \caption{Pairwise correlation plots for coefficients estimated using $\mathcal{D}_\text{III}^1$.}
    \label{SIFig:pairwise_FPU1}
\end{figure}


\begin{figure}
    \centering
    \includegraphics[width=\linewidth]{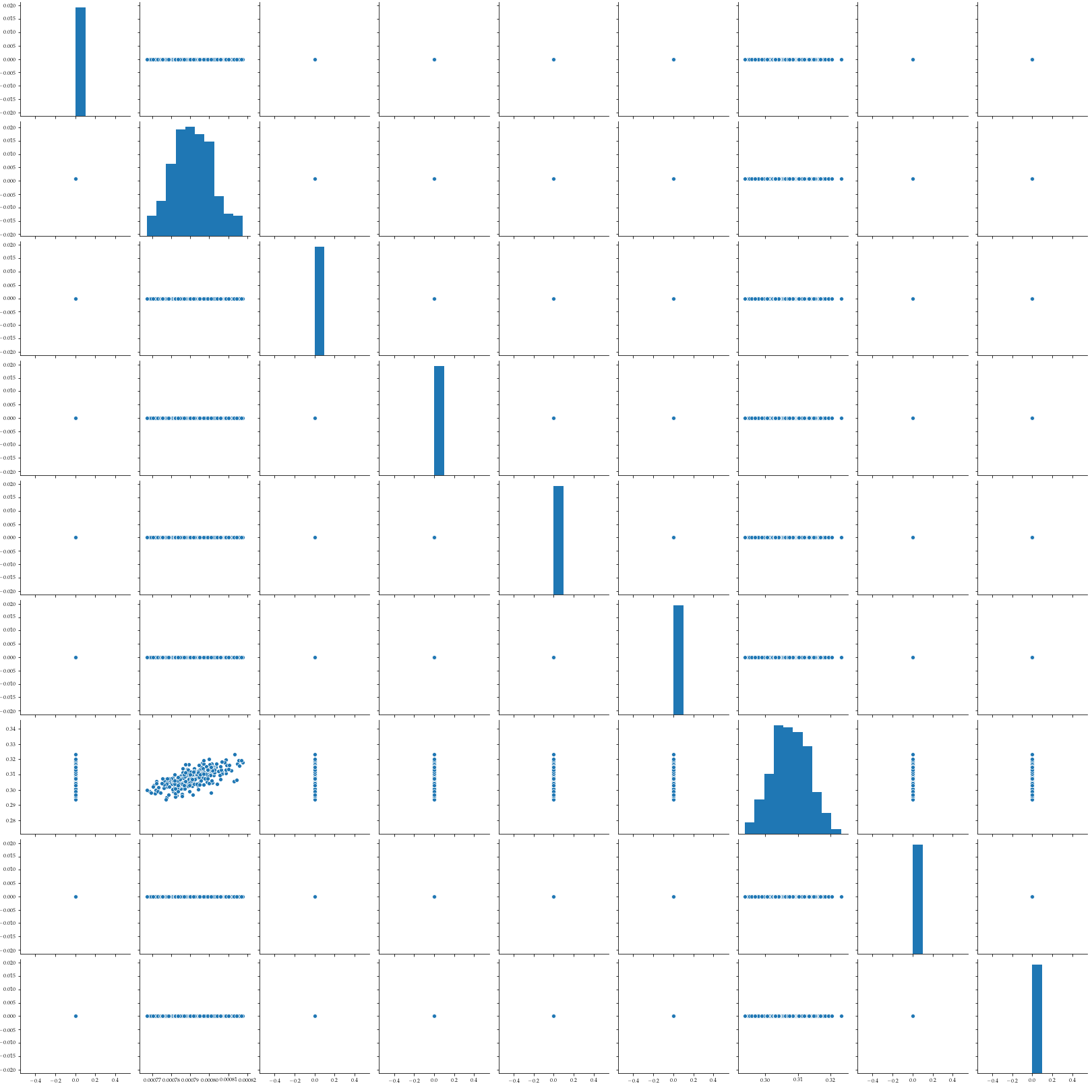}
    \caption{Pairwise correlation plots for coefficients trained on $\mathcal{D}_\text{III}^2$.}
\end{figure}

\clearpage


\section{Denoising approaches}\label{SIsection:denoising}


\subsection{Rescaled Robust PCA}\label{section:rPCA}

Li \textit{et al.}~[22] propose an approach to perform PDE-FIND on noisy data. They denote the data matrix $U \in \mathbb{C}^{n\times m}$, given on a discretized domain $x \in [a,b]$ and $t \in [0,T]$. That is, $U$ is the discretely and noisily sampled measurements from the function $u(x,t)$ that satisfies the target PDE. The authors observe that the underlying data is often low-rank and so they suppose that $U$ and its time derivative $U_t$ can be decomposed as
\begin{equation}
U=Z+E_1, \qquad U_t = \mathcal{D}(Z,Q)\xi +E_2,
\end{equation}
where $Z$ and $\mathcal{D}(Z,Q)$ have low rank and $E_1$, $E_2$ are sparse. Informally, $Z$ represents the \emph{clean} data, $\mathcal{D}(Z,Q)$ the \emph{clean} derivative matrix, and $E_1$, $E_2$ the perturbations of the clean data and derivatives, respectively. The method developed in~[22] aims to  find optimal $Z$ and $\mathcal{D}(Z,Q)$ from the data matrix $U$. This problem is addressed by solving the optimisation problem 
\begin{equation}
\min_{\xi, Z,E_1,E_2}\Vert Z\Vert_{*} + \Vert\mathcal{D}(Z,Q)\overline{\xi}\Vert_{*} + \lambda_1 \mathcal{R}(\overline{\xi})+\lambda_2\Vert E_1 \Vert_1 + \lambda_3 \Vert E_2\Vert_1,
\end{equation}
such that $U = Z+E_1$ and $U_t = \mathcal{D}(Z,Q)\xi + E_2$. Additionally, $\mathcal{R}(\overline{\xi})$ is a sparse regularisation of the parameters $\xi$, and the $\lambda_i$, for $i=1,2,3$, are positive constants. For details on solving this nonconvex optimisation problem, we refer to~[22]. However, essence of the technique is that the original data is first processed with an algorithm called Robust Principal Component Analysis (rPCA) to obtain $Z$, after which Low-rank STRidge (LrSTR) is used to construct $\mathcal{D}(Z,Q)$ from $Z$, and $U_t$ is denoised prior to applying PDE-FIND. The final algorithm, which uses rPCA and LrSTR in tandem, is called Double Low-rank Sparse Regression (DLrSR). While the examples given by Li \textit{et al.} in~[22] show that the method is capable of retrieving the true parameters of the PDEs with remarkable accuracy, the choice of regularisation parameters is non-trivial and makes implementation challenging, since the many degrees of freedom complicate choosing a robust parameter set. As shown in Algorithm 1 of~[22], $\lambda_2$ corresponds to the penalty in rPCA -- one must choose $\lambda_2$ carefully, since different values of this parameter can yield unfavourable predictions when used in tandem with PDE-FIND. 


 \begin{figure}[h!]
	\centering
	\label{tricky-lambda}
	\includegraphics[width=\linewidth]{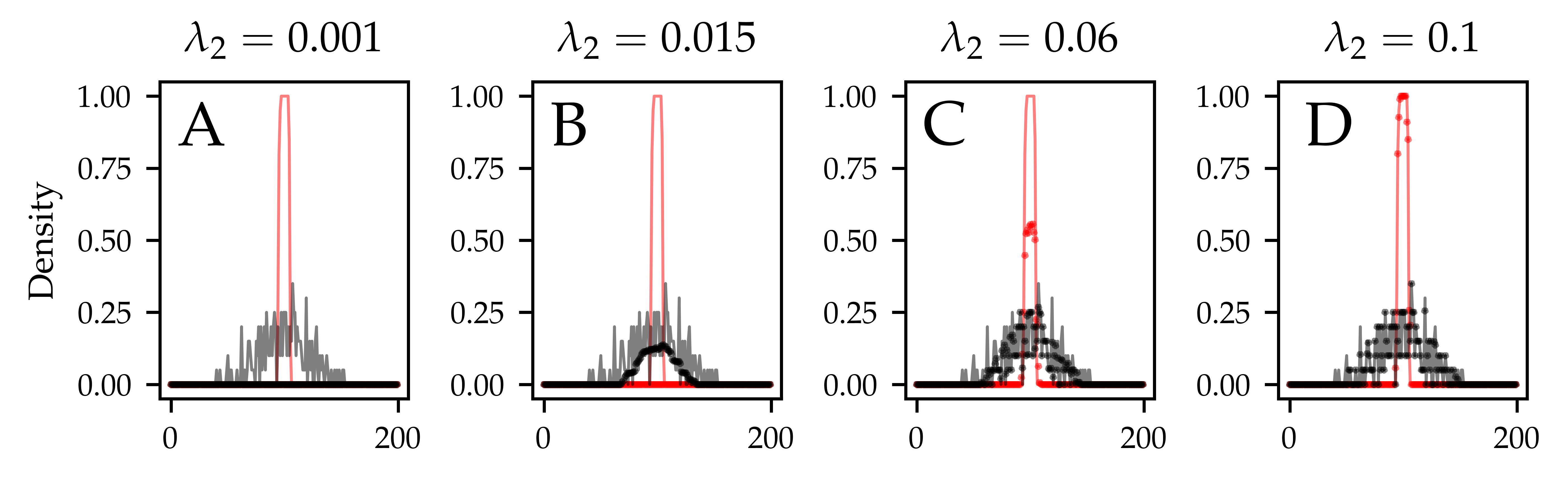}
    \caption{Plots of \emph{denoised} data $Z$ (points) compared to original data (solid lines), plotted at $t=0$ (red) and $t=1000$ (black). A: $\lambda_2 = 0.001$. B: $\lambda_2 = 0.015$. C: $\lambda_2= 0.06$. D: $\lambda_2 = 0.1$. Small values of $\lambda_2$ favour interpreting the entire signal as noise and output $Z=0$. Larger values of $\lambda_2$ do not penalise the noise and return a solution that is identical to the original signal. In the intermediate regime, the order of magnitude of the rPCA output is distorted.}
    \label{bad_rrPCA}
    \end{figure}
    

We start by plotting the output of empirical data processed by rPCA at two different values of $\lambda_2$. As seen in Figure S8, choosing a small value of $\lambda_2$ in rPCA acts in favour of interpreting the entire signal as noise and outputs $Z=0$, effectively rendering the output useless. Larger values of $\lambda_2$ do not penalise the noise and return a solution that is identical to the original signal, as can be seen in the rightmost plot. In the intermediate regime, the order of magnitude of the rPCA output is distorted, as rPCA interprets the data matrix $U$ as sparse and so the maximum of the rPCA output for the densities at $t=0$ and $t=500$ has comparable order of magnitude, effectively obscuring the biological mechanism at play. However, the behaviour of the rPCA output in the intermediate regime is qualitatively good, so we suggest to scale the rPCA output so that it is on the appropriate order of magnitude. Since the maximum of the density is influenced by noise, it should not be the only factor considered in the scaling. Rather, the mean of the data should be included too. Heuristically, we let $Z$ be the output of rPCA and $U$ the original matrix, such that $Z_i$ and $U_i$ denote the $i$-th rows of $Z$ and $U$, respectively (corresponding to the $i$-th time point). Let $\overline{Z_i}$ and $\overline{U_i}$ be the means of the vectors $Z_i$ and $U_i$, respectively, and $Z_i^{\max}$ and $U_i^{\max}$ be their maxima. Then, we re-scale $Z$ to $\tilde{Z}$ such that 
\begin{equation}
\tilde{Z}_i = \frac{1}{2}\cdot\left(\frac{\overline{U_i}}{\overline{Z_i}} + \frac{U_i^{\max}}{Z_i^{\max}}\right)\cdot Z_i. 
\end{equation}
Focusing on intermediate values of $\lambda_2$, we now plot the rescaled rPCA output compared to the original solutions in Figure S9. Based on the improved results in these plots, we use the rescaled output of the DLrSR algorithm in~[22] on $\mathcal{D}^u_1$ going forward, choosing for rPCA the value $\lambda_2 = 0.035$. 


 \begin{figure}[h!]
	\centering
    \includegraphics[width=\linewidth]{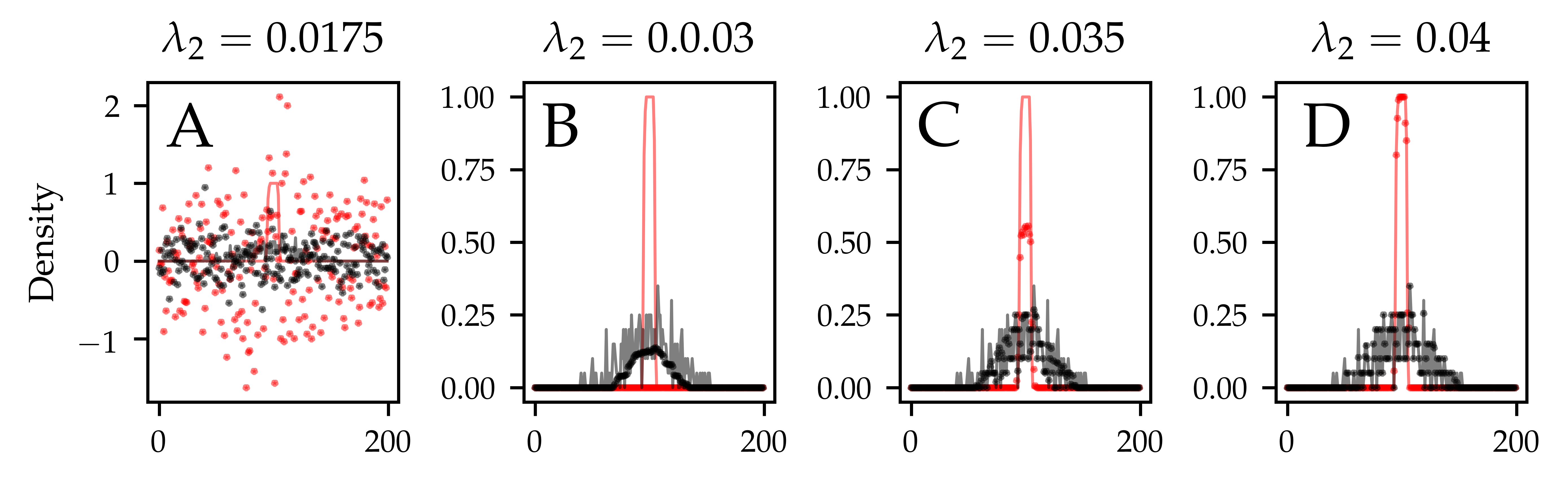}
	\caption{Plots of \emph{rescaled denoised} data, $\tilde{Z}$ (points), compared to original data (solid lines), plotted at $t=0$ (red) and $t=1000$ (black), chosen at values of $\lambda_2$ in the intermediate parameter regime. A: $\lambda_2 = 0.0175$. B: $\lambda_2 = 0.03$. C: $\lambda_2 = 0.035$. D: $\lambda_2 = 0.04$. The order of magnitude of the maximum of the rPCA output is no longer distorted.}
	\label{rescaled_rrPCA}
\end{figure}


Table S1 contains the values of the identification coefficients, $a_i$, found with the rPCA method for Case I, unbiased motility only, using the dataset $\mathcal{D}_\text{I}^1$. It can be seen that no terms in the PDE are conclusively identified, as all the $a_i$ are either small or in the intermediate range. Most saliently, rPCA identifies the diffusion coefficient in only $38\%$ of samples in the dataset. Comparing the rPCA-generated identification coefficients with the corresponding values generated using naive STRidge reveals that the rPCA method has a much higher variability in the sense of variation in the terms that are included and excluded in the resulting PDE model. In addition, Figure S10 shows that the support of the PDE-FIND predictions for the diffusion coefficient obtained using DLrSR on $\mathcal{D}_\text{I}^1$ is similar in size to that obtained with Naive STRidge. This means that rPCA yields estimates for the diffusion coefficient that are no less variable than the original PDE-FIND method. Finally, Figure S10 shows that with rPCA the coefficients are non-trivially correlated. This correlation plot also highlights once again the problem with taking point estimates of the coefficients. For Case I the predicted coefficients for the terms $u_{xx}$ and $uu_{xx}$ are contained in a region of parameter space such that one or the other of the two terms is discovered, or \emph{both} terms are picked up and are not small. This implies that any point estimate from a single data sample is unreliable, and careful consideration needs to be given to the correlations between terms. In summary, we conclude that even when using a state-of-the-art method of denoising observations it is important to statistically quantify uncertainty in the learned coefficients.


\begin{figure}[h!]
	\centering
	\includegraphics[width=\linewidth]{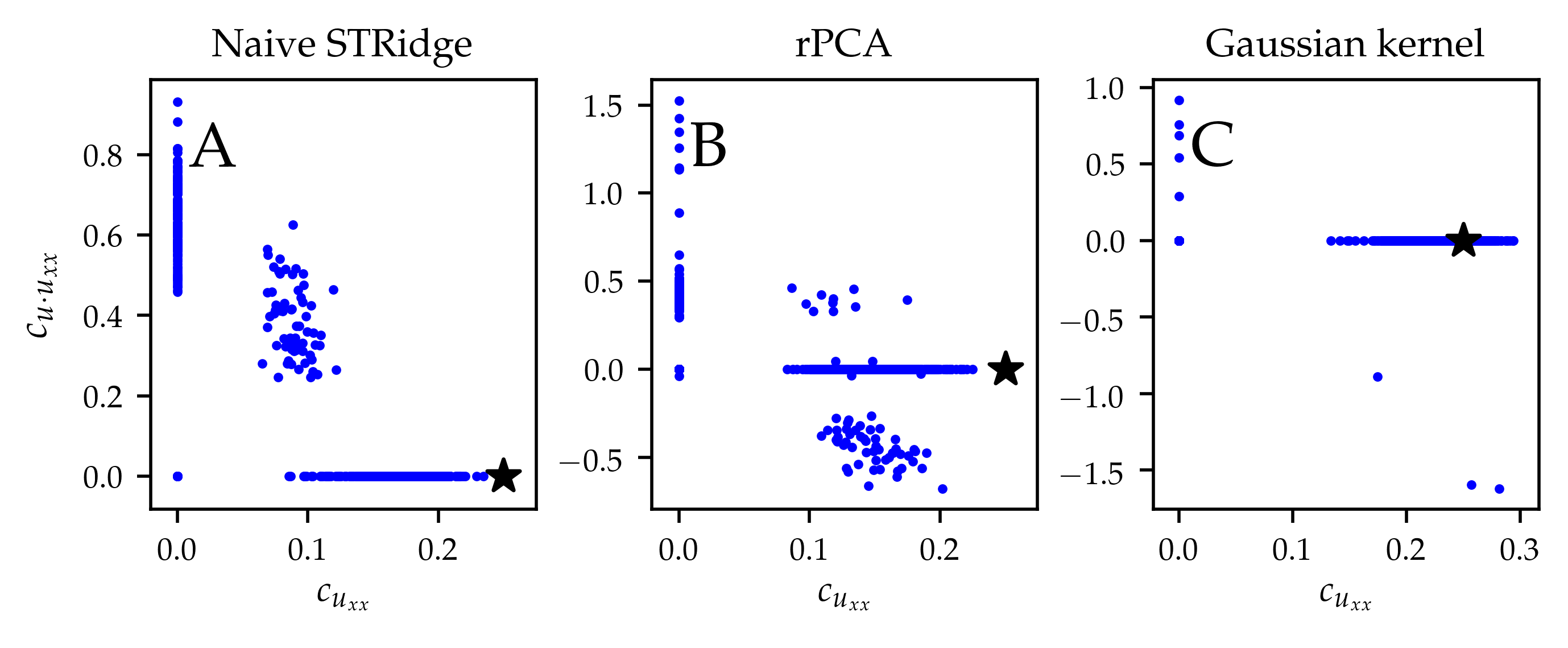}
	\caption{Joint distribution of the $c_{u_{xx}}$ and $c_{u\cdot{u_{xx}}}$ coefficients using PDE-FIND on $\mathcal{D}_\text{I}^1$ (un-biased motion with no proliferation) where each dot is the result from a single sample. The location of the true parameter value is indicated by a black star. A: naive application of STRidge on the ABM data. B: ABM data is pre-processed with rPCA as detailed in Section~\ref{section:rPCA} before the application of PDE-FIND. C: ABM data is pre-processed by smoothing with a Gaussian kernel with $\sigma^2 =2$ before the application of PDE-FIND.}
	\label{correlations}
\end{figure}

\bigskip


\subsection{Convolution with Gaussian kernel: an alternative approach to denoising}

In many applications, noisy data can be successfully smoothed using a suitable kernel. We choose to smooth our data using a Gaussian kernel with standard deviation $\sigma=2$, which is roughly equivalent to the scale used in the spatial discretisation. Summary values for the identification coefficients, $a_i$, are found in Table S1. While the convolution approach is highly successful in recovering one of the relevant terms in the dynamical system, for example the diffusion term in $\mathcal{D}_\text{I}^1$, the bias term in $\mathcal{D}_\text{II}^1$ and the proliferation term in $\mathcal{D}_\text{III}^1$, the method fails to recover the diffusion term regularly in both $\mathcal{D}_\text{II}^1$ and $\mathcal{D}_\text{III}^1$. As a result, we conclude that smoothing using a Gaussian kernel does not result in marked improvements of the method.

\pagebreak


\section{Tuning of the regularisation parameter}\label{SIsection:tuning}

Following Nardini et al. [8], we implement cross-validation to find the optimal value of the regularization parameter in the STRidge algorithm for dataset $\mathcal{D}_\text{I}^1$. As training data, we randomly select ten samples from the ensemble. For each $\lambda \in \{0.05j \,:\, 0\leq j\leq 50\}$, we perform leave-one-out cross validation, that is, we apply PDE-FIND to the left-out data sample, and compute the $L^2$-loss of the integrated solution at $t=250$ with the other nine randomly selected samples. This gives an optimal value of $\lambda=0.5$. To illustrate the role of varying the tuning parameter, Table \ref{CV_table} shows the values of the identification coefficients, $a_i$, at different values of $\lambda$. For these values, Figure 6 of the main text reports the empirical distribution of the PDE-FIND coefficients. Note that $\lambda =0.01$ is the valued used in the initial exploration detailed in the main text.


\bigskip
\begin{table}[h!]
	\centering
	\begin{tabular}{|c|| c| c| c| c| c| c| c| c| c|} 
		\hline
		Value of $\lambda$ & $c_{1}$ & $c_{u}$ & $c_{u^2}$ & $c_{u_x}$ & $c_{u\cdot u_x}$ & $c_{u^2\cdot u_{xx}}$ & $c_{u_{xx}}$ & $c_{u\cdot u_{xx}}$ & $c_{u^2\cdot u_{xx}}$\\ [0.5ex] 
		\hline\hline
		0.01& 0.001 & 0.0 & 0.002 & 0.0 & 0.008 & 0.008 & \cellcolor[gray]{0.7}\textbf{0.826} & 0.199 & 0.05\\
		\hline
		0.05 & 0.0 & 0.0 & 0.011 & 0.001 & 0.01 & 0.019 & \cellcolor[gray]{0.7}\textbf{0.996} & 0.34 & 0.088\\
		\hline
		0.1 & 0.0 & 0.0 & 0.003 & 0.001 & 0.002 & 0.007 & \cellcolor[gray]{0.7}\textbf{0.998} & 0.365 & 0.063\\
		\hline
		0.5 & 0.0 & 0.0 & 0.0 & 0.0 & 0.0 & 0.0 & \cellcolor[gray]{0.7}\textbf{0.996} & 0.098 & 0.013 \\
		\hline\hline
	\end{tabular}
	\caption{Values of the identification parameters, $a_i$, as the parameter $\lambda$ is varied. Coefficients that appear in the corresponding coarse-grained PDEs are highlighted in grey and using \textbf{bold} font.}
	\label{CV_table}
\end{table}


\pagebreak

\section{Comparison of model predictions}\label{SIsection:comparisons}

\begin{figure}[h!]
	\centering
	\includegraphics[width=\linewidth]{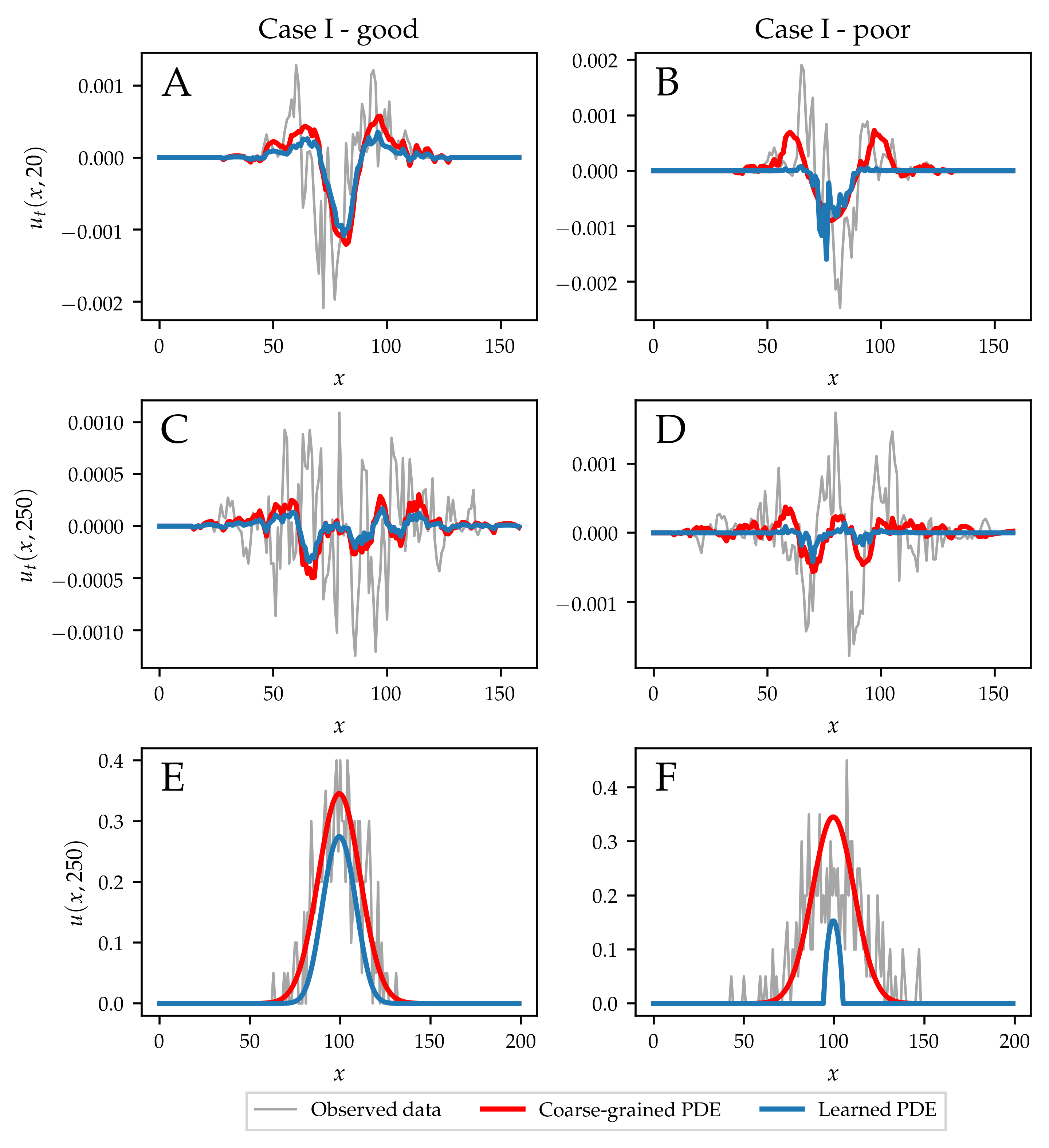}
	\caption{Case I: comparisons of the observed time derivative at $t=20$ (top row) and $t=250$ (middle row) and predicted solutions (bottom row). Left column: mis-specified model with $c_{u_{xx}} = 0.104$ and $c_{u\cdot{u_{xx}}}=0.26$ (blue) together with ground truth (red), compared to observed data (grey). Right column: mis-specified model with $c_{u_{xx}}=0.0$ and $c_{u^2\cdot{u_{xx}}} =1.52$ (blue) together with ground truth (red), compared to observed data (grey).}
	\label{dcomparisons_a}
\end{figure}

\begin{figure}[h!]
	\centering
    \includegraphics[width=\linewidth]{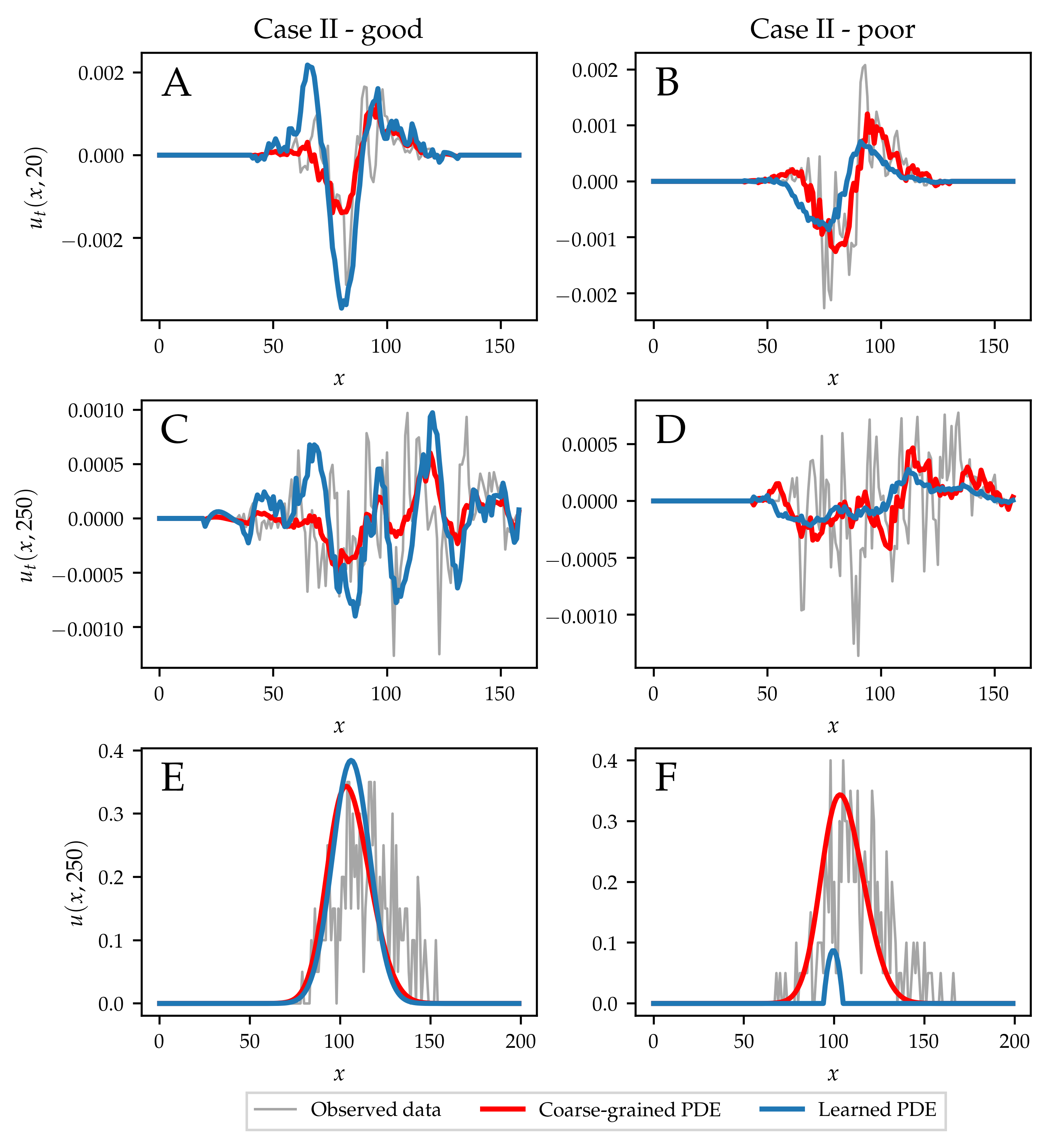}
	\caption{Case II: comparisons of the observed time derivative at $t=20$ (top row) and $t=250$ (middle row) and predicted solutions (bottom row). Left column: mis-specified model with $c_{u_{xx}}=0.20$ and $c_{u_x} =0.026$. Right column: mis-specified model with $c_{u_{xx}}=0.0$ and $c_{u\cdot{u_{xx}}}=0.64$ (blue) together with ground truth (red), compared to observed data (grey).}
	\label{dcomparisons_b}
\end{figure}

\begin{figure}[h!]
	\centering
    \includegraphics[width=\linewidth]{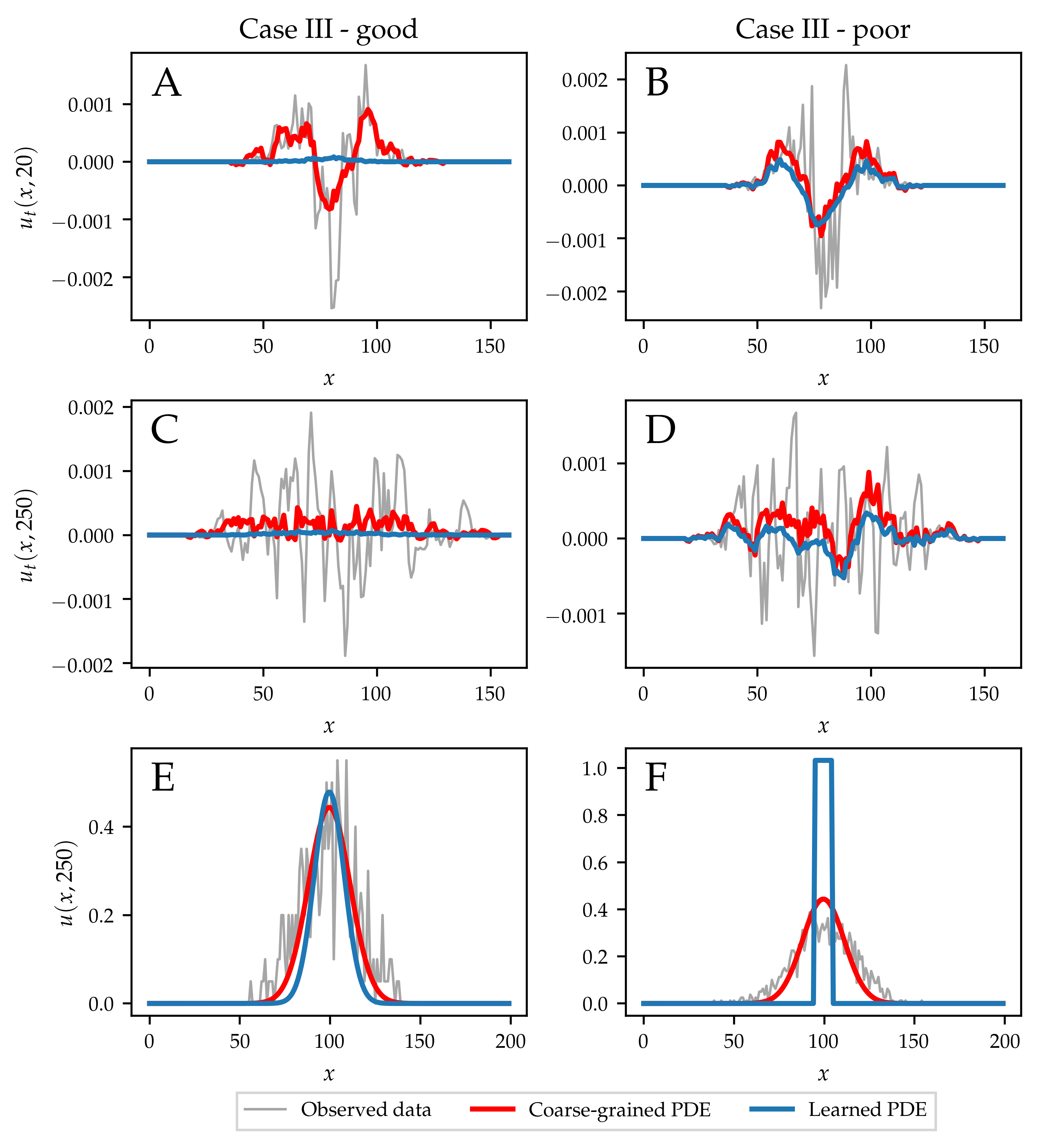}
	\caption{Case III: comparisons of the observed time derivative at $t=20$ (top row) and $t=250$ (middle row) and predicted solutions (bottom row). Left column: mis-specified model with $c_{u_{xx}}=0$ and $c_{u}=0.0013$. Right column: mis-specified model with $c_{u_{xx}}=0.12$ and $c_u = 0.0$ (blue) together with ground truth (red), compared to observed data (grey).}
	\label{dcomparisons_c}
\end{figure}

For noisy data, different linear combinations of the spatial derivatives best match the temporal derivative from different samples when measured according to the $L^2$-loss, and this can have a significant impact on the density profile predicted by the learned equation. In Figures~\ref{dcomparisons_a}--\ref{dcomparisons_c} we demonstrate this for each of Case I (using $\mathcal{D}_\text{I}^1$, Figure~\ref{dcomparisons_a}), Case II (using $\mathcal{D}_\text{II}^1$, Figure~\ref{dcomparisons_b}) and Case 3 (using $\mathcal{D}_\text{III}^1$, Figure~\ref{dcomparisons_c}). In the left-hand column of each figure we show results when the PDE-FIND algorithm accurately predicts evolution of the population density, and in the right-hand column of each figure we show results when PDE-FIND fails to accurately predict evolution of the population density. The top two rows show plots of the temporal derivative estimated from single data samples, the temporal derivative estimated assuming the relevant coarse-grained PDE model in Table 2 of the main text, and the temporal derivative estimated using the (mis-specified) coefficients established via application of PDE-FIND to the data sample. In each case, the top row shows results at $t=20$ and the middle row shows results at $t=250$. The bottom row of each figure shows the density profiles computed via numerical solution of the corresponding PDEs at $t=250$. In five of the six cases, the temporal derivatives estimated assuming the (mis-specified) PDE-FIND coefficients reproduce the observed temporal derivative qualitatively. However, Figures~\ref{dcomparisons_a}--\ref{dcomparisons_c} show that in many cases, these mis-specified PDE coefficients do not yield predictions that are in close agreement with the observed density data. 


\clearpage


\section{Additional information on noisy reference data}

\begin{figure}[h!]
	\centering
	\includegraphics[width=\linewidth]{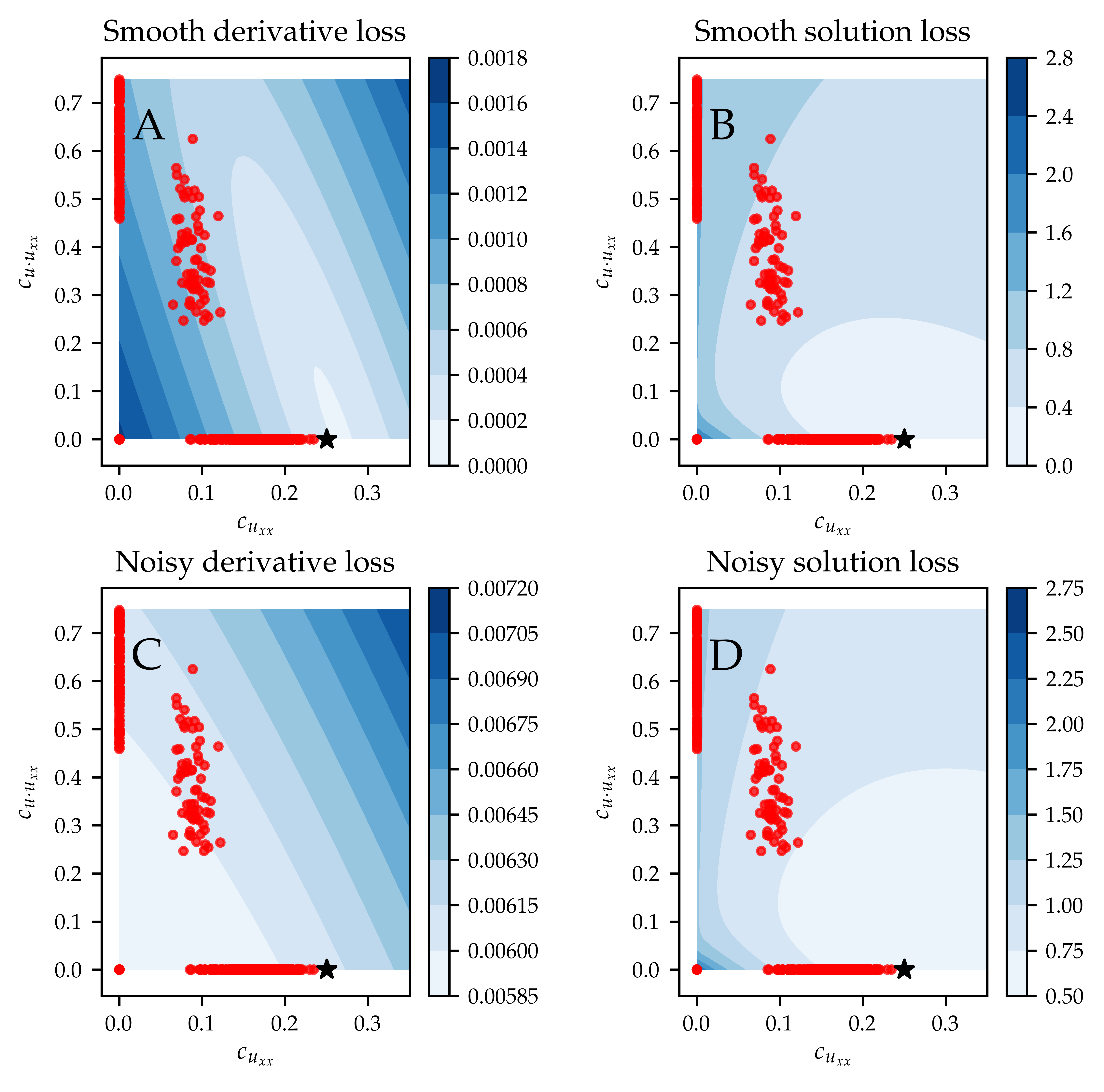}
	\caption{Heatmap showing $L^2$-error for each pair of $c_{u\cdot{u_{xx}}}$ and $c_{u_{xx}}$, as well as the PDE-FIND predictions. A: $L^2$-loss between observed derivative when averaged over all samples in the data set and fitted value of the derivative. B: $L^2$-loss between observed solution when averaged over all samples in data set and fitted model prediction of solution. C: comparison as in A, with respect to a single data sample. D: comparison as in B, with respect to a single data sample.}
	\label{loss_landscape_SI}
\end{figure}

In Section (c) of the main text we identify significant differences between the loss landscapes of the different error metrics for the computed derivatives and spatial solutions of the PDEs. Here, we show that we may also take as a reference point a single, noisy, observation instead of the average of the observations in the ensemble, noting that the quality of the resulting posterior is decreased. As in the main text, we take the dataset $\mathcal{D}_\text{I}^1$ which consists of unbiased motility and no proliferation, and each sample in the dataset consists of an average over $K=1$ simulations from the ABM, and we plot the different loss landscapes in Figure~\ref{loss_landscape_SI}. In the top row we plot results when we average over all $N_s=1000$ samples in the dataset, whereas in the bottom row we plot results when we select a single sample at random from the data set.

The large amount of noise in the single data sample case means that the loss landscape when considering the derivative does not have a minimum around the true parameter value, implying that minimising with respect to this norm does not guarantee convergence to the correct solution. This is very different to the case of low noise, where the loss landscape has a minimum around the true parameter value. On the other hand, the fit of the integrated solutions is a much more consistent summary statistic in the case of a single, highly noisy data sample, and retrieves low measurement error in a region around the true parameter value. When the loss is considered on the averaged data, this region is smaller, and the range of the error is bigger, suggesting that the averaged-data loss function is better able to distinguish between solutions. 


\clearpage


\section{Intractability of the likelihood function}\label{SIsection:likelihood}
In this section, we aim to provide further motivation for the choice of a likelihood free method instead of a classical likelihood based approach. Recall that, in Bayesian statistics, the likelihood $P(\mathcal{D}_\text{obs}\,|\,\boldsymbol{\theta})$ defines the probability density of the observations $\mathcal{D}_{\text{obs}}$ given the model parameters $\boldsymbol{\theta}$. In this context, the observed data $\{U(x,t)\}$ at space points $x = x_1, x_2, \dots, x_N$ and time points $t = t_1, t_2, \dots, t_N$ are obtained from a stochastic ABM. The solution $u(x,t;\boldsymbol{\theta})$ of the PDE model is an approximation of the mean of the ABM data, \textit{i.e.}: 
\begin{equation}
    u(x,t;\boldsymbol{\theta}) \approx \mathbb{E}_{\boldsymbol{\theta}}[(U(x,t)].
\end{equation} 
To define a classical likelihood, one would need to first assume that the mean of the ABM data is exactly given by the PDE solution and describe the distribution of ABM outputs around the PDE model mean by 
\begin{equation}
    U(x_i,t_i) = u(x_i,t_i;\boldsymbol{\theta}) + \epsilon_i,
\end{equation}
where $\epsilon_i$ is a random variable with zero mean (not necessarily identically distributed, but generally assumed independent). One can then obtain a likelihood for the observed data by prescribing a probability distribution for the $\epsilon_i$, which would yield a likelihood for $U(x,t)$ given $u(x,t;\boldsymbol{\theta})$. However, for a general ABM, the distribution for the deviation from the mean (\textit{i.e.} the distribution of the $\epsilon_i$) is unknown. In some cases, one might choose to make a simplifying assumption on the $\epsilon_i$, such as a Gaussian approximation. This may be appropriate when one is familiar with the noise process and the distribution of the data is approximately Gaussian, for example in the case of a very large number of samples, where the Central Limit Theorem can be invoked. However, in the small data limit considered in EQL applications, the assumption that the $\epsilon_i$ are Gaussian is unreasonable. It will depend on the details of the ABM as to the extent to which individual realisations vary from their mean, which is for the purposes of inference, unknown. As we prefer to avoid placing unnecessary assumptions on the process, in the form of assumptions on the $\epsilon_i$, we opt instead for a likelihood-free approach. We additionally stress that classical likelihood models, when considering a likelihood for general PDE data, employ the assumption that the errors $\epsilon_i$ are independent in space and time, and so the joint likelihood of the data given the model,  
\begin{equation}
    P(U(x_1,t_1), \dots, U(x_N,t_N)\vert u(x_1,t_1;\boldsymbol{\theta}),\dots, u(x_N,t_N;\boldsymbol{\theta})),
\end{equation}
can be decomposed as 
\begin{equation}
    P(U(x_1,t_1), \dots, U(x_N,t_N)\vert u(x_1,t_1;\boldsymbol{\theta}),\dots, u(x_N,t_N;\boldsymbol{\theta})) = \prod_i P(U(x_i,t_i)\vert u(x_i,t_i;\boldsymbol{\theta})).
\end{equation}
Since individual realisations of the ABM may have deviations from the mean that have spatial and temporal structure, it is inappropriate to assume that the $\epsilon_i$ are independent. To summarise, a classical likelihood approach can be used, but it requires the introduction of nontrivial additional assumptions and so it will potentially not target the true posterior. We prefer not to take this approach, and rather opt for a method that has been developed specifically to deal with cases where the likelihood is mathematically intractable. 


\section{ABC-rejection implementation}\label{SIsection:ABCrejection}

ABC-rejection can be retrieved from Algorithm 1 by choosing the importance distribution $q$ to be equal to the prior distribution $\pi(\boldsymbol{\theta})$. 

\bigskip


\begin{algorithm}[H]
\label{Algo:ABC-IS}
\SetAlgoLined
\SetAlgoNoLine
\medskip
\KwIn{Data $y_{\text{obs}}$; tolerance $\varepsilon$; prior distribution $\pi(\boldsymbol{\theta})$; importance distribution $\hat{q}(\boldsymbol{\theta})$ proportional to $q(\boldsymbol{\theta})$; model $f(\bullet\,|\,\boldsymbol{\theta})$; stopping condition $S$}
\KwOut{Weighted samples $\{\boldsymbol{\theta}_n,w_n\}_{n=1}^{N}$.}
Set $n=0$. \\
\Repeat{S \text{is true}}{
    Increment $n\leftarrow n+1$\;
    Generate $\boldsymbol{\theta}_n \sim \hat{q}(\boldsymbol{\theta})$\;
    Simulate $y_{\text{sim}} \sim f\left(\bullet\,|\,\boldsymbol{\theta}_n\right)$\;
    Set $w_n = \left[\pi(\boldsymbol{\theta}_n)/q(\boldsymbol{\theta}_n)\right]\cdot \mathbb{I}(d\left(y_{\text{sim}},y_{\text{obs}}\right)< \varepsilon)$
}
Set $N=n$.
 \caption{Importance sampling ABC}
\end{algorithm}


\bigskip
We choose the tolerance, $\varepsilon$, as $0.3$ for Case I and $0.8$ for Case II and Case III, as detailed in the main text. For each posterior distribution, we set the stopping condition to 1500 accepted parameters. 

\pagebreak

\section{Additional experiments} \label{SIsection:additionalexperiments}
In this section we report the hyperparameters used for the various spike-and-slab models employed in the paper and the settings for the uniform prior over all library terms. Recall that to tune the hyperparameters, we have used the exploration subset of the data. In all cases, the mixture parameters in the spike-and-slab models were taken as the previously found identification ratios. Recall that the values for $\alpha, \beta$ in each hypervariance distribution are chosen such that the expected variance of the slab is of the same order of magnitude as the empirical variance of the sample. 
\subsection{Informed spike-and-slab}
In Case I, we use $\mu = 0.17$, $\alpha=3$ and $\beta=78$ for the term $u_{xx}$ and $\mu = 0.64$, $\alpha=3$ and $\beta=78$ for the term $uu_{xx}$. In Case II, we use $\mu = 0.17$, $\alpha=3$, $\beta=78$ for the term $u_{xx}$ and $\mu=-0.03$, $\alpha=3, \beta=5\cdot 10^3$ for the term $u_x$. In Case III, we use $\mu=0.0008, \alpha=3, \beta = 5\cdot 10^6$ for the term $u$, $\mu=-0.00075, \alpha=3, \beta = 5\cdot10^6$ for the term $u^2$ and $\mu = 0.17$, $\alpha=3$ and $\beta=78$ for the term $u_{xx}$.
\subsection{Naive spike-and-slab}
In Case I, we use $\mu=0$, $\alpha=3, \beta=2$ for the term $u_{xx}$ and $\mu=0$, $\alpha=3, \beta=2$ for the term $uu_{xx}$. In Case II, we use $\mu = 0$, $\alpha=3$, $\beta=2$ for the term $u_{xx}$ and $\mu=0$, $\alpha=3, \beta=50$ for the term $u_x$. In Case III, we use $\mu=0, \alpha=3, \beta = 5\cdot 10^6$ for the term $u$, $\mu=0, \alpha=3, \beta = 5\cdot10^6$ for the term $u^2$ and $\mu = 0.17$, $\alpha=3$ and $\beta=2$ for the term $u_{xx}$.
\subsection{Uniform prior on all parameters} 
We report the prior used to compare the performance of Bayes-PDEFIND against using a uniform prior over all parameters. We choose the prior $c_1 \sim \mathcal{U}(-0.05,0.05)$, $c_u\sim\mathcal{U}(-0.05,0.05)$, $c_{u^2} \sim\mathcal{U}(-0.05,0.05)$, $c_{u_x} \sim\mathcal{U}(-0.05,0.05)$, $c_{uu_x} \sim \mathcal{U}(-0.05,0.05)$, $c_{u^2u_x} \sim \mathcal{U}(-0.05,0.05)$, $c_{u_{xx}} \sim\mathcal{U}(-0.5,0.5)$, $c_{uu_{xx}} \sim \mathcal{U}(-0.5,0.5)$, $c_{u^2u_{xx}} \sim \mathcal{U}(-0.5,0.5)$.

\pagebreak

\section{Posterior distributions for Case III}

\begin{figure}[htbp]
	\centering
	\includegraphics[width=\linewidth]{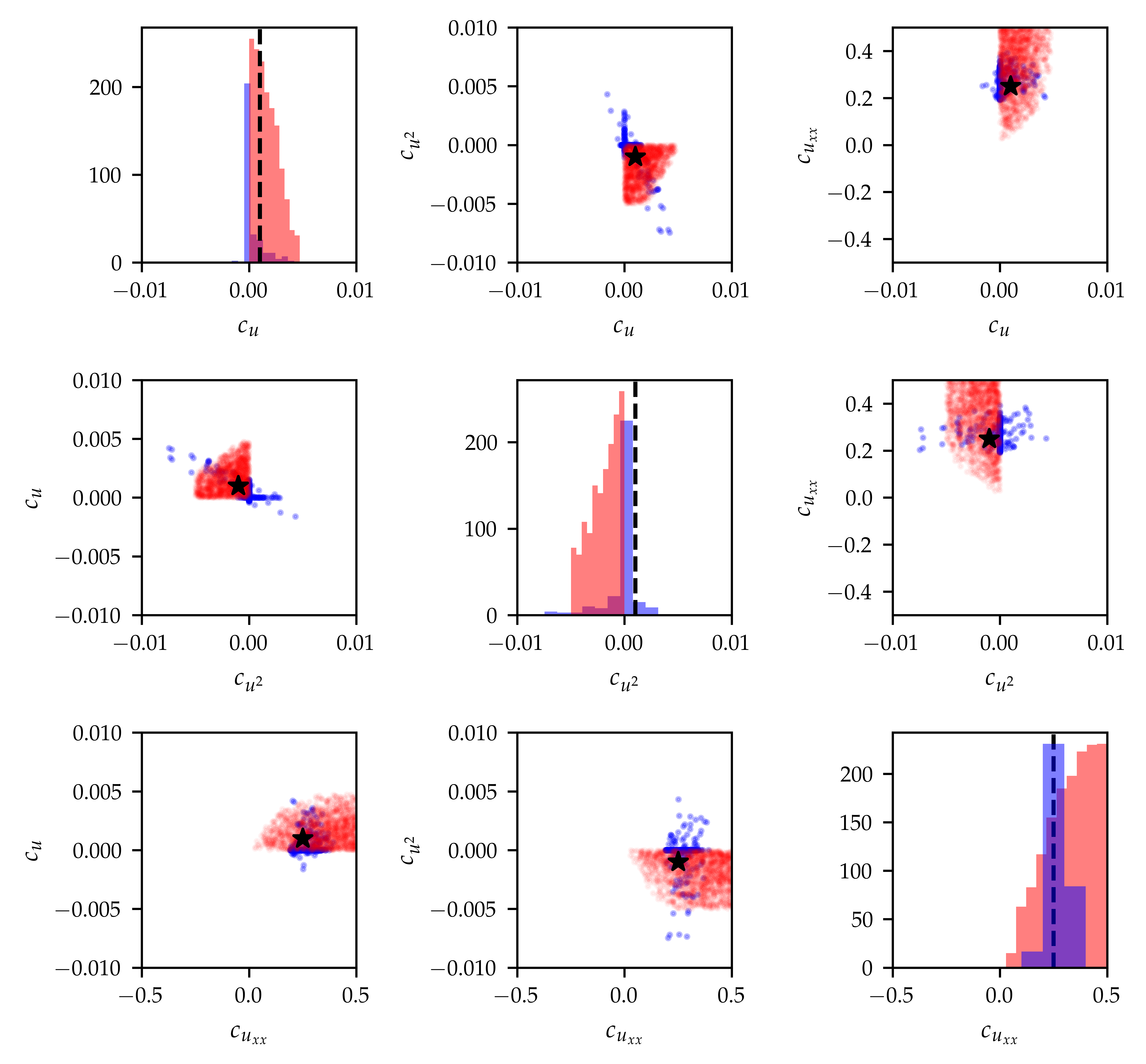}
	\caption{Plot matrix of marginal posterior densities of $c_u$, $c_{u^2}$ and $c_{u_{xx}}$ obtained from ABC-rejection sampling using the spike-and-slab prior (blue) and a uniform prior (red), compared to the true parameter values (black stars / black dashed lines).}
	\label{Figure:CaseIIIposterior_full}
\end{figure}

In Figure~\ref{Figure:CaseIIIposterior_full} we report in plot matrix format the posterior distributions found for the parameters in Case III. 


\clearpage


\bibliography{BIBL}
\bibliographystyle{unsrt}
